\documentclass[%
 preprint,
 amsmath,amssymb,
 aps, physrev,
]{revtex4-2}

\usepackage{graphicx}% Include figure files
\usepackage{dcolumn}% Align table columns on decimal point
\usepackage{bm}% bold math
\usepackage{hyperref}% add hypertext capabilities
\hypersetup{
    colorlinks=true,      % Enable color for links
    linkcolor=blue,       % Color for internal links (e.g., table of contents)
    citecolor=blue,        % Color for citation links
    filecolor=magenta,    % Color for file links
    urlcolor=blue         % Color for external links
}

\usepackage{siunitx}
\renewcommand{\selectlanguage}[1]{}

\begin{document}

% \preprint{APS/123-QED}

%\title{\textbf{Jeans instability in nonthermal EiBI-gravitating magnetized astroclouds.} 
% }% 

%or

\title{\textbf{Effect of kappa-modified polarization force on Jeans instability in nonthermal EiBI-gravitating dust clouds} 
}% 

\author{Dipankar Ray\href{https://orcid.org/0009-0002-2308-7850}{\includegraphics[scale=0.1]{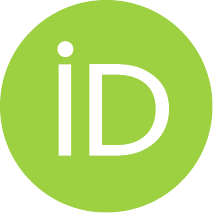}}}
    \email{dray056@tezu.ernet.in}
    \affiliation{Department of Physics, Tezpur Univesity,\\
Napaam, Tezpur, Sonitpur, Assam 784028, India
}

\author{Siddhartha Saikia\href{https://orcid.org/0009-0000-9275-9943}{\includegraphics[scale=0.1]{orcid.pdf}}}
    \email{siddhartha.saikia@unsw.edu.au}
    \affiliation{School of Physics, University of New South Wales,\\ Sydney, NSW 2052, Australia}%

\author{Pralay Kumar Karmakar\href{https://orcid.org/0000-0002-3078-9247}{\includegraphics[scale=0.1]{orcid.pdf}}}
    \email{pkk@tezu.ernet.in}
    \affiliation{Department of Physics, Tezpur Univesity,\\
Napaam, Tezpur, Sonitpur, Assam 784028, India
}
%  \homepage{http://www.Second.institution.edu/~Charlie.Author}
% \affiliation{
%  First affiliation for this author
% }%
% \affiliation{
%  second institution for this author
% \affiliation{%
%  Authors' institution and/or address\\
%  This line break forced with \textbackslash\textbackslash
% }%

% \collaboration{CLEO Collaboration}%\noaffiliation

\date{\today}% It is always \today, today,

\begin{abstract}

A semi-analytic model is developed to study the effects of $\kappa$-distributed lighter constituents and the associated $\kappa$-modified polarization force on the classical Jeans instability in dust molecular clouds (DMCs). The constitutive electrons and ions are considered to follow a nonthermal $\kappa$-velocity distribution law, while the constitutive massive dust grains are treated as the EiBI-gravitating fluids. A linearized quadratic dispersion relation is derived using spherical normal mode analysis. The resulting dispersion relation is analyzed in both the hydrodynamic and kinetic regimes along with their corresponding modified instability criteria. The oscillatory and propagatory mode characteristics are illustratively analyzed. It is seen that the EiBI gravity introduces a new velocity term $(V_\chi)$ in the dispersion relation. In contrast, the nonthermal $\kappa$-distributed constituents significantly enhance the polarization force against their respective Maxwellian counterparts. The $\kappa$-modified polarization force $(F_{p\kappa})$ and the negative EiBI gravity parameter $(-\chi)$ have destabilizing influences unlike the positive EiBI parameter $(+\chi)$. An enhanced polarization interaction parameter $(R_{\kappa})$ and positive EiBI parameter $(+\chi)$ reduce the real normalized frequency $(\Omega_r)$. Consequently, the phase velocity exhibits strong dispersion, increasing with wavenumber until reaching saturation, after which it transitions into a weakly dispersive regime. These findings provide new insights into the formation of smaller astrophysical structures via the non-local Jeans instability in the ultracompact H\textsc{ii} regions of dense DMCs.
% \begin{description}
% Secondary publications and information retrieval purposes.
% You may use the \texttt{description} environment to structure your abstract;
% use the optional argument of the \verb+\item+ command to give the category of each item. 
% \end{description}
\end{abstract}

\maketitle

%\tableofcontents

\section{\label{sec:Introduction}Introduction}
Stars are known to be the fundamental building blocks of galaxies. The study of their formation mechanism provides valuable physical insights about galactic evolution and consequently, of the Universe \cite{peltonen_jwst_2023, gurian_analytic_2025}. The associated complex plasma systems, such as dust molecular clouds (DMCs), serve as ideal sites for astrostructure formation. In recent decades, there has been a significant surge in the interest of studying the DMCs \cite{prajapati_effect_2011, bhakta_effects_2019, kalita_analyzing_2020,dhiman_jeans_2023, sharma_modified_2014,dasgupta_jeans_2019}, vast media containing micron-sized electrically charged dust grains, electrons, ions, and radiation, which are widely distributed throughout the cosmos. Self-gravitating dusty plasmas carry various hydrodynamic instabilities and are subject to the Jeans (gravitational) instability \cite{hopkins_fundamentally_2016}. The Jeans instability, named after Sir Jeans Hopewood James, plays a pivotal role in the formation of planetesimals and stars within interstellar media \cite{eddington_internal_1988,jeans_stability_1901}. This phenomenon takes place when the balance between the inward gravitational force and the outward gas or plasma pressure force is disturbed due to a sudden increase in the size or mass of a DMC beyond the critical Jeans mass-length threshold \cite{jeans_stability_1901}. As a result, this leads to the non-local collapse of large DMCs into smaller cloudlets, paving the way for the initial processes in the formation dynamics of stars and other bounded structures \cite{choudhuri_astrophysics_2010}. For typical DMC parameters, the Jeans length is approximately $10^{14}$ \unit{cm}, significantly larger than the usual length scale of planetesimals (around a few thousand \unit{km}) \cite{gaurav_effect_2018}. Hence, there is a demand for improved theories to elucidate the formation dynamics of such small bounded structures.

The ubiquitous presence of constitutive micron to sub-micron-sized dust particles, such as silicate, graphite, and other intricate derivatives, in molecular clouds has been substantiated by various research studies \cite{pagani_ubiquity_2010,de_marchi_extinction_2014,lefevre_dust_2014,draine_optical_1984}. The evolution of giant DMCs and star formation is profoundly influenced by the presence of these colossal dust grains \cite{soliman_thermodynamics_2024} due to their substantially greater mass and electrical charge compared to the lighter constituents (electrons and ions) \cite{jacobs_linearly_2004, pandey_jeans_1994}. Depending on the dust charge $(q_d)$, dust temperature $(T_d)$, and Wigner-Seitz radius $(a)$, a dusty plasma system behaves as a weakly interacting gas, a strongly coupled liquid, or a solid. This phase transition is governed by the Coulomb coupling parameter $(\Gamma_\text{Cou})$, which represents the ratio between the Coulomb (electrostatic) interaction energy of the neighboring particles and their thermal (random) kinetic energy. It is expressed as $\Gamma_\text{Cou} = q_d^2 / a k_B T_d$ (in cgs units), where $k_B (= 1.38 \times 10^{-16}$ \unit{erg. K^{-1}}) is the Boltzmann constant \cite{shukla_introduction_2002,melzer_physics_2019}. For $\Gamma_\text{Cou} \ll 1$, the plasma system is considered to be in a weakly coupled gaseous phase; whereas, in the range, $ 1 \leq \Gamma_\text{Cou} < \Gamma_c$ ($\Gamma_c \sim 172$ \cite{thomas_plasma_1994}, being the critical value), it remains in a strongly coupled liquid phase with viscoelastic effects. For $\Gamma_\text{Cou} > \Gamma_c$, the system undergos a transition to a strongly coupled solid state or the crystalline state (for $\Gamma_\text{Cou} \gg \Gamma_c$) \cite{shukla_introduction_2002,zobaer_observing_2013}. As a result, the traditional Jeans instability criteria undergo modifications in the presence of such electrically charged dust particles \cite{kalita_analyzing_2020}. Consequently, several modified Jeans instability (Jeans-like) criteria have been developed over time that more effectively elucidate the fragmentation processes in Newtonian self-gravitating DMCs \cite{delzanno_modified_2005,dwivedi_pulsational_1999, lundin_modified_2008,sharma_modified_2014,pandey_jeans_1994, pandey_pulsational_2002}.

Despite its success, the Newtonian gravity and the Einstein’s theory of general relativity (GR) have certain limitations (for more details, see \cite{shankaranarayanan_modified_2022, clifton_modified_2012, sotiriou_fr_2010}). To address these, various authors have explored modified gravity theories in the context of Jeans instability \cite{vainio_jeans_2016, ray_pulsational_2024, de_martino_kinetic_2017, he_jeans_2022, roshan_jeans_2014, das_dynamics_2024, bessiri_jeans_2021, yang_jeans_2023}. These alternative GR theories such as, scalar-tensor theory \cite{brans_machs_1961, faraoni_scalar-tensor_2004}, $f(R)$ gravity \cite{sotiriou_f_2010}, Einstein-Gauss-Bonnet gravity \cite{glavan_einstein-gauss-bonnet_2020}, fourth-order gravity (FOG) \cite{clifton_parameterised_2008}, Eddington-inspired Born-Infeld (EiBI) gravity, etc. \cite{banados_eddingtons_2010} are introduced to address issues related to the rapid cosmic expansion, singularity at the primordial Universe, and the end of stellar life cycle, among other challenges. Among the numerous existing theories of modified gravity, the EiBI gravity (for details, see \cite{pani_eddington-inspired_2012, banados_eddingtons_2010, avelino_eddington-inspired_2012, avelino_eddington-inspired_2012-1, banerjee_stellar_2022} and references therein), introduced by Ba\~nados and Ferreira (2010) \cite{banados_eddingtons_2010}, has successfully predicted singularity-free cosmology \cite{banerjee_stellar_2022} and possesses the capability to avoid singularities arising from gravitational collapse \cite{pani_compact_2011,banerjee_stellar_2022}. Therefore, it emerges as a viable alternative to conventional GR. Accordingly, the EiBI gravity is employed to investigate various astronomical objects, such as cataclysmic variables \cite{banerjee_stellar_2022}, compact objects (neutron stars and white dwarfs) \cite{prasetyo_26text_2021,pani_compact_2011,sham_radial_2012,banerjee_constraints_2017}, Sun and Sun-like objects \cite{casanellas_testing_2012,das_solar_2024}, and so forth. Although the modified gravity effects are more pronounced for strong gravity sources, they often leave their imprint at lower energy scales in the non-relativistic weak-field limit \cite{banerjee_stellar_2022}. The non-relativistic weak-field limit signifies that the source of gravitation is weak and is moving with a speed significantly lower than the speed of light \cite{peron_gravity_2016}. Hence, it becomes very tempting to analyze the stability dynamics of the such self-gravitating DMCs in a non-relativistic weak-field limit of the EiBI gravity.

Numerous satellite measurements and space plasma probe missions have demonstrated that the majority of space plasma particles deviate from the conventional Maxwellian velocity distribution \cite{carovillano_low-energy_1968, livadiotis_first_2011,pierrard_electron_1999,pierrard_kappa_2010, sun_relativistic_2025,li_effect_2025}. Instead, their velocity distribution can be effectively modeled using nonthermal (non-Maxwellian or superthermal) velocity distributions characterized by low-energy cores and high-energy tails. This observation has been corroborated by various studies in the literature that have thermostatically employed the nonthermal kappa $(\kappa)$ velocity distribution function to elucidate the particle distribution in diverse astrophysical regions, including solar winds \cite{pierrard_electron_1999,heerikhuisen_kappadistributed_2015,sarma_solar_2022}, planetary magnetospheres \cite{dialynas_energetic_2009, nicolaou_properties_2014,eyelade_relation_2021}, heliosheath \cite{livadiotis_first_2011, richardson_observations_2022}, nebulae \cite{zhang_electron_2004, nicholls_resolving_2012, zhang_nonthermal_2016,yao_ultraviolet_2022}, and others. Notably, in 2012, Nicholls et al. \cite{nicholls_resolving_2012} conducted a comprehensive analysis of data from H\textsc{ii} regions and planetary nebulae to demonstrate that the $\kappa$ distribution provides significantly more accurate estimates of temperature and metallicity for gaseous nebulae. Introduced by Vasyliunas \cite{carovillano_low-energy_1968} in 1968, the $\kappa$ distribution (a generalized Lorentzian distribution) is characterized by a nonthermal (superthermal) index, $\kappa$, which measures the deviation from the Maxwellian characteristics. Lower the value of $\kappa$, higher is the nonthermality and vice versa. However, in the limit $\kappa \to \infty$, the standard Maxwellian distribution law gets recovered \cite{pierrard_kappa_2010}. Thus, the $\kappa$ distribution provides a more generalized way to describe the space plasma particles that deviates from thermal equilibrium \cite{sethi_effect_2018}.

The deformation of plasma sheaths around charged dust particles often results in the polarization of dust grains \cite{melzer_physics_2019}. In an inhomogeneous plasma, where electron and ion distributions are uneven, this polarization gives rise to polarization force $(F_p)$ that opposes the electrostatic (Coulomb) force exerted by the lighter plasma constituents \cite{hamaguchi_polarization_1994, prajapati_influence_2015, khrapak_influence_2009}. Mathematically, the polarization force can be expressed (for Maxwellian electrons and ions) as $\Vec{F}_p= -q_d^2(\Vec{\nabla}\lambda_D /2\lambda_D^2)$ (in cgs units); where $\lambda_D = \lambda_{Di}/\sqrt{1+ (\lambda_{Di}/\lambda_{De})^2}$ represents the effective plasma Debye length \cite{khrapak_influence_2009, prajapati_effect_2011}. Furthermore, $\lambda_{De(i)} = \sqrt{k_BT_{e(i)} / 4\pi e^2 n_{e(i)}}$ denotes the electron (ion) Debye length, with $T_{e(i)}$ referring to the electron (ion) temperature, $e$ the electron charge unit and $n_{e(i)}$ representing the electron (ion) number density. In a media with negatively charged dust grains, the Debye sheath is usually formed by the ions around the grains. Hence, the mathematical expression for polarization force can be simplified as $\vec{F}_p = - q_d R (n_i/n_{i0})^{1/2}\Vec{\nabla} \phi$, where $R = q_d e/4k_BT_i\lambda_{Di0}$ is the polarization interaction parameter, characterizing the strength of the polarization force. Here, $n_{i0}$ denotes the equilibrium ion number density, and $\phi$ represents the electrostatic potential \cite{asaduzzaman_effects_2012}.

Although several investigations have previously been reported on the polarization force with the Maxwellian lighter constitutive species \cite{bentabet_generalized_2017,dolai_effects_2020,prajapati_effect_2011,prajapati_influence_2015,asaduzzaman_effects_2012,sharma_modified_2014,asaduzzaman_dust-acoustic_2011,khrapak_influence_2009,hamaguchi_polarization_1994}, the nonthermal effects on the said polarization force in astrophysical star-forming environments have been poorly addressed so far in the literature. It is well known that the polarization force undergoes substantial modifications due to the presence of nonthermal electrons (ions) depending on the dust charge and hence, the associated stability dynamics \cite{zobaer_observing_2013,ellabany_effects_2020,li_effect_2025,li_transport-driven_2024,sethi_effect_2018,khaled_modulational_2021}. In a nonthermal plasma system, the constituent electron and ion population densities get significantly modified due to the superthermal nature of the lighter constituents which directly affect the shielding around the dust grains. Therefore, there has been a great need of a refined expression for the polarization force in the presence of all the possible realistic nonthermal effects for a long period of time. The corresponding modified polarization force expression can be derived in the usual symbolism as, $F_{p\kappa}= -z_d e R_\kappa (n_i/n_{i0})^{1/2}(1 - e \phi / (\kappa -3/2) k_B T_i) \vec{\nabla} \phi$, where $R_\kappa = \sigma_k (z_d e^2/4 \lambda_{Di0}k_B T_i)$ is the $\kappa$-modified polarization interaction parameter and $\sigma_\kappa = (\kappa - 1/2)/(\kappa - 3/2)$ is a dimensionless parameter (derivation in Appendix~\ref{sec:appendix}). Clearly, as $\kappa \to \infty$, the usual Maxwellian polarization force $(F_p)$ can be recovered from the above.

By considering all the key factors described above, we investigate the effect of the $\kappa$-modified polarization force on the Jeans instability of a strongly coupled magneto-turbulent DMC in the EiBI gravity framework. Here, we show that these novel key mechanisms combined together may significantly reduce the Jeans length so that smaller bounded structures can form. After an introduction in Sec.~\ref{sec:Introduction}, the paper is arranged in a normal format with Sec.~\ref{sec: Model formalism} explaining the physical DMC model along with its properties. Sec.~\ref{sec: Dispersion properties} outlines the mathematical method for finding out the generalized dispersion relation. It also dwells into the newly found modified Jeans instability criteria in both the hydrodynamic and kinetic regimes. Sec.~\ref{sec:Results and discussions} depicts the main outcomes of our investigation illustratively. Finally, all the derived results along with the future scope and applicability are summarized in Sec.~\ref{sec: Conclusion}.

\section{\label{sec: Model formalism} Physical model formalism}

To study the conjoint effects of several critical factors, such as EiBI gravity, kappa $(\kappa)$ velocity distribution and its corresponding $\kappa$-modified polarization force on the Jeans instability, a viscoelastic magneto-turbulent DMC is considered. It is assumed to be spherical and inhomogeneous in nature, characterized by varying electron and ion densities. The self-gravitating multi-component DMC consists of two nonthermal species: electrons and singly-ionized positive ions along with thermal negatively charged dust grains. The spherical dust grains, having equal radius $(r_d)$, mass $(m_d)$, and surface charge $(q_d)$ are placed in a uniform magnetic field $\vec{B}\ (0, 0, B_{\phi})$ generated by the movement of the charged background medium. These micron to sub-micron-sized grains, substantially more massive and higher electric charge than the other lighter constituents, are strongly correlated with each other because of their larger electric fields and lower temperatures. When modeling a multi-component system, it is neither worthwhile nor possible to track the orbits of each particle individually. Hence, the dust grains, with relatively low thermal energy, are treated as a fluid described by generalized hydrodynamic equations. However, the lighter nonthermal species (electrons and ions) are assumed to follow the $\kappa$ distribution, exhibiting weak coupling due to their smaller electric charges and higher temperatures.
 
The superthermal isotropic, $\kappa$ distribution function in a defined velocity space is expressed in generic notations \cite{mahmood_existence_2023} as
\begin{eqnarray}
    f_{\kappa}(\vec{v}) = &\frac{n_{j0}}{(\pi\kappa\theta^2_{eff})^{\frac{3}{2}}}\frac{\Gamma(\kappa+1)}{\Gamma(\kappa-1/2)}\Bigg(1+\frac{\vec{v}_j^ {\hspace{0.1 em} 2}}{\kappa\theta^2_{eff}} \Bigg)^{-(\kappa+1)}; \label{eq:1} \\
    \theta_{eff} = &\sqrt{\left(\frac{\kappa-3/2}{\kappa}\right)\left(\frac{2k_BT_j}{m_j}\right)}, \label{eq:2}
\end{eqnarray}
    
% \begin{eqnarray}\label{eq:1}
%     f_{\kappa}(\vec{v})=\frac{n_{j0}}{(\pi\kappa\theta^2_{eff})^{\frac{3}{2}}}\frac{\Gamma(\kappa+1)}{\Gamma(\kappa-1/2)}\Bigg(1+\frac{\vec{v}_j^ {\hspace{0.1 em} 2}}{\kappa\theta^2_{eff}} \Bigg)^{-(\kappa+1)};
% \end{eqnarray}

% \begin{eqnarray}\label{eq:2}
%     \theta_{eff}=\sqrt{\left(\frac{\kappa-3/2}{\kappa}\right)\left(\frac{2k_BT_j}{m_j}\right)},
% \end{eqnarray}

where, $\theta_{eff}$ represents the most probable (characteristic) speed, with $n_{j0}\ (j=e,i)$ being the equilibrium number density of the species. The symbols $\vec{v}_j$, $m_j$, and $T_j$ respectively refer to the velocity, mass, and temperature of the particle species. Furthermore, $\Gamma$ denotes the gamma function and $\kappa$ is the nonthermal spectral index that governs the effect of kappa velocity distribution. The spectral index, $\kappa$, provides the measure of deviation of the stationary states from thermal equilibrium \cite{nicholls_resolving_2012, livadiotis_kappa_2015}. It is important to note that $\kappa$ must take values greater than 3/2, as for $\kappa = 3/2$, the most probable velocity (Eq.~\ref{eq:2}) becomes zero, consequently making the distribution function (Eq.~\ref{eq:1}) infinite \cite{pierrard_kappa_2010}. For a very large value of $\kappa\ (\to \infty)$, the $\kappa$ velocity distribution converges to the Maxwellian velocity distribution \cite{eyelade_relation_2021}. However, low $\kappa$-values represent a group of particles (nonthermal) whose speed exceeds the thermal regime. These high velocity particles that form the tail section of the $\kappa$ distribution are often termed as superthermal particles \cite{hakimi_pajouh_influence_2016}. 

The number densities of the $\kappa$-distributed electrons and ions in the non-relativistic regime is obtained by integrating Eq.~\eqref{eq:1} over a velocity space in usual notations \cite{hakimi_pajouh_influence_2016, livadiotis_kappa_2015} respectively as
\begin{eqnarray}
        n_e =& n_{e0}\left[1 - \frac{e\phi}{k_B T_e(\kappa-3/2)}\right]^{-(\kappa - 1/2)}, \label{eq:3}\\
        n_i =& n_{i0}\left[1 + \frac{e\phi}{k_B T_i(\kappa-3/2)}\right]^{-(\kappa - 1/2)}. \label{eq:4}
\end{eqnarray}

% \begin{eqnarray}\label{eq:3}
%     n_e = n_{e0}\left[1 - \frac{e\phi}{k_B T_e(\kappa-3/2)}\right]^{-(\kappa - 1/2)},
% \end{eqnarray}

% \begin{eqnarray}\label{eq:4}
%     n_i = n_{i0}\left[1 + \frac{e\phi}{k_B T_i(\kappa-3/2)}\right]^{-(\kappa - 1/2)}.
% \end{eqnarray}
Here, $\phi$ refers to the electrostatic potential and $e$ denotes the electronic charge. In the limiting condition $\kappa \to \infty$, Eq.~\eqref{eq:3} and Eq.~\eqref{eq:4} approaches to the classical Maxwellian number densities of the respective species.

 The DMC mass conservation can be described with the continuity equation in usual notations as 
\begin{eqnarray}
    \partial_t n_d + \vec{\nabla}.(n_d \vec{v}_d) = 0.
    \label{eq:5}
\end{eqnarray}
Here, $\partial_t \equiv \partial/\partial t $ refers to the time gradient operator, $n_d$ denotes the dust number density, and $\vec{v}_d$ corresponds to the dust velocity.

The momentum transport equation for dust grains along with the effects of viscoelasticity, $\kappa$-modified polarization force, and turbulence in the presence of magnetic field can be written as
\begin{eqnarray}
    \Bigg(1 + \tau_m \partial_t \Bigg) \Bigg[\partial_t \vec{v}_d - \frac{1}{c} \frac{(\vec{J} \times \vec{B})}{m_d n_d} + && \frac{\vec{\nabla} P}{m_d n_d} - \left(\frac{z_de}{m_d} \right) \vec{\nabla} \phi + \left( \frac{z_de}{m_d} \right) R_\kappa \Bigg\{1 - \frac{e \phi}{k_B T_i} \left({\frac{\kappa + 1/2}{\kappa - 3/2}}\right)\Bigg\} \nonumber \\&& \times \vec{\nabla} \phi + \vec{\nabla} \psi \Bigg] = \frac{\eta}{m_d n_d} (\nabla^2 \vec{v}_d) + \left( \frac{\zeta + \eta/3 }{m_d n_d} \right) \vec{\nabla } (\vec{\nabla} \cdot \vec{v}_d); 
    \label{eq:6}
\end{eqnarray} 

\begin{eqnarray}
    P = (P_d + P_{turb}^d) = n_{d0}k_B T_{d0} + n_{d0}k_B T_{d0} \log\left(\frac{n_d}{n_{d0}}\right).\label{eq:7}
\end{eqnarray}

%\begin{eqnarray}\label{eq:6}
%    \begin{aligned}
%        &\Bigg(1 + \tau_m \partial_t \Bigg) \Bigg[\partial_t \vec{v}_d - \frac{1}{c} \frac{(\vec{J} \times \vec{B})}{m_d n_d} + \frac{\vec{\nabla} P}{m_d n_d} + \left(\frac{q_d}{m_d} \right) \vec{\nabla} \phi + \left( \frac{z_de}{m_d} \right) R_\kappa \Bigg\{ 1 - \left(\frac{\kappa + 1/2}{\kappa - 3/2} \right) \times \\ 
%        & \left(\frac{e \phi}{k_B T_i} \right) \Bigg\} \vec{\nabla} \phi + \vec{\nabla} \psi \Bigg] = \frac{\eta}{m_d n_d} (\nabla^2 \vec{v}_d) + \left( \frac{\zeta + \eta/3 }{m_d n_d} \right) \vec{\nabla } (\vec{\nabla} \cdot \vec{v}_d);
%    \end{aligned}
%\end{eqnarray}

%\begin{eqnarray}\label{eq:7}
%    P = (P_d + P_{turb}^d) = n_{d0}k_B T_{d0} + n_{d0}k_B T_{d0} \log\left(\frac{n_d}{n_{d0}}\right).
%\end{eqnarray}
Here, $(1 + \tau_m \partial_t)$ represents the Frankel term, that serves as the viscoelastic operator with $\tau_m$ being the relaxation time \cite{dhiman_radiation_2024, sharma_jeans_2015}. The different terms $m_d$, $q_d$, $c$, $\vec{J}$, $\vec{B}$, $P$, $P_d$, $P_{turb}^d$, $R_\kappa$, $\eta$, and $\zeta$ represent the dust mass, dust surface charge, speed of light (in vacuum), current density, magnetic field, total pressure, dust thermal pressure, dust turbulence pressure, $\kappa$-modified polarization interaction parameter, shear viscosity coefficient, and bulk viscosity coefficient, respectively.
\\
The magnetic induction equation can be expressed in customary notations as \cite{bhakta_effects_2019} 
\begin{eqnarray}
    \partial_t \vec{B}+ \vec{\nabla}\times(\vec{v}_d\times \vec{B})=0.
    \label{eq:8}
\end{eqnarray}
Finally, the DMC model is closed with electrostatic and EiBI-modified gravitational Poisson equations as follows \cite{prajapati_influence_2015,banerjee_stellar_2022}
\begin{eqnarray}
    \nabla^2\phi&=4\pi e(n_e-n_i)-4\pi q_d n_d,\label{eq:9}
\end{eqnarray}
\begin{eqnarray}
    \nabla^2\psi&=4\pi G \rho_d + \left(\frac{\chi}{4}\right)\nabla^2 \rho_d. \label{eq:10}
\end{eqnarray}
Here, $\rho_d = m_d(n_d-n_{d0})$ is the effective material density of the dusty plasma system corrected after the so-called Jeans swindle. It is noteworthy that the equilibrium dust number density ($n_{d0}$) functions as the Jeans swindle as the \textit{ad hoc} homogenization approximation for the inhomogeneous equilibrium of the considered self-gravitating system \cite{falco_why_2013,kiessling_jeans_2003}. The symbols $G$, $\phi$, and $\psi$ stand for the universal gravitational constant, electrostatic potential, and gravitational potential, respectively. In Eq.~\eqref{eq:10}, the variable $\chi$ acts as the EiBI gravity parameter that introduces a correction to the Newtonian gravitational Poisson equation. In the limiting condition, $\chi \to 0$, Eq.~\eqref{eq:10} simplifies to the standard gravitational Poisson equation describing the associated potential distribution sourced by the material density fields in the considered self-gravitating plasma system.

As the self-gravitating DMCs are considered to be in a spherically symmteric geometry, for analytic simplicity, the generalized hydrodynamic equations Eq.~\eqref{eq:5} - Eq.~\eqref{eq:10} are formulated in spherical coordinates $(r,\theta,\phi)$ rather than planner coordinates $(x,y,z)$, accounting for spherical symmetry $(r, 0, 0)$ with all standard notations as 
\begin{eqnarray}\label{eq:11}
   \partial_t n_d + r^{-2}\partial_r(r^2 n_d v_d) = 0,
\end{eqnarray}
\begin{eqnarray}
       &&\Bigg(1 + \tau_m \partial_t \Bigg) \Bigg[\partial_t v_d + \left(\frac{B_\phi}{4 \pi r m_d n_d} \right) \partial_r (r B_\phi) + \frac{1}{m_d n_d} \partial_r P - \Bigg\{1 - R_\kappa \Bigg\{1 - \left({\frac{\kappa + 1/2}{\kappa - 3/2}}\right)\Bigg\} \times \nonumber\\&& \frac{e \phi}{k_B T_i} \Bigg\} \Bigg(\frac{z_de}{m_d}\Bigg) \partial_r \phi + \partial_r \psi \Bigg] = \frac{\eta}{m_d n_d} \Bigg[\frac{1}{r^2} \Bigg(\partial_r (r^2 \partial_r v_d) -2 v_d \Bigg) \Bigg] + \left(\frac{\zeta + \eta/3 }{m_d n_d} \right) \times \nonumber \\&& \partial_r \Bigg[\frac{1}{r^2} \partial_r (r^2 \partial_r v_d)\Bigg],
       \label{eq:12}
\end{eqnarray}
\begin{eqnarray}
   \partial_t B_{\phi}+r^{-1}\partial_r(r v_d B_{\phi})=0,
   \label{eq:2.13}
\end{eqnarray}
\begin{eqnarray}
   r^{-2}\partial_r(r^2 \partial_r \phi)=4\pi e(n_e-n_i)-4\pi q_d n_d,
   \label{eq:14}
\end{eqnarray}
\begin{eqnarray}
   r^{-2}\partial_r (r^2 \partial_r \psi)=4\pi Gm_d(n_d-n_{d0})+\left(\frac{\chi}{4}\right)r^{-2}\partial_r[r^2 m_d\partial_r(n_d-n_{d0})].
   \label{eq:15}
\end{eqnarray}

Eqs.~\eqref{eq:11}-\eqref{eq:15} are the required spherical hydrodynamic equations governing the EiBI-gravitating magneto-turbulent DMC. In the above equations, $\partial_r \equiv \partial/\partial_r$ represents the space gradient operator. It is worth mentioning that the term $1/r$, $r$ being the radial distance, appearing in the spherical governing equations is due to the curvature effect. For $r\to\infty$, the usual planner geometric equations can be recovered.

\section{\label{sec: Dispersion properties}Dispersion properties}
\subsection{Linear normal mode analysis}
To investigate the stability dynamics of the EiBI-gravitating magneto-turbulent DMC, a standard spherical normal mode analysis is employed \cite{ray_pulsational_2024, das_dynamics_2024}. A infinitesimal perturbation $(f_1)$ is introduced to each of the equilibrium physical quantities $(f_0)$ as follows \cite{kalita_analyzing_2020}
\begin{eqnarray}
   f(r,t) = f_0 + f_1 = f_0 + r^{-1}f_{10}\exp[-i(\omega t - kr)];
   \label{eq:16}
\end{eqnarray}
\begin{eqnarray}
   f = [n_d\; n_e\; n_i\; v_d\; \phi\; \psi]^{\text{T}},
   \label{eq:17}
\end{eqnarray}
\begin{eqnarray}
   f_0 = [n_{d0}\; n_{e0}\; n_{i0}\; 0\; 0\; 0]^{\text{T}},
   \label{eq:18}
\end{eqnarray}
\begin{eqnarray}
   f_1 = [n_{d1}\; n_{e1}\; n_{i1}\; v_{d1}\; \phi_1\; \psi_1]^{\text{T}}.
   \label{eq:19}
\end{eqnarray}
Here, T represents the transpose operation. It is important to highlight here that $f_0$ and $r^{-1} f_{10}$ share identical dimensions with no disparity. Additionally, the perturbation $f_1$ is assumed to be sinusoidal in nature, given by $f_1 = r^{-1} f_{10} \exp{(-i(\omega t - k r))}$, where, $f_{10}$, $\omega $, and $k$ represent the perturbation amplitude, angular frequency, and wavenumber, respectively. In the linearization process, higher-order terms are neglected due to their insignificant contribution. The fluctuation dynamics is then transformed from the real coordinate space $(r,t)$ to the wave space $(k, \omega)$ using the spherical Fourier analysis, which transforms the linear differential operators as $\partial_r \rightarrow (ik-r^{-1})$, $\partial_t \rightarrow (-i\omega)$, and $\partial_r^2 \rightarrow(-k^2+2 r^{-2}-2ik r^{-1})$ \cite{kalita_jeans_2021,das_dynamics_2024}. Finally, employing the perturbation and Fourier analysis, the Eqs.~\eqref{eq:3}, \eqref{eq:4}, and \eqref{eq:11} - \eqref{eq:15} can be linearized respectively as
\begin{eqnarray}
   n_{e1}=n_{e0}\sigma_{\kappa}\Bigg(\frac{e\phi_1}{k_B T_e}\Bigg),
   \label{eq:20}
\end{eqnarray}
\begin{eqnarray}
   n_{i1}=-n_{i0}\sigma_{\kappa}\Bigg(\frac{e\phi_1}{k_B T_i}\Bigg),
   \label{eq:21}
\end{eqnarray}
\begin{eqnarray}
   n_{d1}=\Bigg[\frac{n_{d0}(ik+r^{-1})}{i\omega}\Bigg]v_{d1},
   \label{eq:22}
\end{eqnarray}
\begin{eqnarray}
    && \Bigg(1-i\omega \tau_m\Bigg)\Bigg[-i\omega v_{d1}+i\Bigg(\frac{B_{\phi_0}B_{\phi_1}}{4\pi m_d n_{d0}}\Bigg)k+\Bigg(\frac{V^2_{Td}}{n_{d0}}\Bigg)\Bigg(ik-r^{-1}\Bigg)n_{d1}-(1-R_\kappa) \nonumber \\&& \times \Bigg(\frac{z_de}{m_d}\Bigg)(ik-r^{-1})\phi_1 + (ik-r^{-1})\psi_1 \Bigg]+\Bigg(\frac{\zeta+4\eta/3}{m_d n_{d0}}\Bigg)\Bigg(k^2+2r^{-2}\Bigg)v_{d1}=0,
    \label{eq:23}
\end{eqnarray}
\begin{eqnarray}
   B_{\phi_1} = B_{\phi_0}\Bigg(\frac{k}{\omega}\Bigg)v_{d1},
   \label{eq:24}
\end{eqnarray}
\begin{eqnarray}
   \phi_1 = \Bigg[\frac{4\pi q_d \lambda_{Dk}^2 n_{d0}(ik+r^{-1})}{i\omega(1+k^2\lambda_{Dk}^2)} \Bigg] v_{d1},
   \label{eq:25}
\end{eqnarray}
\begin{eqnarray}
   \psi_1=-\Bigg[\Bigg(\frac{4\pi G}{k^2}-\frac{\chi}{4}\Bigg)\Bigg(\frac{ik+r^{-1}}{i\omega}\Bigg)m_d n_{d0} \Bigg]v_{d1}.
   \label{eq:26}   
\end{eqnarray}
Here, $\lambda_{D\kappa}= \sqrt{\left(k_B/4 \pi e^2\right)\left(T_e T_i/ (T_i n_{e0}\sigma_{\kappa} + T_e n_{i0}\sigma_{\kappa})\right)}$ is the $\kappa$-modified effective plasma Debye length with $\sigma_{\kappa} = (\kappa-1/2)/(\kappa -3/2)$. Interestingly, for a large value of $\kappa \ (\to \infty)$, $\sigma_{\kappa} \to 1$, hence the $\kappa$-modified Debye length $(\lambda_{D\kappa})$ converges to the usual plasma Debye length $(\lambda_D)$. As mentioned earlier the feasible value of $\kappa$ should be greater than 3/2; otherwise, the Debye length approaches to infinity $(\text{for } \kappa = 3/2)$ or becomes imaginary $(\text{for }\kappa < 3/2)$, which is not pragmatic.

Finally, substitution of the perturbed values of $n_{d1}$, $B_{\phi1}$, $\phi_1$, and $\psi_1$ in Eq.~\eqref{eq:23} followed by the use of elimination methods yields an unnormalized quadratic dispersion relation as
\begin{eqnarray}
    \omega^2 + \frac{i\omega}{(1-i\omega\tau_m)} && \Bigg(\frac{1}{m_dn_{d0}} \Bigg) \Bigg[\Bigg(\zeta+\frac{4\eta}{3}\Bigg)\Bigg(k^2+2r^{-2} \Bigg) \Bigg] - V_A^2k^2 - \Bigg[V^2_{td} \nonumber \\&& + V^2_{da}\Bigg(\frac{1-R_\kappa}{1+k^2\lambda^2_{Dk}}\Bigg) - \Bigg(\frac{\omega_{Jd}}{k}\Bigg)^2 + V_{\chi}^2\Bigg]\Bigg(k^2+r^{-2}\Bigg)=0.
    \label{eq:27}
\end{eqnarray}
% \begin{eqnarray}\label{eq:28}
%     V_A^2 = \frac{B_{\phi 0}}{4 \pi m_d n_{d0}},
% \end{eqnarray}
% \begin{eqnarray}\label{eq:29}
%     V_{td}^2 = \frac{2k_B T_d}{m_d},
% \end{eqnarray}
% \begin{eqnarray}\label{eq:30}
%     V_{da}^2 = \omega_{pd}\lambda_{D\kappa},
% \end{eqnarray}
% \begin{eqnarray}\label{eq:3.16}
%     V_{Jd}^2 = \frac{\omega_{Jd}}{k},
% \end{eqnarray}
% \begin{eqnarray}\label{eq:3.17}
%     V_{\chi}^2 = \frac{\chi m_d n_{d0}}{4},
% \end{eqnarray}
Here, $V_A = \sqrt{B_{\phi 0}/(4 \pi m_d n_{d0})}$, $V_{td} = \sqrt{2k_B T_d/m_d}$, $V_{da} = \sqrt{\omega_{pd}\lambda_{D\kappa}}$, and $V_{\chi} = \sqrt{(\chi m_d n_{d0}) / 4}$ denote the magnitudes of the Alfv\'en velocity, dust thermal velocity, dust acoustic speed, and EiBI-induced velocity, respectively. In addition, $\omega_{Jd} = \sqrt{4 \pi G m_d n_{d0}}$ and $\omega_{pd} = \sqrt{4 \pi q_d^2 n_{d0}/m_d}$ respectively refer to the Jeans dust angular frequency and dust plasma oscillation frequency. Eq.~\eqref{eq:27} is the required linearized dispersion relation (quadratic in nature) for the viscoelastic magneto-turbulent DMC in the presence of $\kappa$-distributed electrons and ions, $\kappa$-modified polarization force, and EiBI gravity. Clearly, Eq.~\eqref{eq:27} shows the inclusion of EiBI gravity in this analysis introduces a new velocity term $V_{\chi}$ (we name it as EiBI-induced fluid velocity) in the dispersion relation. Additionally, the $\kappa$-distributed lighter species significantly modifies the polarization force. Depending on $\omega$ and $\tau_m$, the obtained dispersion relation (Eq.~\ref{eq:27}) can be written in the low-frequency (hydrodynamic) regime $(\omega \tau_m \ll 1$, Eq.~\ref{eq:28}), and the high-frequency (kinetic) regime $(\omega \tau_m \gg 1$, Eq.~\ref{eq:29}), respectively, as 
\begin{eqnarray}
    \omega^2 + i \omega \Bigg(\frac{1}{m_dn_{d0}} \Bigg) \Bigg[ \Bigg(\zeta+\frac{4\eta}{3}\Bigg)\Bigg(k^2 + 2r^{-2} \Bigg)&&\Bigg]
    - V_A^2k^2-\Bigg[V^2_{td} +  V^2_{da} \Bigg(\frac{1-R_\kappa}{1+k^2\lambda^2_{Dk}}\Bigg) \nonumber \\&& + V_{\chi}^2 -\Bigg(\frac{\omega_{Jd}}{k}\Bigg)^2\Bigg]\Bigg(k^2+r^{-2}\Bigg) = 0,
    \label{eq:28}
\end{eqnarray}

\begin{eqnarray}
    \omega^2-V_A^2k^2-\Bigg[V^2_{td} + V^2_{da}\Bigg(\frac{1-R_\kappa}{1+k^2\lambda^2_{Dk}}\Bigg) + V_{\chi}^2 - &&  \Bigg(\frac{\omega_{Jd}}{k}\Bigg)^2\Bigg]\Bigg(k^2+r^{-2}\Bigg) - \Bigg(\frac{1}{m_dn_{d0} \tau_m} 
    \Bigg) \nonumber \\ && \times \Bigg[\Bigg(\zeta+\frac{4\eta}{3}\Bigg)\Bigg(k^2+2r^{-2}\Bigg)\Bigg]=0.
    \label{eq:29}
\end{eqnarray}
\subsection{Instability criterion analysis}
To derive the modified Jeans length and mass condition, along with the effects of newly considered parameters, triggering the collapse of the DMC, Eq.~\eqref{eq:27} is converted to planner geometry by taking the approximation $r \to \infty$ as follows

\begin{widetext}
    \begin{equation}
    \omega^2 + \frac{i\omega}{(1-i\omega\tau_m)} \Bigg(\frac{\zeta + 4\eta/3}{m_dn_{d0}}\Bigg)k^2 - 
    \Bigg[V_A^2 + V^2_{td} + V^2_{da}\Bigg(\frac{1-R_\kappa}{1+k^2\lambda^2_{Dk}}\Bigg)  + V_{\chi}^2\Bigg]k^2 + \omega_{Jd}^2 = 0.
    \label{eq:30}
    \end{equation}
\end{widetext}

\subsubsection{Analysis in hydrodynamic regime}

In the hydrodynamic limit, $\omega \tau_m \ll 1$, the planner dispersion relation (Eq.~\ref{eq:30}) takes the form 
\begin{widetext}
    \begin{equation}
        \omega^2 + i\omega \Bigg(\frac{1}{m_dn_{d0}}\Bigg)\Bigg(\zeta+\frac{4\eta}{3}\Bigg)k^2 -\Bigg[V_A^2 + V^2_{td} + V^2_{da}\Bigg(\frac{1-R_\kappa}{1+k^2\lambda^2_{Dk}}\Bigg) + V_{\chi}^2\Bigg]k^2 + \omega_{Jd}^2 = 0.
    \label{eq:37}
    \end{equation}
\end{widetext}

The critical instability criteria essential for the collapse of the DMC can be deduced with the help of Routh-Hurwitz stability criterion \cite{irwin_engineering_2022,hassanien_routh-hurwitz_2021}, given as
\begin{eqnarray}
    \Bigg[V_A^2 + V^2_{td} + V^2_{da}\Bigg(\frac{1-R_\kappa}{1+k^2\lambda^2_{Dk}}\Bigg) + V_{\chi}^2\Bigg]k^2 - \omega_{Jd}^2 < 0.
    \label{eq:32}
\end{eqnarray}
Since $\lambda_{D\kappa}/{\lambda_{J}} \sim 10^{-8}\ (\ll 1)$, the term $(1 + k^2 \lambda_{D\kappa}^2)$ can be reasonably ignored in Eq.~\eqref{eq:32}. Therefore, Eq.~\eqref{eq:32} yields the following critical Jeans wavenumber
\begin{eqnarray}
    k_{Jc1} = \Bigg[\frac{4 \pi G m_d n_{d0}}{V_A^2 + V_{td}^2+ V_{da}^2 (1 - R_\kappa) + V_{\chi}^2} \Bigg]^\frac{1}{2}.
    \label{eq:33}
\end{eqnarray}
Here, $k_{Jc1}$ is the critical Jeans wavenumber modified due to the collective effects of $\kappa$-modified polarization force, nonthermal electrons and ions, magnetic field, and EiBI gravity. It is noteworthy that for $k \ge k_{Jc1}$, the instability decays out resulting in a stable DMC.

The critical Jeans length can be expressed as 

\begin{eqnarray}
    L_{Jc1} = \frac{2 \pi}{k_{Jc1}} \approx \Bigg[\frac{V_A^2 + V_{td}^2+ V_{da}^2 (1 - R_\kappa) + V_{\chi}^2}{G m_d n_{d0}}\Bigg]^\frac{1}{2}.
    \label{eq:34}
\end{eqnarray}
Any self-gravitating interstellar cloud whose length exceeds this critical limit becomes unstable due to Jeans instability, leading to fragmentation. As evident from Eq.~\eqref{eq:34}, the magnetic field, $B$ (as $V_A \propto B$), dust temperature, $T_d$ (as $V_{td} \propto \sqrt{T_d}$), and EiBI parameter, $\chi$ (as $V_{\chi} \propto \sqrt{\chi}$), all increase the $L_{Jc}$-value, thereby acting as stabilizing factors. However, with an increase in the polarization interaction parameter ($R_\kappa$), the polarization force increases, which subsequently decreases the Jeans length and, thus, acts as a destabilizing factor.

Considering the spherical nature of the DMC, the corresponding critical Jeans mass can be estimated \cite{sharma_modified_2014} as
 \begin{eqnarray}
    M_{Jc} = \frac{4}{3} \pi L_{Jc}^3 m_d n_{d0}.
    \label{eq:35}
 \end{eqnarray}
Substituting $L_{Jc1}$ from Eq.~\eqref{eq:34} in Eq.~\eqref{eq:35}, one obtains the critical Jeans mass as 
\begin{eqnarray}
    M_{Jc1} =  \frac{4 \pi}{3} \Bigg[\frac{V_A^2 + V_{td}^2 + V_{da}^2 (1 - R_\kappa) + V_{\chi}^2}{G m_{d}^{1/3} n_{d0}^{1/3}}\Bigg]^\frac{3}{2}.
    \label{eq:36}
\end{eqnarray}
The critical Jeans mass $(M_{Jc})$ provides a maximum mass limit above which the internal pressure can no longer balance the force of gravity resulting in gravitational collapse.

\subsubsection{Analysis in kinetic regime}
In the kinetic limit, $\omega \tau_m \gg 1$, the planner dispersion relation (Eq.~\ref{eq:30}) takes the form 
\begin{eqnarray}
    \omega^2 - \Bigg[V_A^2 + V^2_{td} + V^2_{da}\Bigg(\frac{1-R_\kappa}{1+k^2\lambda^2_{Dk}}\Bigg) + V_{\chi}^2 + V_\text{com}^2\Bigg]k^2 + \omega_{Jd}^2 = 0.
    \label{eq:31}
\end{eqnarray}

Apart from the usual notations used earlier, here $V_\text{com} = \sqrt{(\zeta + 4\eta/3)/(m_d n_{d0} \tau_m)}$ is the compressional velocity arising due to the viscoelasticity coefficients $\zeta$ and $\eta$. Within a viscoelastic medium, the fluid compressibility introduces this extra wave mode in addition to the existing ones. The compressional velocity term increases with an increase in the viscoelastic coefficients $(\zeta \text{ and } \eta)$ and decreases with an increase in the relaxation time $(\tau_m)$.

Analogous to the hydrodynamic regime, the critical Jeans length and mass are obtained for kinetic regime as follows
\begin{eqnarray}
    L_{Jc2} = \Bigg[\frac{V_A^2 + V_{td}^2+ V_{da}^2 (1 - R_\kappa) + V_\text{com}^2 + V_{\chi}^2}{G m_d n_{d0}}\Bigg]^\frac{1}{2},
    \label{eq:38}
\end{eqnarray}

\begin{eqnarray}
    M_{Jc2} =  \frac{4 \pi}{3} \Bigg[\frac{V_A^2 + V_{td}^2 + V_{da}^2 (1 - R_\kappa) + V_\text{com}^2 + V_{\chi}^2}{G m_{d}^{1/3} n_{d0}^{1/3}}\Bigg]^\frac{3}{2}.
    \label{eq:39}
\end{eqnarray}
It is clear from Eq.~\eqref{eq:38} and Eq.~\eqref{eq:39} that in the kinetic limit, both the critical Jeans mass and length are dependent of the viscoelastic coefficients ($\zeta$ and $\eta$), which was not the case for hydrodynamic regime. Therefore, the compressional viscoelastic mode is entirely attenuated within the hydrodynamic regime and has no contribution to the DMC stability dynamics. Nevertheless, the effects of other parameters remain the same as in both the limiting case.

The ratio of the critical Jeans length scales in the hydrodynamic and kinetic regimes is derived and presented as
\begin{eqnarray}
    \alpha_{HK} = \Bigg[\frac{V_A^2 + V_{td}^2 + V_{da}^2 (1 - R_\kappa) + V_\chi^2}{V_A^2 + V_{td}^2+ V_{da}^2 (1 - R_\kappa) + V_{\text{com}}^2 + V_\chi^2}\Bigg]^{\frac{1}{2}}. 
    \label{eq:40}
\end{eqnarray}
Similarly, the corresponding ratio of the critical Jeans masses in the hydrodynamic and kinetic regimes is obtained as 
\begin{eqnarray}
        \mu_{HK} = \Bigg[\frac{V_A^2 + V_{td}^2 + V_{da}^2 (1 - R_\kappa) + V_{\chi}^2}{V_A^2 + V_{td}^2 + V_{da}^2 (1 - R_\kappa) + V_{\text{com}}^2 + V_{\chi}^2}\Bigg]^\frac{3}{2}.
        \label{eq:41}
\end{eqnarray}
\subsection{Normalized dispersion analysis}
In order to carry out the Jeans instability analysis in a scale-invariant form, the derived dispersion relations in the hydrodynamic (Eq.~\ref{eq:28}) and kinetic (Eq.~\ref{eq:29}) regimes are normalized with a standard astrophysical normalization scheme \cite{ray_pulsational_2024,kalita_jeans_2021}, cast respectively as
\begin{eqnarray}
    \Omega^2 + i \Omega \Bigg(\frac{\omega_{Jd}}{m_dn_{d0} {V_{da}^2}}\Bigg)&& \Bigg[\Bigg(\zeta+\frac{4\eta}{3}\Bigg) \Bigg(K^2 + \xi^{-2}\Bigg)\Bigg] - \Bigg(\frac{V_A}{V_{da}}\Bigg)^2 K^2 - \Bigg[\Bigg(\frac{V_{td}}{V_{da}}\Bigg)^2 + V^2_{da} \times \nonumber \\&& \Bigg(\frac{1-R_\kappa}{V_{da}^2 + K^2 \omega_{Jd}^2 \lambda^2_{Dk}}\Bigg) + \Bigg(\frac{V_{\chi}}{V_{da}}\Bigg)^2 - \Bigg(\frac{1}{K}\Bigg)^2 \Bigg]\Bigg(K^2+\xi^{-2}\Bigg) = 0,
    \label{eq:42}
\end{eqnarray}

\begin{eqnarray}
    \Omega^2 - \Bigg(\frac{V_A}{V_{da}}\Bigg)^2 K^2 - && \Bigg[\Bigg(\frac{V_{td}}{V_{da}}\Bigg)^2 + V^2_{da} \Bigg(\frac{1-R_\kappa}{V_{da}^2 + K^2 \omega_{Jd}^2 \lambda^2_{Dk}}\Bigg) + \Bigg(\frac{V_{\chi}}{V_{da}}\Bigg)^2 - \Bigg(\frac{1}{K}\Bigg)^2 \nonumber \\&& + \Bigg(\frac{1}{m_dn_{d0} \tau_m V_{da}^2} \Bigg)\Bigg(\zeta + \frac{4\eta}{3}\Bigg)\Bigg]\Bigg(K^2+\xi^{-2}\Bigg)=0.
    \label{eq:43}
\end{eqnarray}

In this adopted normalization scheme, $\xi = r / \lambda_{Jd}$, $ K = k / k_{Jd} = k (V_{da} / \omega_{Jd})$, and $\Omega = \omega / \omega_{Jd}$ represent the normalized (dimensionless) radial distance, angular wavenumber, and angular frequency scaled with the Jeans wavelength $(\lambda_{Jd})$, Jeans angular wavenumber $(k_{Jd})$, and Jeans angular frequency $(\omega_{Jd})$ as the normalizing parameters, respectively.

\section{Results and discussions}\label{sec:Results and discussions}

\begin{figure}[htbp]
    \centering
    \includegraphics[width=1\linewidth]{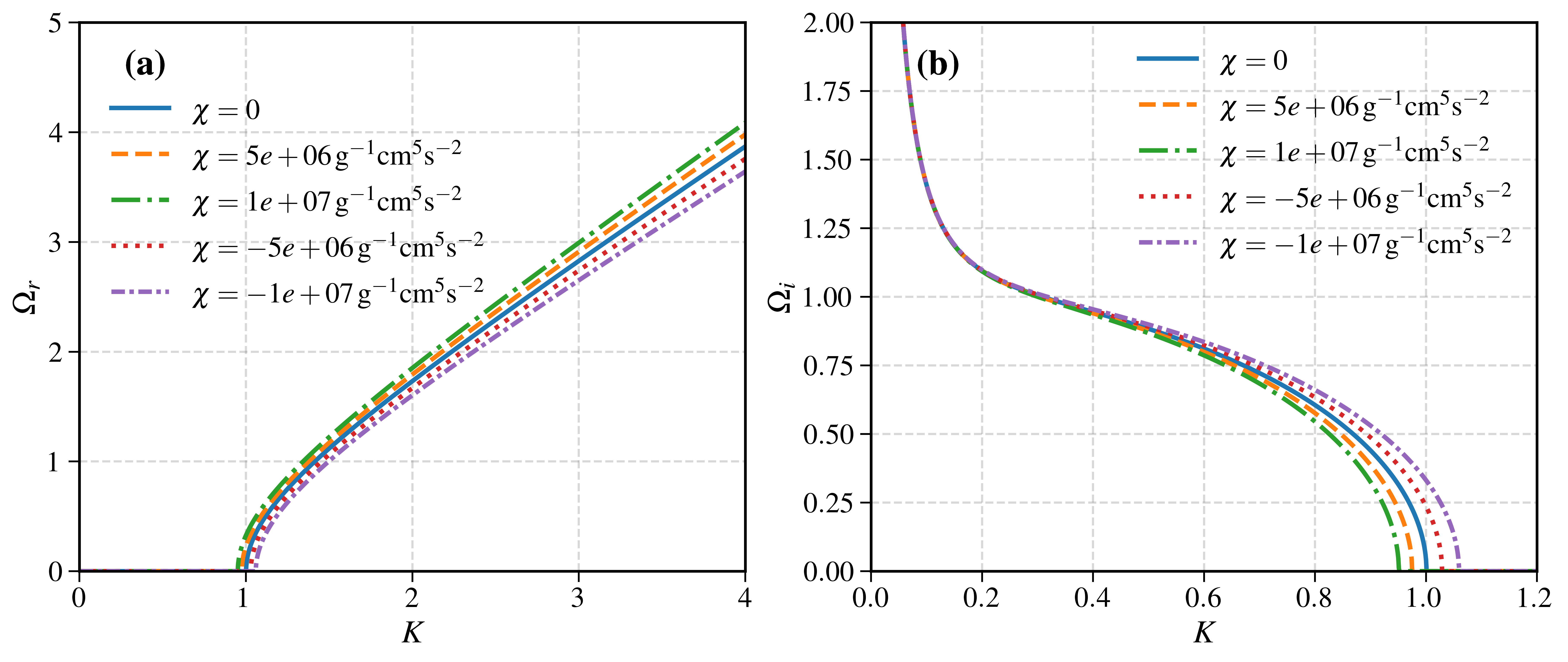}
    % \caption{Profile of variation of the real frequency $(\Omega_r)$ and imaginary frequency $(\Omega_i)$ of the first (a and b) and second (c and d) root of the hydrodynamic dispersion relation, respectively with the angular wavenumber $(K)$ for different values of the EiBI gravity parameter $(\chi)$ represented on the Jeans scale. The various curves link to: (i) $\chi = 0$ (blue solid line, taken as the Newtonian reference line), (ii) $\chi = 5 \times10^6$ \unit{g^{-1}.cm^5.s^{-2}} (yellow dashed line), (iii) $\chi = 1 \times 10^7$ \unit{g^{-1}.cm^5.s^{-2}} (green dash-dotted line), (iv) $\chi = -5 \times10^6$ \unit{g^{-1}.cm^5.s^{-2}} (red dotted line), and (v) $\chi= -1\times10^7$ \unit{g^{-1}.cm^5.s^{-2}} (purple tightly dash-dotted line).}
    \caption{Profile of the Jeans-scaled (a) real frequency part $(\Omega_r)$ and (b) imaginary frequency part $(\Omega_i)$ with variation in the angular wavenumber $(K)$ for different values of $\chi$. The various lines link to: (i) $\chi = 0$ (blue solid), (ii) $\chi = 5 \times10^6$ \unit{g^{-1}.cm^5.s^{-2}} (yellow dashed), (iii) $\chi = 1 \times 10^7$ \unit{g^{-1}.cm^5.s^{-2}} (green loosely dash-dotted), (iv) $\chi = -5 \times10^6$ \unit{g^{-1}.cm^5.s^{-2}} (red dotted), and (v) $\chi= -1\times10^7$ \unit{g^{-1}.cm^5.s^{-2}} (purple tightly dash-dotted).}
    \label{fig:1}
\end{figure}

\begin{figure}[htbp]
    \centering
    \includegraphics[width=1\linewidth]{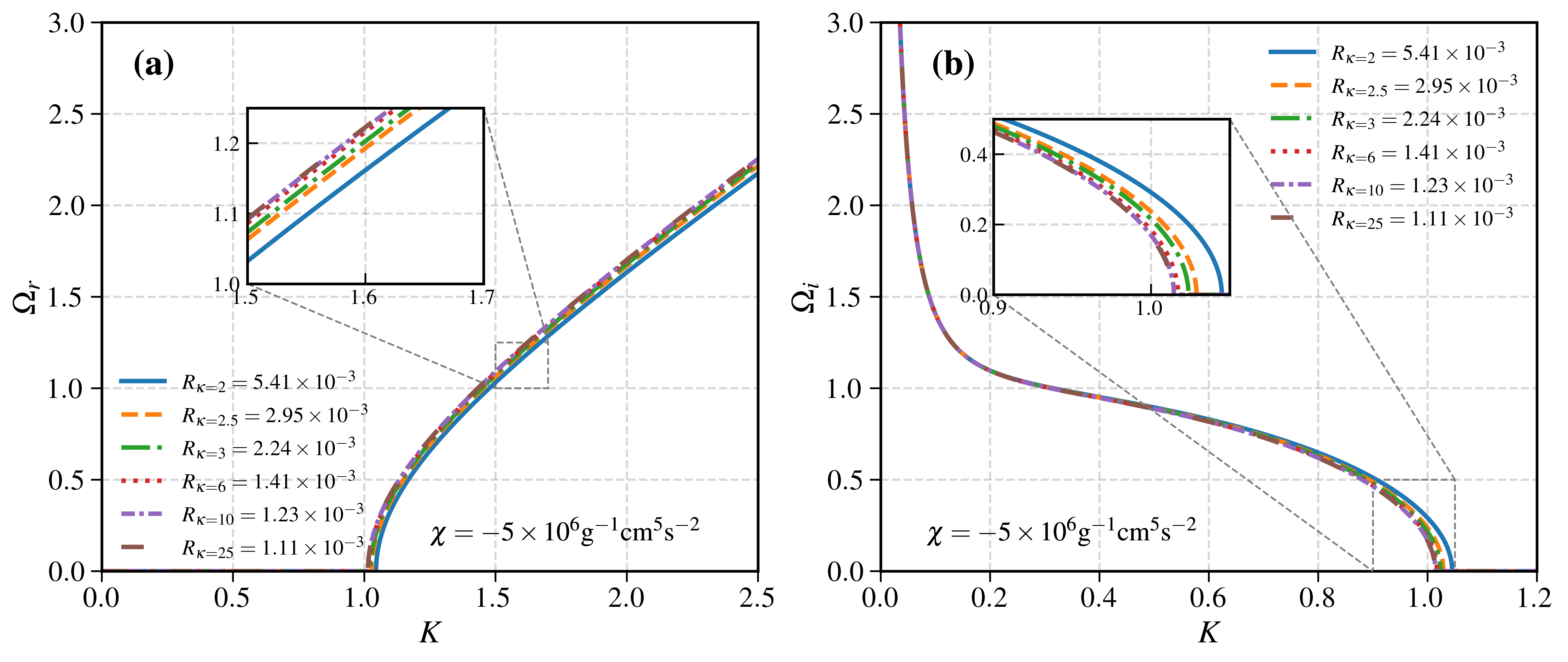}
    \caption{Same as Fig.~\ref{fig:1}, but for different values of $\kappa$ and corresponding $R_\kappa$. The various lines link to: (i) $R_{\kappa = 2} = 5.41 \times 10^{-3}$ (blue solid), (ii) $R_{\kappa = 2.5} = 2.95 \times 10^{-3}$ (yellow dashed), (iii) $R_{\kappa = 3} = 2.24 \times 10^{-3}$ (green loosely dash-dotted), (iv) $R_{\kappa = 6} = 1.41 \times 10^{-3}$ (red dotted), (v) $R_{\kappa = 25} = 1.11 \times 10^{-3}$ (purple tightly dash-dotted), and (v) $R_{\text{Max}} = 1.04 \times 10^{-3}$ (brown loosely dashed).}
    \label{fig:2}
\end{figure}

\begin{figure}[htbp]
    \centering
    \includegraphics[width=1\linewidth]{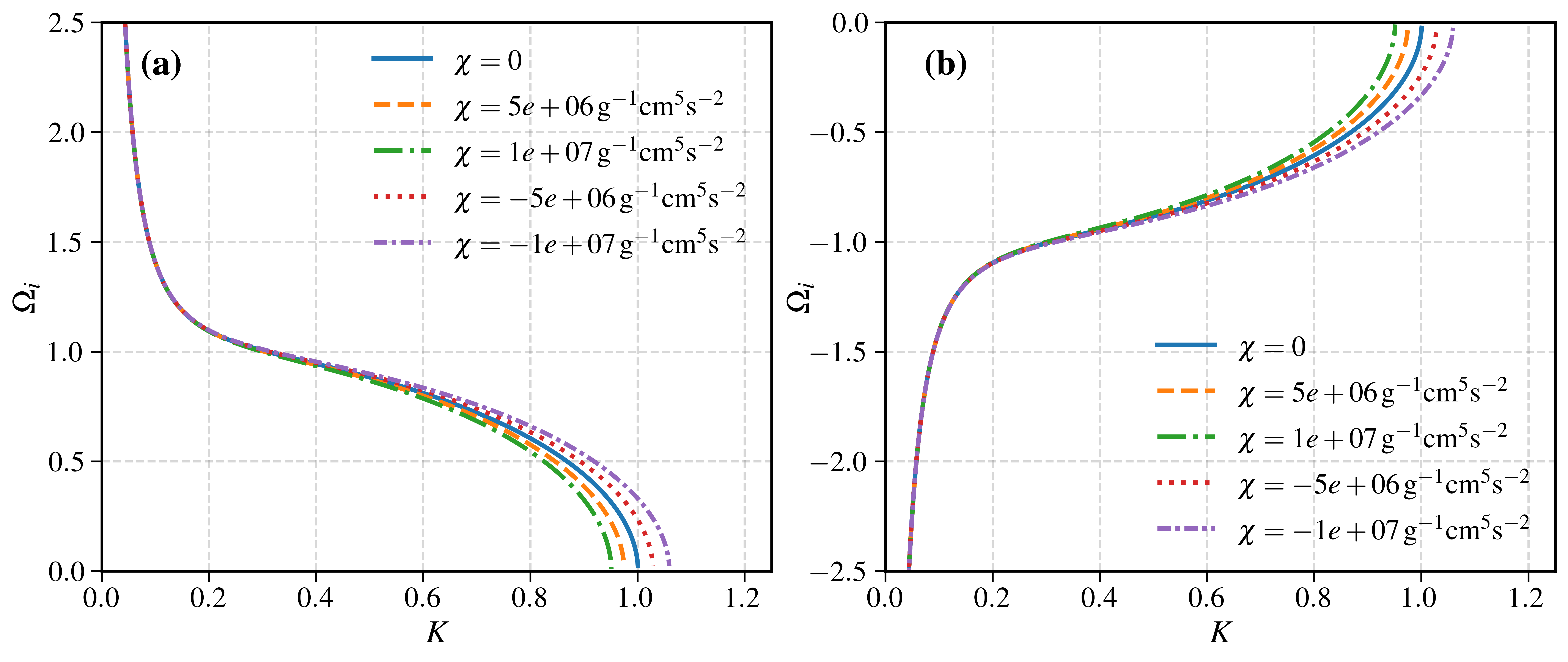}
    \caption{Same as Fig.~\ref{fig:1}, but for different values of $\chi$, assuming the kinetic limit $(\omega\tau_m \gg 1)$. The various curves link to: (i) $\chi = 0$ (blue solid), (ii) $\chi = 5\times10^6$ \unit{g^{-1}.cm^5.s^{-2}} (yellow dashed), (iii) $\chi = 1 \times 10^7$ \unit{g^{-1}.cm^5.s^{-2}} (green loosely dash-dotted), (iv) $\chi = -5 \times10^6$ \unit{g^{-1}.cm^5.s^{-2}} (red dotted), and (v) $\chi= -1\times10^7$ \unit{g^{-1}.cm^5.s^{-2}} (purple tightly dash-dotted).}
    \label{fig:3}
\end{figure}

\begin{figure}[htbp]
    \centering
    \begin{minipage}{0.48\linewidth}
        \centering
        \includegraphics[width=\linewidth]{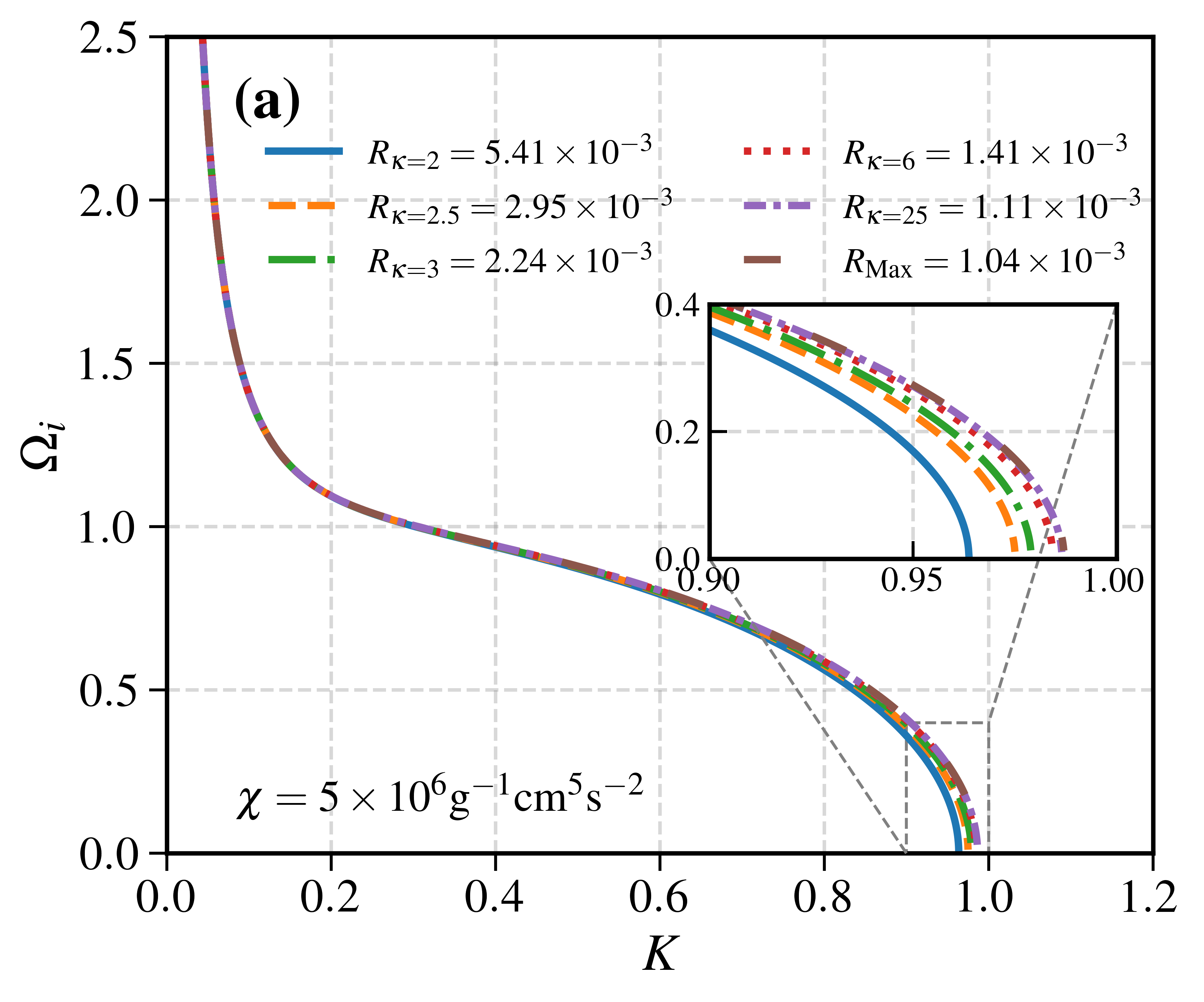}
        \label{fig:4a}
    \end{minipage}
    \hfill
    \begin{minipage}{0.48\linewidth}
        \centering
        \includegraphics[width=\linewidth]{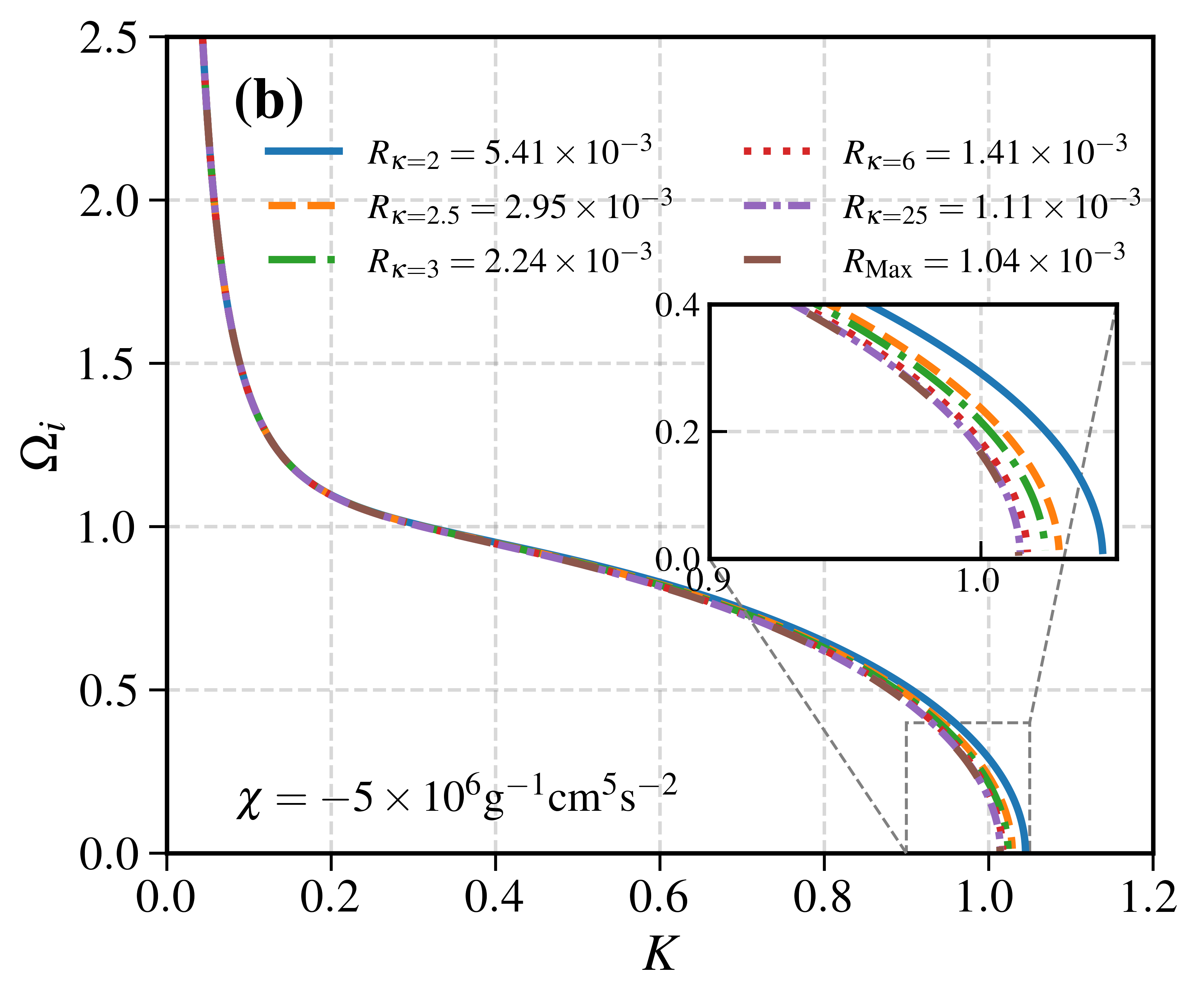}
        \label{fig:4b}
    \end{minipage}
    \caption{Same as Fig.~\ref{fig:2}, but for (a) $\chi = 5 \times 10^{6}$ \unit{g^{-1}.cm^{5}.s^{-2}} and (b) $\chi = - 5 \times 10^{6}$ \unit{g^{-1}.cm^{5}.s^{-2}}, assuming the kinetic limit $(\omega\tau_m \gg 1)$. 
    The various curves correspond to: 
    (i) $R_{\kappa = 2} = 5.41 \times 10^{-3}$ (blue solid), 
    (ii) $R_{\kappa = 2.5} = 2.95 \times 10^{-3}$ (yellow dashed), 
    (iii) $R_{\kappa = 3} = 2.24 \times 10^{-3}$ (green loosely dash-dotted), 
    (iv) $R_{\kappa = 6} = 1.41 \times 10^{-3}$ (red dotted), 
    (v) $R_{\kappa = 25} = 1.11 \times 10^{-3}$ (purple tightly dash-dotted), and 
    (vi) $R_{\text{Max}} = 1.04 \times 10^{-3}$ (brown loosely dashed).}
    \label{fig:4}
\end{figure}

\begin{figure}
    \centering
        \includegraphics[width = 0.8\textwidth]{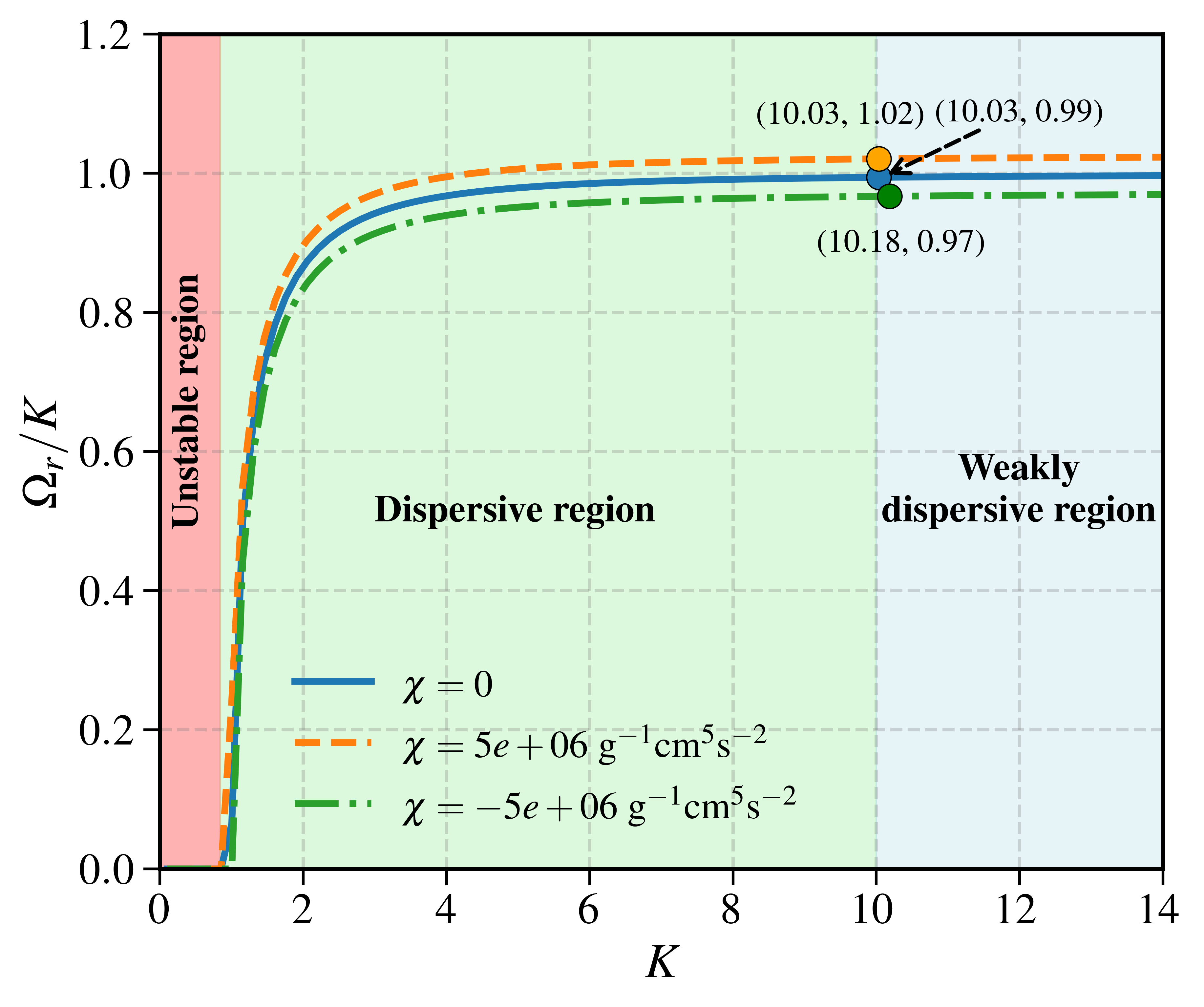}
    \caption{Profile of variation of the Jeans-scaled phase velocity $(\Omega_r / K)$ with Jeans-scaled angular wavenumber $(K)$ for different values of $\chi$ at $\kappa = 2.5$. The various curves link to: (i) $\chi = 0$ (blue solid), (ii) $\chi = 5 \times 10^{6}$ \unit{g^{-1}.cm^{5}.s^{-2}} (yellow dashed), and (iii) $\chi = -5 \times 10^{6}$ \unit{g^{-1}.cm^{5}.s^{-2}} (green dash-dotted).}
    \label{fig:5}
\end{figure}

A theoretical model formalism is developed to study the collective influence of the $\kappa$-distributed lighter species (electron and ions) and thereby, the $\kappa$-modified polarization force on the Jeans instability of a self-gravitating magneto-turbulent DMC in EiBI gravity fabric. A quadratic linearized dispersion relation (Eq.~\ref{eq:27}) is derived from the DMC governing equations following the standard spherical normal mode analysis. Depending on the wave frequency, the derived dispersion relation is converted to hydrodynamic regime (Eq.~\ref{eq:28}) and kinetic regime (Eq.~\ref{eq:29}) followed by a standard normalization procedure. Subsequently, a numerical illustrative platform is utilized to further examine the instability dynamics and the influence of various included parameters embedded in the derived dispersion relation. Here, we especially consider the ultracompact H\textsc{ii} regions coined by Wood and Churchwell \cite{wood_morphologies_1989}. These hot photoionized nebulae are characterized by diameters $\le 0.01$ \unit{pc} and electron densities $\ge 10^4$ \unit{cm^{-3}}. The typical equilibrium parametric values are considered as follows, $m_e = 9.1 \times 10^{-28}$ \unit{g}, $m_i = 1.6 \times 10^{-24}$ \unit{g}, $m_d = 4 \times 10^{-12}$ \unit{g}, $e = 4.80 \times 10^{-10}$ \unit{esu}, $q_d = - 200 e$, $r_d = 1.5 \times 10^{-4}$ \unit{cm}, $n_{e0}= 4 \times 10^6$ \unit{cm^{-3}}, $n_{i0} = 7 \times 10^6$ \unit{cm^{-3}}, $n_{d0} = 10^3$ \unit{cm^{-3}}, $T_e = 10^{4}$ \unit{K}, $T_i = 10^{3}$ \unit{K}, $T_d = 30$ \unit{K}, $\zeta = 10^{-14}$ \unit{g.cm^{-2}.s^{-1}}, $\eta = 10^{-15}$ \unit{g.cm^{-2}.s^{-1}}, $\tau_m = 10^{-2}$ \unit{s}, and $B_{\phi0} = 10^{-6}$ \unit{G} \cite{wood_morphologies_1989, lefevre_dust_2014,anderson_physical_2010,shukla_jeans_2006, kalita_adapted_2021}. The nonthermal spectral parameter is varied in the range $\kappa = 2 - \infty$ and the EiBI parameter is chosen as $\chi < 3\times10^{10}$ \unit{g^{-1}.cm^5.s^{-2}} (in SI, $3\times10^3$ \unit{kg^{-1}.m^5.s^{-2}}) as calculated from atomic constraint by Avelino \cite{avelino_eddington-inspired_2012}. It is important to note that the equilibrium values considered here are average values as the equilibrium values of such vast and exotic systems varies over a range. Using these parametric values the coupling parameter is found as $\Gamma_\text{Cou} \sim 77\ (< \Gamma_c)$, which shows the system is indeed in strongly coupled liquid phase with viscoelastic effect. The other physical parameters are found as dust thermal velocity, $V_{td} = 0.03$ \unit{cm.s^{-1}}, $\kappa$-modified plasma Debye length, $\lambda_{D\kappa} \sim 0.04$ \unit{cm} (for $\kappa = 2$), and dust plasma oscillation frequency, $\omega_{pd} = 17.03$ \unit{rad. s^{-1}}. The $\lambda_{D\kappa}$ is found to be 2 order smaller than the plasma Debye length ($\lambda_D \sim 0.08$ \unit{cm} ) for Maxwellian lighter particles. Both the hydrodynamic (Eq.~\ref{eq:42}) and kinetic (Eq.~\ref{eq:43}) dispersion relations are graphically analyzed and illustratively depicted in Figs.~\ref{fig:1}-\ref{fig:5}. 

We numerically construct Fig.~\ref{fig:1} to illustrate the considered nonthermal Jeans instability in the hydrodynamic regime $(\omega \tau_m \ll 1)$. The quadratic hydrodynamic dispersion relation (Eq.~\ref{eq:42}) obviously has two non-trivial solutions, out of which the first solution shows a reduction in the instability growth rate $(\Omega_i)$ as wavenumber $(K)$ decreases, thereby indicating its stability effects for longer wavelength disturbances. This observation deviates from the classical Jeans stability criterion, traditionally applicable to gaseous nebulae under the Newtonian gravity framework. Hence, the first root of the hydrodynamic dispersion relation can be reasonably neglected. However, for the second root, $\Omega_i$ decreases with increasing $K$, suggesting that the system stabilizes at shorter wavelengths. It is indeed in full conformity with previously reported studies on the Newtonian gravitational formalism portraying self-gravitating fluids on infinite spatial scales. Thus, the second root (unstable) is used for further analysis. Accordingly, the variation of the real frequency part $(\Omega_r)$ and the imaginary frequency part $(\Omega_i)$ of the second root (Figs.~\ref{fig:1}a - \ref{fig:1}b) are plotted against the angular wavenumber $(K)$ for different values of the EiBI gravity parameter $(\chi)$. The $\chi$-value is varied in the range $\chi = - 1 \times 10^{7} \text{ to } 1 \times 10^{7}$ $\unit{g^{-1}. cm^{5}. s^{-2}}$, which is well below said the limit set by Avelino \cite{avelino_eddington-inspired_2012}. Notably, $\chi = 0$ corresponds to the Newtonian case. The nonthermal index $(\kappa)$ and the normalized radial distance $(\xi)$ are kept fixed at $2$ and $10$, respectively. Clearly, the system stabilizes quickly with an increasing $\chi$-value and vice versa. The plots further reveal that the instability persists for perturbation wavenumbers exceeding the standard critical Jeans value ($k_J$) for $\chi < 0$. It implies that the Jeans length for the EiBI-modified systems with a negative $\chi$-value is lower compared to the Newtonian picture. Consequently, smaller structures can form in such EiBI-gravitating astroclouds. However, the system remains stable for perturbations with $k<k_J$ when $\chi>0$, implying that the EiBI gravity can either enhance or suppress instability compared to the Newtonian scenario, depending on the sign and magnitude of $\chi$. It strengthens the practical reliability of the classical Jeans instability analysis in the realistic EiBI-gravity framework.

Fig.~\ref{fig:2} depicts the effect of the $\kappa$-distributed lighter constituents and associated $\kappa$-modified polarization force on the EiBI-modified Jeans instability in the hydrodynamic limit $(\omega \tau_m \ll 1)$. Here, the variation of $\Omega_r$ and $\Omega_i$ of the second dispersion root are plotted against $K$ for different values of $\kappa$ and hence, the corresponding $\kappa$-modified polarization interaction parameter $(R_\kappa)$. The chosen values of $\kappa$ are $2, 2.5, 3, 6, 10, 25,$ and $\infty$, which yields respective $R_\kappa$ values of $5.41 \times 10^{-3}, 2.95 \times 10^{-3}, 2.24 \times 10^{-3}, 1.41 \times 10^{-3}, 1.23 \times 10^{-3}, 1.11 \times 10^{-3},$ and $1.04 \times 10^{-3}$. It can be clearly seen that, with an increase in the $\kappa$-value, the $R_\kappa$-value decreases, i.e., the higher is the non-Maxwellian characteristic (lower $\kappa$-value), the larger is the polarization force. Notably, the polarization force for $\kappa = 2$ is nearly five times stronger than its Maxwellian counterpart. This is due to the fact that as the non-Maxwellian lighter particles possess higher nonthermal energies, they tend to polarize the dust grains to a higher extent. The other parameters are kept the same as before except $\chi = 5 \times10^{6}$ \unit{g^{-1}.cm^{5}.s^{-2}}. Similar to Fig.~\ref{fig:1}, it is observed that $\Omega_r$ increases and $\Omega_i$ decreases with increasing $K$-value. Additionally, $\Omega_i$ decreases more rapidly with increasing $\kappa$-value (smaller $R_\kappa$), indicating that a stronger polarization force reduces system stability. Therefore, the polarization force acts as a destabilizing and decelerating agent. This destabilizing effect arises because the polarization force counteracts with the electrostatic one, which, along with the thermal pressure, balances the inward gravitational pull \cite{ray_pulsational_2024, hamaguchi_polarization_1994, dolai_effects_2020}. Besides, even for the Maxwellian case, a modification to the critical Jeans length is observed, which is due to the EiBI gravity. As a strong polarization decreases (increases) the critical Jeans length (wavenumber), the combined effect of the polarization force and EiBI gravity can support the self-gravitational fragmentation of cloudlets (smaller) originating from source clouds (larger).

Fig.~\ref{fig:3} resembles Fig.~\ref{fig:1}, but it pertains to distinct values of $\chi$ in the kinetic limit $(\omega\tau_m \gg 1)$. An important characteristic of the kinetic dispersion relation (Eq.~\ref{eq:43}) is that it is associated with a vanishing propagatory component $(\Omega_r)$ and a non-vanishing instability growth rate $(\Omega_i)$, which signifies that the wave undergoes either exponential decay or growth over time in the absence of wave propagation \cite{ray_pulsational_2024, karmakar_evolutionary_2017}. Therefore, in Fig.~\ref{fig:3}, the $\Omega_i$ of the (a) first and (b) second root of the kinetic dispersion relation is plotted against $K$ for different values of $\chi$. Interestingly, it is seen that for the kinetic regime the instability corresponds to the first root, unlike the second in the case of hydrodynamic regime. Similar influences of the EiBI gravity are evident in the kinetic regime as in the hydrodynamic regime. Therefore, it is established from both the hydrodynamic and kinetic perspectives that systems characterized by a positive EiBI parameter demonstrate enhanced stability in comparison to those with a negative EiBI parameter. Hence, the positive EiBI parameter is acting as a stabilizing agent, whereas the negative one is acting as a destabilizing agent.
% Fig.~\ref{fig:3} shows the influence of the kappa-distributed electrons and ions on the Jeans instability in the hydrodynamic limit. Here, the variation of $\Omega_r$ and $\Omega_i$ of the second root is plotted, respectively, against $K$ for different values $\kappa$. Furthermore, $\kappa$ is varied in the range $\kappa = 2.5 \text{ to } \infty$, keeping the other parametric values the same as before. In this case, $\kappa \to \infty$ corresponds to the Maxwellian reference. It should be noted that lower $\kappa$-values imply higher nonthermality of the system. Clearly, the instability growth rate $(\Omega_i)$ decreases with an increase in the $\kappa$-value, indicating that the nonthermal (non-Maxwellian) system exhibits greater stability compared to the Maxwellian system. Therefore, the nonthermal spectral parameter $(\kappa)$ acts as a stabilizing agent. 
% \begin{figure}[htbp]
%     \centering
%      \includegraphics[width=1\linewidth]{figure3.png}
%     \caption{Same as Fig.~\ref{fig:1}, but for different values of nonthermal spectral parameter $(\kappa)$. The various curves respectively link to: (a) $\kappa \to \infty $ (blue solid line, taken as the Maxwellian reference line), (b) $\kappa = 2.5$ (yellow dashed line), (c) $\kappa = 3$ (green dash-dotted line), (d) $\kappa = 5$ (red dotted line), (e) $\kappa = 7$ (purple tightly dash-dotted line), (f) $\kappa = 15$ (brown loosely dashed line), (g) $\kappa = 18$ (pink dashed line), and (h) $\kappa = 25$ (grey tightly dashed line).}
%     \label{fig:3}
% \end{figure}

As in fig.~\ref{fig:4}, we depict the instability properties similar to Fig.~\ref{fig:2}, but here only the first root of the kinetic dispersion relation is presented with (a) $\chi = 5 \times 10^{6}$ \unit{g^{-1}.cm^{5}.s^{-2}} and (b) $\chi = - 5 \times 10^{6}$ \unit{g^{-1}.cm^{5}.s^{-2}}. The instability growth rate is found to decrease for a larger $K$-value (shorter wavelengths) as observed in Figs.~\ref{fig:1}-\ref{fig:3}. However, it is found that $\Omega_i$ for various polarization interaction parameters is dependent on the polarity of $\chi$. For negative $\chi$-values, the $\kappa$-modified polarization force behaves as the conventional polarization force and acts as a destabilizing agent. Conversely, for positive $\chi$-values, stabilizing influences on the cloud are observed against the above scenarios.

Fig.~\ref{fig:5} illustrates the phase velocity $(\Omega_r/K)$ of the propagating wave, arising due to the perturbations in the system, as a function of $K$ on the Jeans scale for different values of $\chi$. Here, we set $\kappa = 2.5$ (fixed), while keeping the other parameters remain unchanged. The figure reveals that the phase velocity shows dispersive characteristics at lower $K$-values (longer wavelengths). As $K$ increases, the phase velocity gradually rises until it reaches a critical point, beyond which it saturates. Notably, for very large wavelengths $(K \lesssim 0.8)$, the wave collapses, and the propagation ceases, marking the unstable region where wave motion is not sustained. It is also found that the positive $\chi$-values enhance the phase velocity against the Newtonian case, indicating influence of the EiBI gravity on the wave dynamics. Additionally, the phase velocity at the saturation point $(10.03, 1.02)$ for $\chi = 5 \times 10^{6}$ \unit{g^{-1}.cm^{5}.s^{-1}} is found to be $\sim 0.43$ \unit{cm.s^{-1}}. It is noteworthy that these findings are well consistent with the results already presented in Figs.~\ref{fig:1}-\ref{fig:4}.

\section{\label{sec: Conclusion}Conclusions}
In this paper, the combined effect of $\kappa$-distributed lighter constituents (electrons and ions) and corresponding $\kappa$-modified polarization force on the Jeans instability of a strongly coupled $(\Gamma_{\text{Cou}}\gg 1)$ magneto-turbulent dust molecular cloud (DMC) is investigated in the EiBI gravity framework. A quadratic dispersion relation is derived using the spherical normal mode analysis of the linearized perturbed DMC governing equations. The resulting dispersion relation is analyzed in both the hydrodynamic ($\omega\tau_m \ll 1$) and kinetic ($\omega \tau_m \gg 1$) regimes along with their respective modified instability criteria. The dispersion relations are then normalized with appropriately chosen Jeans parameters and analyzed numerically along with graphs to study the oscillatory and propagating characteristics.

The classical Jeans instability criterion is found to be significantly altered due to the inclusion of the aforementioned parameters into the system. In both the limiting regimes, the EiBI gravity introduces a new fluid velocity term, the EiBI-induced velocity $(V_\chi)$, in the dispersion relation which may support novel wave modes. However, the compressional velocity term $(V_{\text{com}})$ that arises due to the bulk and shear viscosity, is only present in the kinetic limit and completely vanishes in the hydrodynamic limit. The $\kappa$-distributed (nonthermal) lighter constituents enhance the polarization force by approximately five times compared to the Maxwellian (thermal) constituents, with $\kappa = 2$. A thorough analysis of both regimes reveals that the $\kappa$-modified polarization force ($F_{p\kappa}$) and the negative EiBI parameter ($-\chi$) exert destabilizing effects on the system, while the positive EiBI parameter ($+\chi$) serves as a DMC stabilizer. Additionally, the $-\chi$ reduces the critical Jeans length relative to the Newtonian scenario, potentially facilitating the fragmentation of smaller clouds. An increase in the polarization interaction parameter $(R_{\kappa})$ and $+\chi$ are also found to decrease the real normalized frequency $(\Omega_r)$. Consequently, the phase velocity ($\Omega_r/K$) of the resulting wave exhibits a variation with different parameters. It progressively increases with increasing $K$, indicating a highly dispersive nature, until it reaches a critical point after which it becomes saturated, showing a weakly dispersive characteristic. This suggests that longer gravitational-like perturbations tend to disturb the system at higher magnitude compared to shorter acoustic-like disturbances. Furthermore, no wave propagation is noticed after a critical value of $K \lesssim 0.8$, which defines the boundary of the unstable region. 

The critical Jeans length, considering the parameters of ultracompact H\textsc{ii} regions, is determined to be approximately $1.5 \times 10^{7}$ \unit{cm}, which is significantly smaller than the typical Jeans length of approximately $10^{14}$ \unit{cm}. Therefore, this calculated Jeans length provides a plausible explanation for the formation of smaller self-gravitationally bounded structures, which cannot be adequately explained by the traditional Jeans theory. Although the parameters employed are tailored to the H\textsc{ii} regions, the theoretical model developed here is applicable to other plasmic regions of self-gravitating objects, regardless of whether modified gravity scenarios or Newtonian gravity (as a limiting condition) are considered.

It is worth mentioning that several effects, such as radiation pressure exerted by newly born stars and cosmic rays, as well as effects of rotation, are not taken into account in this study. Hence, there exists a potential for further refinement of the current model to enhance its realistic applicability in a broader horizon of sophisticated astrophysical environments.

\appendix
\section{\label{sec:appendix}Polarization force for kappa distributed ions}

In a complex DMC, the polarization force can be expressed as \cite{hamaguchi_polarization_1994}
\begin{eqnarray}\label{eq:A.1}
    F_p=-\left(\frac{q_d^2}{2}\right)\left(\frac{\nabla \lambda_D}{\lambda_D^2}\right),
\end{eqnarray}
where, $\lambda_D$ is the effective plasma Debye length, given as \cite{hamaguchi_polarization_1994}
\begin{eqnarray}\label{eq:A.2}
    \lambda_D = \lambda_{Di}\left(1+\frac{\lambda_{Di}^2}{\lambda_{De}^2} \right)^{-1/2}.
\end{eqnarray}
Here, $\lambda_{Di(e)} = (k_B T_{i(e)} / 4\pi e^2 n_{i(e)})^{1/2}$ is the electron (ion) Debye length, where $T_{e(i)}$ is the electron (ion) temperature (in K) and $n_{e(i)}$ is the electron (ion) number density. A complex (dusty) plasma system with negatively charged dust grains $(q_d = - z_d e)$ is characterized by $n_i \gg n_e$ and $T_e \gg T_i$ \cite{ashrafi_polarization_2014}. Therefore, using the condition $n_eT_i \ll n_iT_e$ \cite{dolai_effects_2020} in Eq.~\eqref{eq:A.2}, the effective Debye length can be written as $\lambda_{D} \approx \lambda_{Di} = (k_B T_{i} / 4\pi e^2 n_{i})^{1/2}$. Taking gradients on both sides, we arrive at
\begin{eqnarray}\label{eq:A.3}
    \nabla \lambda_D = \nabla \left(\frac{k_B T_i}{4 \pi e^2 n_i} \right)^{1/2}=-\frac{1}{2}\left(\frac{k_B T_i}{4 \pi e^2 n_i} \right)^{1/2}\left(\frac{\nabla n_i}{n_i}\right)=-\frac{\lambda_D}{2}\left(\frac{\nabla n_i}{n_i}\right)
\end{eqnarray}
For kappa distribution law, $n_{i}$ is given by Eq.~\eqref{eq:4}, the gradient of which results in
\begin{eqnarray}\label{eq:A.4}
    \nabla n_i= \left(-\kappa+\frac{1}{2}\right)n_{i0}\left[1+\frac{e\phi}{k_B T_i (\kappa -3/2)} \right]^{-\kappa-1/2}\left[\frac{e\nabla \phi}{k_B T_i (\kappa-3/2)} \right].
\end{eqnarray}
Substituting Eq.~\eqref{eq:A.4} in Eq.~\eqref{eq:A.3} yields
\begin{eqnarray}\label{eq:A.5}
    \nabla \lambda_D=-\frac{1}{2}\lambda_{Di}\left(-\kappa+\frac{1}{2}\right)\left[1-\frac{e\phi}{k_B T_i (\kappa -3/2)}\right]\left[\frac{e\nabla\phi}{k_B T_i (\kappa -3/2)} \right].
\end{eqnarray}
Finally substituting Eq.~\eqref{eq:A.5} in Eq.~\eqref{eq:A.1}, we get
\begin{eqnarray}\label{eq:A.6}
    F_{p\kappa}=-z_deR_\kappa\left(\frac{n_i}{n_{i0}} \right)^{1/2}\left[1-\frac{e\phi}{k_B T_i (\kappa-3/2)}\right]\nabla \phi,
\end{eqnarray}
where, $R_\kappa$ is the $\kappa$-modified Polarization interaction parameter, given as
\begin{eqnarray}\label{eq:A.7}
    R_\kappa = \sigma_k\left(\frac{z_d e^2}{4 \lambda_{Di0}k_B T_i}\right). 
\end{eqnarray}
Here, $\sigma_k= (\kappa-1/2)/(\kappa-3/2)$ is a dimensionless parameter and $\lambda_{Di0} = (k_B T_i/4\pi e^2 n_{i0})^{1/2}$ is the equilibrium ion Debye length. For the Maxwellian case, $\kappa\rightarrow \infty$, $\sigma_\kappa\rightarrow 1$, which reduces $R_\kappa$ into $R$, thereby the $\kappa$-modified polarization force $(F_{p\kappa})$ converges to the usual polarization force, $F_p = -q_dR(n_i/n_{i0})^{1/2}\nabla \phi$ \cite{ashrafi_polarization_2014}.

It is clear from Eq.~\eqref{eq:A.6} that the $\kappa$-modified polarization force is directly proportional to the dust charge $(q_d)$, the modified polarization parameter $(R_\kappa)$, gradient of the electrostatic potential $(\nabla \phi)$, and inversely proportional to the ion temperature $(T_i)$ of a plasma system.

\acknowledgments
We express our sincere gratitude to Tezpur University for their invaluable support and resources. We also extend our special appreciation to all the members of the Astrophysical Plasma and Nonlinear Dynamics Research Laboratory (APNDRL) for their insightful discussions and collaborative efforts. Additionally, we acknowledge the support received from institutions such as IUCAA, Pune.

% \nocite{*}

\bibliography{apssamp}% Produces the bibliography via BibTeX.

\begin{thebibliography}{99}%
\makeatletter
\providecommand \@ifxundefined [1]{%
 \@ifx{#1\undefined}
}%
\providecommand \@ifnum [1]{%
 \ifnum #1\expandafter \@firstoftwo
 \else \expandafter \@secondoftwo
 \fi
}%
\providecommand \@ifx [1]{%
 \ifx #1\expandafter \@firstoftwo
 \else \expandafter \@secondoftwo
 \fi
}%
\providecommand \natexlab [1]{#1}%
\providecommand \enquote  [1]{``#1''}%
\providecommand \bibnamefont  [1]{#1}%
\providecommand \bibfnamefont [1]{#1}%
\providecommand \citenamefont [1]{#1}%
\providecommand \href@noop [0]{\@secondoftwo}%
\providecommand \href [0]{\begingroup \@sanitize@url \@href}%
\providecommand \@href[1]{\@@startlink{#1}\@@href}%
\providecommand \@@href[1]{\endgroup#1\@@endlink}%
\providecommand \@sanitize@url [0]{\catcode `\\12\catcode `\$12\catcode `\&12\catcode `\#12\catcode `\^12\catcode `\_12\catcode `\%12\relax}%
\providecommand \@@startlink[1]{}%
\providecommand \@@endlink[0]{}%
\providecommand \url  [0]{\begingroup\@sanitize@url \@url }%
\providecommand \@url [1]{\endgroup\@href {#1}{\urlprefix }}%
\providecommand \urlprefix  [0]{URL }%
\providecommand \Eprint [0]{\href }%
\providecommand \doibase [0]{https://doi.org/}%
\providecommand \selectlanguage [0]{\@gobble}%
\providecommand \bibinfo  [0]{\@secondoftwo}%
\providecommand \bibfield  [0]{\@secondoftwo}%
\providecommand \translation [1]{[#1]}%
\providecommand \BibitemOpen [0]{}%
\providecommand \bibitemStop [0]{}%
\providecommand \bibitemNoStop [0]{.\EOS\space}%
\providecommand \EOS [0]{\spacefactor3000\relax}%
\providecommand \BibitemShut  [1]{\csname bibitem#1\endcsname}%
\let\auto@bib@innerbib\@empty
%</preamble>
\bibitem [{\citenamefont {Peltonen}\ \emph {et~al.}(2023)\citenamefont {Peltonen}, \citenamefont {Rosolowsky}, \citenamefont {Williams}, \citenamefont {Koch}, \citenamefont {Dolphin}, \citenamefont {Chastenet}, \citenamefont {Dalcanton}, \citenamefont {Ginsburg}, \citenamefont {Johnson}, \citenamefont {Leroy}, \citenamefont {Richardson}, \citenamefont {Sandstrom}, \citenamefont {Sarbadhicary}, \citenamefont {Smercina}, \citenamefont {Wainer},\ and\ \citenamefont {Williams}}]{peltonen_jwst_2023}%
  \BibitemOpen
  \bibfield  {author} {\bibinfo {author} {\bibfnamefont {J.}~\bibnamefont {Peltonen}}, \bibinfo {author} {\bibfnamefont {E.}~\bibnamefont {Rosolowsky}}, \bibinfo {author} {\bibfnamefont {T.~G.}\ \bibnamefont {Williams}}, \bibinfo {author} {\bibfnamefont {E.~W.}\ \bibnamefont {Koch}}, \bibinfo {author} {\bibfnamefont {A.}~\bibnamefont {Dolphin}}, \bibinfo {author} {\bibfnamefont {J.}~\bibnamefont {Chastenet}}, \bibinfo {author} {\bibfnamefont {J.~J.}\ \bibnamefont {Dalcanton}}, \bibinfo {author} {\bibfnamefont {A.}~\bibnamefont {Ginsburg}}, \bibinfo {author} {\bibfnamefont {L.~C.}\ \bibnamefont {Johnson}}, \bibinfo {author} {\bibfnamefont {A.~K.}\ \bibnamefont {Leroy}}, \bibinfo {author} {\bibfnamefont {T.}~\bibnamefont {Richardson}}, \bibinfo {author} {\bibfnamefont {K.~M.}\ \bibnamefont {Sandstrom}}, \bibinfo {author} {\bibfnamefont {S.~K.}\ \bibnamefont {Sarbadhicary}}, \bibinfo {author} {\bibfnamefont {A.}~\bibnamefont {Smercina}}, \bibinfo {author} {\bibfnamefont {T.}~\bibnamefont {Wainer}},\ and\
  \bibinfo {author} {\bibfnamefont {B.~F.}\ \bibnamefont {Williams}},\ }\bibfield  {title} {{\selectlanguage {en}\bibinfo {title} {\textit{{JWST}} reveals star formation across a spiral arm in {M33}}},\ }\href {https://doi.org/10.1093/mnras/stad3879} {\bibfield  {journal} {\bibinfo  {journal} {Monthly Notices of the Royal Astronomical Society}\ }\textbf {\bibinfo {volume} {527}},\ \bibinfo {pages} {10668} (\bibinfo {year} {2023})}\BibitemShut {NoStop}%
\bibitem [{\citenamefont {Gurian}\ \emph {et~al.}(2025)\citenamefont {Gurian}, \citenamefont {Liu}, \citenamefont {Jeong}, \citenamefont {Hosokawa}, \citenamefont {Hirano},\ and\ \citenamefont {Yoshida}}]{gurian_analytic_2025}%
  \BibitemOpen
  \bibfield  {author} {\bibinfo {author} {\bibfnamefont {J.}~\bibnamefont {Gurian}}, \bibinfo {author} {\bibfnamefont {B.}~\bibnamefont {Liu}}, \bibinfo {author} {\bibfnamefont {D.}~\bibnamefont {Jeong}}, \bibinfo {author} {\bibfnamefont {T.}~\bibnamefont {Hosokawa}}, \bibinfo {author} {\bibfnamefont {S.}~\bibnamefont {Hirano}},\ and\ \bibinfo {author} {\bibfnamefont {N.}~\bibnamefont {Yoshida}},\ }\bibfield  {title} {{\selectlanguage {en}\bibinfo {title} {An analytic model of gravitational collapse induced by radiative cooling: instability scale, density profile, and mass infall rate}},\ }\href {https://doi.org/10.1093/mnras/staf012} {\bibfield  {journal} {\bibinfo  {journal} {Monthly Notices of the Royal Astronomical Society}\ }\textbf {\bibinfo {volume} {537}},\ \bibinfo {pages} {580} (\bibinfo {year} {2025})}\BibitemShut {NoStop}%
\bibitem [{\citenamefont {Prajapati}(2011)}]{prajapati_effect_2011}%
  \BibitemOpen
  \bibfield  {author} {\bibinfo {author} {\bibfnamefont {R.}~\bibnamefont {Prajapati}},\ }\bibfield  {title} {{\selectlanguage {en}\bibinfo {title} {Effect of polarization force on the {Jeans} instability of self-gravitating dusty plasma}},\ }\href {https://doi.org/10.1016/j.physleta.2011.05.020} {\bibfield  {journal} {\bibinfo  {journal} {Physics Letters A}\ }\textbf {\bibinfo {volume} {375}},\ \bibinfo {pages} {2624} (\bibinfo {year} {2011})}\BibitemShut {NoStop}%
\bibitem [{\citenamefont {Bhakta}\ \emph {et~al.}(2019)\citenamefont {Bhakta}, \citenamefont {Chhajlani},\ and\ \citenamefont {Prajapati}}]{bhakta_effects_2019}%
  \BibitemOpen
  \bibfield  {author} {\bibinfo {author} {\bibfnamefont {S.}~\bibnamefont {Bhakta}}, \bibinfo {author} {\bibfnamefont {R.~K.}\ \bibnamefont {Chhajlani}},\ and\ \bibinfo {author} {\bibfnamefont {R.~P.}\ \bibnamefont {Prajapati}},\ }\bibfield  {title} {{\selectlanguage {en}\bibinfo {title} {Effects of radiation pressure and polarization force on {Jeans} instability in magnetized strongly coupled dusty plasma}},\ }\href {https://doi.org/10.1088/1402-4896/aafc71} {\bibfield  {journal} {\bibinfo  {journal} {Physica Scripta}\ }\textbf {\bibinfo {volume} {94}},\ \bibinfo {pages} {045603} (\bibinfo {year} {2019})}\BibitemShut {NoStop}%
\bibitem [{\citenamefont {Kalita}\ and\ \citenamefont {Karmakar}(2020)}]{kalita_analyzing_2020}%
  \BibitemOpen
  \bibfield  {author} {\bibinfo {author} {\bibfnamefont {D.}~\bibnamefont {Kalita}}\ and\ \bibinfo {author} {\bibfnamefont {P.~K.}\ \bibnamefont {Karmakar}},\ }\bibfield  {title} {{\selectlanguage {en}\bibinfo {title} {Analyzing the instability dynamics of spherical complex astroclouds in a magnetized meanfluidic fabric}},\ }\href {https://doi.org/10.1063/1.5143267} {\bibfield  {journal} {\bibinfo  {journal} {Physics of Plasmas}\ }\textbf {\bibinfo {volume} {27}},\ \bibinfo {pages} {022902} (\bibinfo {year} {2020})}\BibitemShut {NoStop}%
\bibitem [{\citenamefont {Dhiman}\ and\ \citenamefont {Mahajan}(2023)}]{dhiman_jeans_2023}%
  \BibitemOpen
  \bibfield  {author} {\bibinfo {author} {\bibfnamefont {J.~S.}\ \bibnamefont {Dhiman}}\ and\ \bibinfo {author} {\bibfnamefont {M.}~\bibnamefont {Mahajan}},\ }\bibfield  {title} {{\selectlanguage {en}\bibinfo {title} {Jeans instability in strongly coupled clumpy molecular cloud with dissipative effects}},\ }\href {https://doi.org/10.1007/s12036-022-09894-9} {\bibfield  {journal} {\bibinfo  {journal} {Journal of Astrophysics and Astronomy}\ }\textbf {\bibinfo {volume} {44}},\ \bibinfo {pages} {8} (\bibinfo {year} {2023})}\BibitemShut {NoStop}%
\bibitem [{\citenamefont {Sharma}(2014)}]{sharma_modified_2014}%
  \BibitemOpen
  \bibfield  {author} {\bibinfo {author} {\bibfnamefont {P.}~\bibnamefont {Sharma}},\ }\bibfield  {title} {{\selectlanguage {en}\bibinfo {title} {Modified {Jeans} instability of strongly coupled inhomogeneous magneto dusty plasma in the presence of polarization force}},\ }\href {https://doi.org/10.1209/0295-5075/107/15001} {\bibfield  {journal} {\bibinfo  {journal} {EPL (Europhysics Letters)}\ }\textbf {\bibinfo {volume} {107}},\ \bibinfo {pages} {15001} (\bibinfo {year} {2014})}\BibitemShut {NoStop}%
\bibitem [{\citenamefont {Dasgupta}\ and\ \citenamefont {Karmakar}(2019)}]{dasgupta_jeans_2019}%
  \BibitemOpen
  \bibfield  {author} {\bibinfo {author} {\bibfnamefont {S.}~\bibnamefont {Dasgupta}}\ and\ \bibinfo {author} {\bibfnamefont {P.~K.}\ \bibnamefont {Karmakar}},\ }\bibfield  {title} {{\selectlanguage {en}\bibinfo {title} {The {Jeans} instability in viscoelastic spherical astrophysical fluid media}},\ }\href {https://doi.org/10.1007/s10509-019-3706-x} {\bibfield  {journal} {\bibinfo  {journal} {Astrophysics and Space Science}\ }\textbf {\bibinfo {volume} {364}},\ \bibinfo {pages} {213} (\bibinfo {year} {2019})}\BibitemShut {NoStop}%
\bibitem [{\citenamefont {Hopkins}\ and\ \citenamefont {Lee}(2016)}]{hopkins_fundamentally_2016}%
  \BibitemOpen
  \bibfield  {author} {\bibinfo {author} {\bibfnamefont {P.~F.}\ \bibnamefont {Hopkins}}\ and\ \bibinfo {author} {\bibfnamefont {H.}~\bibnamefont {Lee}},\ }\bibfield  {title} {{\selectlanguage {en}\bibinfo {title} {The fundamentally different dynamics of dust and gas in molecular clouds}},\ }\href {https://doi.org/10.1093/mnras/stv2745} {\bibfield  {journal} {\bibinfo  {journal} {Monthly Notices of the Royal Astronomical Society}\ }\textbf {\bibinfo {volume} {456}},\ \bibinfo {pages} {4174} (\bibinfo {year} {2016})}\BibitemShut {NoStop}%
\bibitem [{\citenamefont {Eddington}(1988)}]{eddington_internal_1988}%
  \BibitemOpen
  \bibfield  {author} {\bibinfo {author} {\bibfnamefont {A.~S.}\ \bibnamefont {Eddington}},\ }\href@noop {} {{\selectlanguage {eng}\emph {\bibinfo {title} {The internal constitution of the stars}}}},\ \bibinfo {edition} {1st}\ ed.,\ Cambridge science classics\ (\bibinfo  {publisher} {Cambridge University Press},\ \bibinfo {address} {Cambridge},\ \bibinfo {year} {1988})\BibitemShut {NoStop}%
\bibitem [{\citenamefont {Jeans}(1901)}]{jeans_stability_1901}%
  \BibitemOpen
  \bibfield  {author} {\bibinfo {author} {\bibfnamefont {J.~H.}\ \bibnamefont {Jeans}},\ }\bibfield  {title} {{\selectlanguage {en}\bibinfo {title} {The stability of a spherical {Nebula}}},\ }\href {https://doi.org/10.1098/rspl.1901.0072} {\bibfield  {journal} {\bibinfo  {journal} {Proceedings of the Royal Society of London}\ }\textbf {\bibinfo {volume} {68}},\ \bibinfo {pages} {454} (\bibinfo {year} {1901})}\BibitemShut {NoStop}%
\bibitem [{\citenamefont {Choudhuri}(2010)}]{choudhuri_astrophysics_2010}%
  \BibitemOpen
  \bibfield  {author} {\bibinfo {author} {\bibfnamefont {A.~R.}\ \bibnamefont {Choudhuri}},\ }\href@noop {} {{\selectlanguage {eng}\emph {\bibinfo {title} {Astrophysics for physicists}}}}\ (\bibinfo  {publisher} {Cambridge University Press},\ \bibinfo {address} {Cambridge, UK},\ \bibinfo {year} {2010})\ \bibinfo {note} {oCLC: 642205334}\BibitemShut {NoStop}%
\bibitem [{\citenamefont {Gaurav}\ and\ \citenamefont {Avinash}(2018)}]{gaurav_effect_2018}%
  \BibitemOpen
  \bibfield  {author} {\bibinfo {author} {\bibfnamefont {S.}~\bibnamefont {Gaurav}}\ and\ \bibinfo {author} {\bibfnamefont {K.}~\bibnamefont {Avinash}},\ }\bibfield  {title} {{\selectlanguage {en}\bibinfo {title} {Effect of ion drag on {Jean}'s instability in dusty plasma}},\ }\href {https://doi.org/10.1063/1.5058284} {\bibfield  {journal} {\bibinfo  {journal} {Physics of Plasmas}\ }\textbf {\bibinfo {volume} {25}},\ \bibinfo {pages} {114503} (\bibinfo {year} {2018})}\BibitemShut {NoStop}%
\bibitem [{\citenamefont {Pagani}\ \emph {et~al.}(2010)\citenamefont {Pagani}, \citenamefont {Steinacker}, \citenamefont {Bacmann}, \citenamefont {Stutz},\ and\ \citenamefont {Henning}}]{pagani_ubiquity_2010}%
  \BibitemOpen
  \bibfield  {author} {\bibinfo {author} {\bibfnamefont {L.}~\bibnamefont {Pagani}}, \bibinfo {author} {\bibfnamefont {J.}~\bibnamefont {Steinacker}}, \bibinfo {author} {\bibfnamefont {A.}~\bibnamefont {Bacmann}}, \bibinfo {author} {\bibfnamefont {A.}~\bibnamefont {Stutz}},\ and\ \bibinfo {author} {\bibfnamefont {T.}~\bibnamefont {Henning}},\ }\bibfield  {title} {{\selectlanguage {en}\bibinfo {title} {The {Ubiquity} of {Micrometer}-{Sized} {Dust} {Grains} in the {Dense} {Interstellar} {Medium}}},\ }\href {https://doi.org/10.1126/science.1193211} {\bibfield  {journal} {\bibinfo  {journal} {Science}\ }\textbf {\bibinfo {volume} {329}},\ \bibinfo {pages} {1622} (\bibinfo {year} {2010})}\BibitemShut {NoStop}%
\bibitem [{\citenamefont {De~Marchi}\ and\ \citenamefont {Panagia}(2014)}]{de_marchi_extinction_2014}%
  \BibitemOpen
  \bibfield  {author} {\bibinfo {author} {\bibfnamefont {G.}~\bibnamefont {De~Marchi}}\ and\ \bibinfo {author} {\bibfnamefont {N.}~\bibnamefont {Panagia}},\ }\bibfield  {title} {{\selectlanguage {en}\bibinfo {title} {The extinction law inside the 30 {Doradus} nebula}},\ }\href {https://doi.org/10.1093/mnras/stu1694} {\bibfield  {journal} {\bibinfo  {journal} {Monthly Notices of the Royal Astronomical Society}\ }\textbf {\bibinfo {volume} {445}},\ \bibinfo {pages} {93} (\bibinfo {year} {2014})}\BibitemShut {NoStop}%
\bibitem [{\citenamefont {Lefèvre}\ \emph {et~al.}(2014)\citenamefont {Lefèvre}, \citenamefont {Pagani}, \citenamefont {Juvela}, \citenamefont {Paladini}, \citenamefont {Lallement}, \citenamefont {Marshall}, \citenamefont {Andersen}, \citenamefont {Bacmann}, \citenamefont {McGehee}, \citenamefont {Montier}, \citenamefont {Noriega-Crespo}, \citenamefont {Pelkonen}, \citenamefont {Ristorcelli},\ and\ \citenamefont {Steinacker}}]{lefevre_dust_2014}%
  \BibitemOpen
  \bibfield  {author} {\bibinfo {author} {\bibfnamefont {C.}~\bibnamefont {Lefèvre}}, \bibinfo {author} {\bibfnamefont {L.}~\bibnamefont {Pagani}}, \bibinfo {author} {\bibfnamefont {M.}~\bibnamefont {Juvela}}, \bibinfo {author} {\bibfnamefont {R.}~\bibnamefont {Paladini}}, \bibinfo {author} {\bibfnamefont {R.}~\bibnamefont {Lallement}}, \bibinfo {author} {\bibfnamefont {D.~J.}\ \bibnamefont {Marshall}}, \bibinfo {author} {\bibfnamefont {M.}~\bibnamefont {Andersen}}, \bibinfo {author} {\bibfnamefont {A.}~\bibnamefont {Bacmann}}, \bibinfo {author} {\bibfnamefont {P.~M.}\ \bibnamefont {McGehee}}, \bibinfo {author} {\bibfnamefont {L.}~\bibnamefont {Montier}}, \bibinfo {author} {\bibfnamefont {A.}~\bibnamefont {Noriega-Crespo}}, \bibinfo {author} {\bibfnamefont {V.-M.}\ \bibnamefont {Pelkonen}}, \bibinfo {author} {\bibfnamefont {I.}~\bibnamefont {Ristorcelli}},\ and\ \bibinfo {author} {\bibfnamefont {J.}~\bibnamefont {Steinacker}},\ }\bibfield  {title} {\bibinfo {title} {Dust properties inside molecular clouds
  from coreshine modeling and observations},\ }\href {https://doi.org/10.1051/0004-6361/201424081} {\bibfield  {journal} {\bibinfo  {journal} {Astronomy \& Astrophysics}\ }\textbf {\bibinfo {volume} {572}},\ \bibinfo {pages} {A20} (\bibinfo {year} {2014})}\BibitemShut {NoStop}%
\bibitem [{\citenamefont {Draine}\ and\ \citenamefont {Lee}(1984)}]{draine_optical_1984}%
  \BibitemOpen
  \bibfield  {author} {\bibinfo {author} {\bibfnamefont {B.~T.}\ \bibnamefont {Draine}}\ and\ \bibinfo {author} {\bibfnamefont {H.~M.}\ \bibnamefont {Lee}},\ }\bibfield  {title} {{\selectlanguage {en}\bibinfo {title} {Optical properties of interstellar graphite and silicate grains}},\ }\href {https://doi.org/10.1086/162480} {\bibfield  {journal} {\bibinfo  {journal} {The Astrophysical Journal}\ }\textbf {\bibinfo {volume} {285}},\ \bibinfo {pages} {89} (\bibinfo {year} {1984})}\BibitemShut {NoStop}%
\bibitem [{\citenamefont {Soliman}\ \emph {et~al.}(2024)\citenamefont {Soliman}, \citenamefont {Hopkins},\ and\ \citenamefont {Grudić}}]{soliman_thermodynamics_2024}%
  \BibitemOpen
  \bibfield  {author} {\bibinfo {author} {\bibfnamefont {N.~H.}\ \bibnamefont {Soliman}}, \bibinfo {author} {\bibfnamefont {P.~F.}\ \bibnamefont {Hopkins}},\ and\ \bibinfo {author} {\bibfnamefont {M.~Y.}\ \bibnamefont {Grudić}},\ }\bibfield  {title} {\bibinfo {title} {Thermodynamics of {Giant} {Molecular} {Clouds}: {The} {Effects} of {Dust} {Grain} {Size}},\ }\href {https://doi.org/10.3847/1538-4357/ad8087} {\bibfield  {journal} {\bibinfo  {journal} {The Astrophysical Journal}\ }\textbf {\bibinfo {volume} {975}},\ \bibinfo {pages} {284} (\bibinfo {year} {2024})}\BibitemShut {NoStop}%
\bibitem [{\citenamefont {Jacobs}\ and\ \citenamefont {Shukla}(2004)}]{jacobs_linearly_2004}%
  \BibitemOpen
  \bibfield  {author} {\bibinfo {author} {\bibfnamefont {G.}~\bibnamefont {Jacobs}}\ and\ \bibinfo {author} {\bibfnamefont {P.~K.}\ \bibnamefont {Shukla}},\ }\bibfield  {title} {\bibinfo {title} {Linearly {Coupled} {Jeans}–{Alfvén} {Modes} in {Self}-{Gravitating} {Astrophysical} {Dusty} {Plasmas}},\ }\href {https://doi.org/10.1238/Physica.Regular.070a00262} {\bibfield  {journal} {\bibinfo  {journal} {Physica Scripta}\ }\textbf {\bibinfo {volume} {70}},\ \bibinfo {pages} {262} (\bibinfo {year} {2004})}\BibitemShut {NoStop}%
\bibitem [{\citenamefont {Pandey}\ \emph {et~al.}(1994)\citenamefont {Pandey}, \citenamefont {Avinash},\ and\ \citenamefont {Dwivedi}}]{pandey_jeans_1994}%
  \BibitemOpen
  \bibfield  {author} {\bibinfo {author} {\bibfnamefont {B.~P.}\ \bibnamefont {Pandey}}, \bibinfo {author} {\bibfnamefont {K.}~\bibnamefont {Avinash}},\ and\ \bibinfo {author} {\bibfnamefont {C.~B.}\ \bibnamefont {Dwivedi}},\ }\bibfield  {title} {{\selectlanguage {en}\bibinfo {title} {Jeans instability of a dusty plasma}},\ }\href {https://doi.org/10.1103/PhysRevE.49.5599} {\bibfield  {journal} {\bibinfo  {journal} {Physical Review E}\ }\textbf {\bibinfo {volume} {49}},\ \bibinfo {pages} {5599} (\bibinfo {year} {1994})}\BibitemShut {NoStop}%
\bibitem [{\citenamefont {Shukla}\ and\ \citenamefont {Mamun}(2002)}]{shukla_introduction_2002}%
  \BibitemOpen
  \bibfield  {author} {\bibinfo {author} {\bibfnamefont {P.~K.}\ \bibnamefont {Shukla}}\ and\ \bibinfo {author} {\bibfnamefont {A.~A.}\ \bibnamefont {Mamun}},\ }\href@noop {} {{\selectlanguage {eng}\emph {\bibinfo {title} {Introduction to dusty plasma physics}}}},\ Series in plasma physics\ (\bibinfo  {publisher} {Institute of Physics Pub},\ \bibinfo {address} {Bristol},\ \bibinfo {year} {2002})\BibitemShut {NoStop}%
\bibitem [{\citenamefont {Melzer}(2019)}]{melzer_physics_2019}%
  \BibitemOpen
  \bibfield  {author} {\bibinfo {author} {\bibfnamefont {A.}~\bibnamefont {Melzer}},\ }\href {https://doi.org/10.1007/978-3-030-20260-6} {{\selectlanguage {en}\emph {\bibinfo {title} {Physics of {Dusty} {Plasmas}: {An} {Introduction}}}}},\ \bibinfo {series} {Lecture {Notes} in {Physics}}, Vol.\ \bibinfo {volume} {962}\ (\bibinfo  {publisher} {Springer International Publishing},\ \bibinfo {address} {Cham},\ \bibinfo {year} {2019})\BibitemShut {NoStop}%
\bibitem [{\citenamefont {Thomas}\ \emph {et~al.}(1994)\citenamefont {Thomas}, \citenamefont {Morfill}, \citenamefont {Demmel}, \citenamefont {Goree}, \citenamefont {Feuerbacher},\ and\ \citenamefont {Möhlmann}}]{thomas_plasma_1994}%
  \BibitemOpen
  \bibfield  {author} {\bibinfo {author} {\bibfnamefont {H.}~\bibnamefont {Thomas}}, \bibinfo {author} {\bibfnamefont {G.~E.}\ \bibnamefont {Morfill}}, \bibinfo {author} {\bibfnamefont {V.}~\bibnamefont {Demmel}}, \bibinfo {author} {\bibfnamefont {J.}~\bibnamefont {Goree}}, \bibinfo {author} {\bibfnamefont {B.}~\bibnamefont {Feuerbacher}},\ and\ \bibinfo {author} {\bibfnamefont {D.}~\bibnamefont {Möhlmann}},\ }\bibfield  {title} {{\selectlanguage {en}\bibinfo {title} {Plasma {Crystal}: {Coulomb} {Crystallization} in a {Dusty} {Plasma}}},\ }\href {https://doi.org/10.1103/PhysRevLett.73.652} {\bibfield  {journal} {\bibinfo  {journal} {Physical Review Letters}\ }\textbf {\bibinfo {volume} {73}},\ \bibinfo {pages} {652} (\bibinfo {year} {1994})}\BibitemShut {NoStop}%
\bibitem [{\citenamefont {Zobaer}\ and\ \citenamefont {Mamun}(2013)}]{zobaer_observing_2013}%
  \BibitemOpen
  \bibfield  {author} {\bibinfo {author} {\bibfnamefont {M.~S.}\ \bibnamefont {Zobaer}}\ and\ \bibinfo {author} {\bibfnamefont {A.~A.}\ \bibnamefont {Mamun}},\ }\bibfield  {title} {{\selectlanguage {en}\bibinfo {title} {Observing the effects of the polarization force in strongly coupled dusty plasmas with suprathermal electrons}},\ }\href {https://doi.org/10.1007/s10509-013-1508-0} {\bibfield  {journal} {\bibinfo  {journal} {Astrophysics and Space Science}\ }\textbf {\bibinfo {volume} {347}},\ \bibinfo {pages} {145} (\bibinfo {year} {2013})}\BibitemShut {NoStop}%
\bibitem [{\citenamefont {Delzanno}\ and\ \citenamefont {Lapenta}(2005)}]{delzanno_modified_2005}%
  \BibitemOpen
  \bibfield  {author} {\bibinfo {author} {\bibfnamefont {G.~L.}\ \bibnamefont {Delzanno}}\ and\ \bibinfo {author} {\bibfnamefont {G.}~\bibnamefont {Lapenta}},\ }\bibfield  {title} {{\selectlanguage {en}\bibinfo {title} {Modified {Jeans} {Instability} for {Dust} {Grains} in a {Plasma}}},\ }\href {https://doi.org/10.1103/PhysRevLett.94.175005} {\bibfield  {journal} {\bibinfo  {journal} {Physical Review Letters}\ }\textbf {\bibinfo {volume} {94}},\ \bibinfo {pages} {175005} (\bibinfo {year} {2005})}\BibitemShut {NoStop}%
\bibitem [{\citenamefont {Dwivedi}\ \emph {et~al.}(1999)\citenamefont {Dwivedi}, \citenamefont {Sen},\ and\ \citenamefont {Bujarbarua}}]{dwivedi_pulsational_1999}%
  \BibitemOpen
  \bibfield  {author} {\bibinfo {author} {\bibfnamefont {C.~B.}\ \bibnamefont {Dwivedi}}, \bibinfo {author} {\bibfnamefont {A.~K.}\ \bibnamefont {Sen}},\ and\ \bibinfo {author} {\bibfnamefont {S.}~\bibnamefont {Bujarbarua}},\ }\bibfield  {title} {{\selectlanguage {en}\bibinfo {title} {Pulsational mode of gravitational collapse and its impact on the star formation}},\ }\href@noop {} {\bibfield  {journal} {\bibinfo  {journal} {Astronomy \& Astrophysics}\ }\textbf {\bibinfo {volume} {345}},\ \bibinfo {pages} {1049} (\bibinfo {year} {1999})}\BibitemShut {NoStop}%
\bibitem [{\citenamefont {Lundin}\ \emph {et~al.}(2008)\citenamefont {Lundin}, \citenamefont {Marklund},\ and\ \citenamefont {Brodin}}]{lundin_modified_2008}%
  \BibitemOpen
  \bibfield  {author} {\bibinfo {author} {\bibfnamefont {J.}~\bibnamefont {Lundin}}, \bibinfo {author} {\bibfnamefont {M.}~\bibnamefont {Marklund}},\ and\ \bibinfo {author} {\bibfnamefont {G.}~\bibnamefont {Brodin}},\ }\bibfield  {title} {{\selectlanguage {en}\bibinfo {title} {Modified {Jeans} {Instability} {Criteria} for {Magnetized} {Systems}}},\ }\href {https://doi.org/10.1063/1.2956641} {\bibfield  {journal} {\bibinfo  {journal} {Physics of Plasmas}\ }\textbf {\bibinfo {volume} {15}},\ \bibinfo {pages} {072116} (\bibinfo {year} {2008})},\ \bibinfo {note} {arXiv:0805.2287 [physics]}\BibitemShut {NoStop}%
\bibitem [{\citenamefont {Pandey}\ \emph {et~al.}(2002)\citenamefont {Pandey}, \citenamefont {Vranješ}, \citenamefont {Poedts},\ and\ \citenamefont {Shukla}}]{pandey_pulsational_2002}%
  \BibitemOpen
  \bibfield  {author} {\bibinfo {author} {\bibfnamefont {B.~P.}\ \bibnamefont {Pandey}}, \bibinfo {author} {\bibfnamefont {J.}~\bibnamefont {Vranješ}}, \bibinfo {author} {\bibfnamefont {S.}~\bibnamefont {Poedts}},\ and\ \bibinfo {author} {\bibfnamefont {P.~K.}\ \bibnamefont {Shukla}},\ }\bibfield  {title} {\bibinfo {title} {The {Pulsational} {Mode} in the {Presence} of {Dust} {Charge} {Fluctuations}},\ }\href {https://doi.org/10.1238/Physica.Regular.065a00513} {\bibfield  {journal} {\bibinfo  {journal} {Physica Scripta}\ }\textbf {\bibinfo {volume} {65}},\ \bibinfo {pages} {513} (\bibinfo {year} {2002})}\BibitemShut {NoStop}%
\bibitem [{\citenamefont {Shankaranarayanan}\ and\ \citenamefont {Johnson}(2022)}]{shankaranarayanan_modified_2022}%
  \BibitemOpen
  \bibfield  {author} {\bibinfo {author} {\bibfnamefont {S.}~\bibnamefont {Shankaranarayanan}}\ and\ \bibinfo {author} {\bibfnamefont {J.~P.}\ \bibnamefont {Johnson}},\ }\bibfield  {title} {{\selectlanguage {en}\bibinfo {title} {Modified theories of gravity: {Why}, how and what?}},\ }\href {https://doi.org/10.1007/s10714-022-02927-2} {\bibfield  {journal} {\bibinfo  {journal} {General Relativity and Gravitation}\ }\textbf {\bibinfo {volume} {54}},\ \bibinfo {pages} {44} (\bibinfo {year} {2022})}\BibitemShut {NoStop}%
\bibitem [{\citenamefont {Clifton}\ \emph {et~al.}(2012)\citenamefont {Clifton}, \citenamefont {Ferreira}, \citenamefont {Padilla},\ and\ \citenamefont {Skordis}}]{clifton_modified_2012}%
  \BibitemOpen
  \bibfield  {author} {\bibinfo {author} {\bibfnamefont {T.}~\bibnamefont {Clifton}}, \bibinfo {author} {\bibfnamefont {P.~G.}\ \bibnamefont {Ferreira}}, \bibinfo {author} {\bibfnamefont {A.}~\bibnamefont {Padilla}},\ and\ \bibinfo {author} {\bibfnamefont {C.}~\bibnamefont {Skordis}},\ }\bibfield  {title} {{\selectlanguage {en}\bibinfo {title} {Modified gravity and cosmology}},\ }\href {https://doi.org/10.1016/j.physrep.2012.01.001} {\bibfield  {journal} {\bibinfo  {journal} {Physics Reports}\ }\textbf {\bibinfo {volume} {513}},\ \bibinfo {pages} {1} (\bibinfo {year} {2012})}\BibitemShut {NoStop}%
\bibitem [{\citenamefont {Sotiriou}\ and\ \citenamefont {Faraoni}(2010{\natexlab{a}})}]{sotiriou_fr_2010}%
  \BibitemOpen
  \bibfield  {author} {\bibinfo {author} {\bibfnamefont {T.~P.}\ \bibnamefont {Sotiriou}}\ and\ \bibinfo {author} {\bibfnamefont {V.}~\bibnamefont {Faraoni}},\ }\bibfield  {title} {{\selectlanguage {en}\bibinfo {title} {f({R}) {Theories} {Of} {Gravity}}},\ }\href {https://doi.org/10.1103/RevModPhys.82.451} {\bibfield  {journal} {\bibinfo  {journal} {Reviews of Modern Physics}\ }\textbf {\bibinfo {volume} {82}},\ \bibinfo {pages} {451} (\bibinfo {year} {2010}{\natexlab{a}})},\ \bibinfo {note} {arXiv:0805.1726 [astro-ph, physics:gr-qc, physics:hep-th]}\BibitemShut {NoStop}%
\bibitem [{\citenamefont {Vainio}\ and\ \citenamefont {Vilja}(2016)}]{vainio_jeans_2016}%
  \BibitemOpen
  \bibfield  {author} {\bibinfo {author} {\bibfnamefont {J.}~\bibnamefont {Vainio}}\ and\ \bibinfo {author} {\bibfnamefont {I.}~\bibnamefont {Vilja}},\ }\bibfield  {title} {{\selectlanguage {en}\bibinfo {title} {Jeans analysis of {Bok} globules in f({R}) gravity}},\ }\href {https://doi.org/10.1007/s10714-016-2120-8} {\bibfield  {journal} {\bibinfo  {journal} {General Relativity and Gravitation}\ }\textbf {\bibinfo {volume} {48}},\ \bibinfo {pages} {129} (\bibinfo {year} {2016})}\BibitemShut {NoStop}%
\bibitem [{\citenamefont {Ray}\ and\ \citenamefont {Karmakar}(2024)}]{ray_pulsational_2024}%
  \BibitemOpen
  \bibfield  {author} {\bibinfo {author} {\bibfnamefont {D.}~\bibnamefont {Ray}}\ and\ \bibinfo {author} {\bibfnamefont {P.~K.}\ \bibnamefont {Karmakar}},\ }\bibfield  {title} {\bibinfo {title} {Pulsational mode stability in complex {EiBI}-gravitating polarized astroclouds with (r,q)-distributed electrons},\ }\href {https://doi.org/10.1088/1475-7516/2024/12/014} {\bibfield  {journal} {\bibinfo  {journal} {Journal of Cosmology and Astroparticle Physics}\ }\textbf {\bibinfo {volume} {2024}}\bibinfo  {number} { (12)},\ \bibinfo {pages} {014}}\BibitemShut {NoStop}%
\bibitem [{\citenamefont {De~Martino}\ and\ \citenamefont {Capolupo}(2017)}]{de_martino_kinetic_2017}%
  \BibitemOpen
\bibfield  {number} {  }\bibfield  {author} {\bibinfo {author} {\bibfnamefont {I.}~\bibnamefont {De~Martino}}\ and\ \bibinfo {author} {\bibfnamefont {A.}~\bibnamefont {Capolupo}},\ }\bibfield  {title} {{\selectlanguage {en}\bibinfo {title} {Kinetic theory of {Jean} instability in {Eddington}-inspired {Born}–{Infeld} gravity}},\ }\href {https://doi.org/10.1140/epjc/s10052-017-5300-0} {\bibfield  {journal} {\bibinfo  {journal} {The European Physical Journal C}\ }\textbf {\bibinfo {volume} {77}},\ \bibinfo {pages} {715} (\bibinfo {year} {2017})}\BibitemShut {NoStop}%
\bibitem [{\citenamefont {He}(2022)}]{he_jeans_2022}%
  \BibitemOpen
  \bibfield  {author} {\bibinfo {author} {\bibfnamefont {K.-R.}\ \bibnamefont {He}},\ }\bibfield  {title} {{\selectlanguage {en}\bibinfo {title} {Jeans analysis with kappa-deformed {Kaniadakis} distribution in f ({R}) gravity}},\ }\href {https://doi.org/10.1088/1402-4896/ac485e} {\bibfield  {journal} {\bibinfo  {journal} {Physica Scripta}\ }\textbf {\bibinfo {volume} {97}},\ \bibinfo {pages} {025601} (\bibinfo {year} {2022})}\BibitemShut {NoStop}%
\bibitem [{\citenamefont {Roshan}\ and\ \citenamefont {Abbassi}(2014)}]{roshan_jeans_2014}%
  \BibitemOpen
  \bibfield  {author} {\bibinfo {author} {\bibfnamefont {M.}~\bibnamefont {Roshan}}\ and\ \bibinfo {author} {\bibfnamefont {S.}~\bibnamefont {Abbassi}},\ }\bibfield  {title} {{\selectlanguage {en}\bibinfo {title} {Jeans analysis in modified gravity}},\ }\href {https://doi.org/10.1103/PhysRevD.90.044010} {\bibfield  {journal} {\bibinfo  {journal} {Physical Review D}\ }\textbf {\bibinfo {volume} {90}},\ \bibinfo {pages} {044010} (\bibinfo {year} {2014})}\BibitemShut {NoStop}%
\bibitem [{\citenamefont {Das}\ and\ \citenamefont {Karmakar}(2024{\natexlab{a}})}]{das_dynamics_2024}%
  \BibitemOpen
  \bibfield  {author} {\bibinfo {author} {\bibfnamefont {M.}~\bibnamefont {Das}}\ and\ \bibinfo {author} {\bibfnamefont {P.~K.}\ \bibnamefont {Karmakar}},\ }\bibfield  {title} {{\selectlanguage {en}\bibinfo {title} {Dynamics of pulsational mode in the {EiBI} gravity fabric}},\ }\href {https://doi.org/10.1016/j.cjph.2023.11.032} {\bibfield  {journal} {\bibinfo  {journal} {Chinese Journal of Physics}\ }\textbf {\bibinfo {volume} {87}},\ \bibinfo {pages} {232} (\bibinfo {year} {2024}{\natexlab{a}})}\BibitemShut {NoStop}%
\bibitem [{\citenamefont {Bessiri}\ \emph {et~al.}(2021)\citenamefont {Bessiri}, \citenamefont {Ourabah},\ and\ \citenamefont {Zerguini}}]{bessiri_jeans_2021}%
  \BibitemOpen
  \bibfield  {author} {\bibinfo {author} {\bibfnamefont {A.}~\bibnamefont {Bessiri}}, \bibinfo {author} {\bibfnamefont {K.}~\bibnamefont {Ourabah}},\ and\ \bibinfo {author} {\bibfnamefont {T.~H.}\ \bibnamefont {Zerguini}},\ }\bibfield  {title} {\bibinfo {title} {Jeans instability in {Eddington}-inspired {Born}-{Infeld} ({EiBI}) gravity: a quantum approach},\ }\href {https://doi.org/10.1088/1402-4896/ac1cd2} {\bibfield  {journal} {\bibinfo  {journal} {Physica Scripta}\ }\textbf {\bibinfo {volume} {96}},\ \bibinfo {pages} {125208} (\bibinfo {year} {2021})}\BibitemShut {NoStop}%
\bibitem [{\citenamefont {Yang}\ \emph {et~al.}(2023)\citenamefont {Yang}, \citenamefont {Tan}, \citenamefont {Chen},\ and\ \citenamefont {Liu}}]{yang_jeans_2023}%
  \BibitemOpen
  \bibfield  {author} {\bibinfo {author} {\bibfnamefont {Q.}~\bibnamefont {Yang}}, \bibinfo {author} {\bibfnamefont {L.}~\bibnamefont {Tan}}, \bibinfo {author} {\bibfnamefont {H.}~\bibnamefont {Chen}},\ and\ \bibinfo {author} {\bibfnamefont {S.}~\bibnamefont {Liu}},\ }\bibfield  {title} {{\selectlanguage {en}\bibinfo {title} {Jeans instability analysis of viscoelastic astrofluids in {Eddington}-{Inspired}-{Born}–{Infeld}({EiBI}) gravity}},\ }\href {https://doi.org/10.1016/j.newast.2022.101947} {\bibfield  {journal} {\bibinfo  {journal} {New Astronomy}\ }\textbf {\bibinfo {volume} {99}},\ \bibinfo {pages} {101947} (\bibinfo {year} {2023})}\BibitemShut {NoStop}%
\bibitem [{\citenamefont {Brans}\ and\ \citenamefont {Dicke}(1961)}]{brans_machs_1961}%
  \BibitemOpen
  \bibfield  {author} {\bibinfo {author} {\bibfnamefont {C.}~\bibnamefont {Brans}}\ and\ \bibinfo {author} {\bibfnamefont {R.~H.}\ \bibnamefont {Dicke}},\ }\bibfield  {title} {{\selectlanguage {en}\bibinfo {title} {Mach's {Principle} and a {Relativistic} {Theory} of {Gravitation}}},\ }\href {https://doi.org/10.1103/PhysRev.124.925} {\bibfield  {journal} {\bibinfo  {journal} {Physical Review}\ }\textbf {\bibinfo {volume} {124}},\ \bibinfo {pages} {925} (\bibinfo {year} {1961})}\BibitemShut {NoStop}%
\bibitem [{\citenamefont {Faraoni}(2004)}]{faraoni_scalar-tensor_2004}%
  \BibitemOpen
  \bibfield  {author} {\bibinfo {author} {\bibfnamefont {V.}~\bibnamefont {Faraoni}},\ }\bibfield  {title} {{\selectlanguage {en}\bibinfo {title} {Scalar-{Tensor} {Gravity}}},\ }in\ \href {https://doi.org/10.1007/978-1-4020-1989-0_1} {{\selectlanguage {en}\emph {\bibinfo {booktitle} {Cosmology in {Scalar}-{Tensor} {Gravity}}}}}\ (\bibinfo  {publisher} {Springer Netherlands},\ \bibinfo {address} {Dordrecht},\ \bibinfo {year} {2004})\ pp.\ \bibinfo {pages} {1--53}\BibitemShut {NoStop}%
\bibitem [{\citenamefont {Sotiriou}\ and\ \citenamefont {Faraoni}(2010{\natexlab{b}})}]{sotiriou_f_2010}%
  \BibitemOpen
  \bibfield  {author} {\bibinfo {author} {\bibfnamefont {T.~P.}\ \bibnamefont {Sotiriou}}\ and\ \bibinfo {author} {\bibfnamefont {V.}~\bibnamefont {Faraoni}},\ }\bibfield  {title} {{\selectlanguage {en}\bibinfo {title} {f ( {R} ) theories of gravity}},\ }\href {https://doi.org/10.1103/RevModPhys.82.451} {\bibfield  {journal} {\bibinfo  {journal} {Reviews of Modern Physics}\ }\textbf {\bibinfo {volume} {82}},\ \bibinfo {pages} {451} (\bibinfo {year} {2010}{\natexlab{b}})}\BibitemShut {NoStop}%
\bibitem [{\citenamefont {Glavan}\ and\ \citenamefont {Lin}(2020)}]{glavan_einstein-gauss-bonnet_2020}%
  \BibitemOpen
  \bibfield  {author} {\bibinfo {author} {\bibfnamefont {D.}~\bibnamefont {Glavan}}\ and\ \bibinfo {author} {\bibfnamefont {C.}~\bibnamefont {Lin}},\ }\bibfield  {title} {{\selectlanguage {en}\bibinfo {title} {Einstein-{Gauss}-{Bonnet} {Gravity} in {Four}-{Dimensional} {Spacetime}}},\ }\href {https://doi.org/10.1103/PhysRevLett.124.081301} {\bibfield  {journal} {\bibinfo  {journal} {Physical Review Letters}\ }\textbf {\bibinfo {volume} {124}},\ \bibinfo {pages} {081301} (\bibinfo {year} {2020})}\BibitemShut {NoStop}%
\bibitem [{\citenamefont {Clifton}(2008)}]{clifton_parameterised_2008}%
  \BibitemOpen
  \bibfield  {author} {\bibinfo {author} {\bibfnamefont {T.}~\bibnamefont {Clifton}},\ }\bibfield  {title} {{\selectlanguage {en}\bibinfo {title} {The {Parameterised} {Post}-{Newtonian} {Limit} of {Fourth}-{Order} {Theories} of {Gravity}}},\ }\href {https://doi.org/10.1103/PhysRevD.77.024041} {\bibfield  {journal} {\bibinfo  {journal} {Physical Review D}\ }\textbf {\bibinfo {volume} {77}},\ \bibinfo {pages} {024041} (\bibinfo {year} {2008})},\ \bibinfo {note} {arXiv:0801.0983 [astro-ph, physics:gr-qc]}\BibitemShut {NoStop}%
\bibitem [{\citenamefont {Bañados}\ and\ \citenamefont {Ferreira}(2010)}]{banados_eddingtons_2010}%
  \BibitemOpen
  \bibfield  {author} {\bibinfo {author} {\bibfnamefont {M.}~\bibnamefont {Bañados}}\ and\ \bibinfo {author} {\bibfnamefont {P.~G.}\ \bibnamefont {Ferreira}},\ }\bibfield  {title} {{\selectlanguage {en}\bibinfo {title} {Eddington’s {Theory} of {Gravity} and {Its} {Progeny}}},\ }\href {https://doi.org/10.1103/PhysRevLett.105.011101} {\bibfield  {journal} {\bibinfo  {journal} {Physical Review Letters}\ }\textbf {\bibinfo {volume} {105}},\ \bibinfo {pages} {011101} (\bibinfo {year} {2010})}\BibitemShut {NoStop}%
\bibitem [{\citenamefont {Pani}\ \emph {et~al.}(2012)\citenamefont {Pani}, \citenamefont {Delsate},\ and\ \citenamefont {Cardoso}}]{pani_eddington-inspired_2012}%
  \BibitemOpen
  \bibfield  {author} {\bibinfo {author} {\bibfnamefont {P.}~\bibnamefont {Pani}}, \bibinfo {author} {\bibfnamefont {T.}~\bibnamefont {Delsate}},\ and\ \bibinfo {author} {\bibfnamefont {V.}~\bibnamefont {Cardoso}},\ }\bibfield  {title} {{\selectlanguage {en}\bibinfo {title} {Eddington-inspired {Born}-{Infeld} gravity: {Phenomenology} of nonlinear gravity-matter coupling}},\ }\href {https://doi.org/10.1103/PhysRevD.85.084020} {\bibfield  {journal} {\bibinfo  {journal} {Physical Review D}\ }\textbf {\bibinfo {volume} {85}},\ \bibinfo {pages} {084020} (\bibinfo {year} {2012})}\BibitemShut {NoStop}%
\bibitem [{\citenamefont {Avelino}(2012{\natexlab{a}})}]{avelino_eddington-inspired_2012}%
  \BibitemOpen
  \bibfield  {author} {\bibinfo {author} {\bibfnamefont {P.}~\bibnamefont {Avelino}},\ }\bibfield  {title} {{\selectlanguage {en}\bibinfo {title} {Eddington-inspired {Born}-{Infeld} gravity: nuclear physics constraints and the validity of the continuous fluid approximation}},\ }\href {https://doi.org/10.1088/1475-7516/2012/11/022} {\bibfield  {journal} {\bibinfo  {journal} {Journal of Cosmology and Astroparticle Physics}\ }\textbf {\bibinfo {volume} {2012}}\bibinfo  {number} { (11)},\ \bibinfo {pages} {022}}\BibitemShut {NoStop}%
\bibitem [{\citenamefont {Avelino}(2012{\natexlab{b}})}]{avelino_eddington-inspired_2012-1}%
  \BibitemOpen
\bibfield  {number} {  }\bibfield  {author} {\bibinfo {author} {\bibfnamefont {P.~P.}\ \bibnamefont {Avelino}},\ }\bibfield  {title} {{\selectlanguage {en}\bibinfo {title} {Eddington-inspired {Born}-{Infeld} gravity: {Astrophysical} and cosmological constraints}},\ }\href {https://doi.org/10.1103/PhysRevD.85.104053} {\bibfield  {journal} {\bibinfo  {journal} {Physical Review D}\ }\textbf {\bibinfo {volume} {85}},\ \bibinfo {pages} {104053} (\bibinfo {year} {2012}{\natexlab{b}})}\BibitemShut {NoStop}%
\bibitem [{\citenamefont {Banerjee}\ \emph {et~al.}(2022)\citenamefont {Banerjee}, \citenamefont {Garain}, \citenamefont {Paul}, \citenamefont {Shaikh},\ and\ \citenamefont {Sarkar}}]{banerjee_stellar_2022}%
  \BibitemOpen
  \bibfield  {author} {\bibinfo {author} {\bibfnamefont {P.}~\bibnamefont {Banerjee}}, \bibinfo {author} {\bibfnamefont {D.}~\bibnamefont {Garain}}, \bibinfo {author} {\bibfnamefont {S.}~\bibnamefont {Paul}}, \bibinfo {author} {\bibfnamefont {R.}~\bibnamefont {Shaikh}},\ and\ \bibinfo {author} {\bibfnamefont {T.}~\bibnamefont {Sarkar}},\ }\bibfield  {title} {{\selectlanguage {en}\bibinfo {title} {A {Stellar} {Constraint} on {Eddington}-inspired {Born}–{Infeld} {Gravity} from {Cataclysmic} {Variable} {Binaries}}},\ }\href {https://doi.org/10.3847/1538-4357/ac324f} {\bibfield  {journal} {\bibinfo  {journal} {The Astrophysical Journal}\ }\textbf {\bibinfo {volume} {924}},\ \bibinfo {pages} {20} (\bibinfo {year} {2022})}\BibitemShut {NoStop}%
\bibitem [{\citenamefont {Pani}\ \emph {et~al.}(2011)\citenamefont {Pani}, \citenamefont {Cardoso},\ and\ \citenamefont {Delsate}}]{pani_compact_2011}%
  \BibitemOpen
  \bibfield  {author} {\bibinfo {author} {\bibfnamefont {P.}~\bibnamefont {Pani}}, \bibinfo {author} {\bibfnamefont {V.}~\bibnamefont {Cardoso}},\ and\ \bibinfo {author} {\bibfnamefont {T.}~\bibnamefont {Delsate}},\ }\bibfield  {title} {{\selectlanguage {en}\bibinfo {title} {Compact {Stars} in {Eddington} {Inspired} {Gravity}}},\ }\href {https://doi.org/10.1103/PhysRevLett.107.031101} {\bibfield  {journal} {\bibinfo  {journal} {Physical Review Letters}\ }\textbf {\bibinfo {volume} {107}},\ \bibinfo {pages} {031101} (\bibinfo {year} {2011})}\BibitemShut {NoStop}%
\bibitem [{\citenamefont {Prasetyo}\ \emph {et~al.}(2021)\citenamefont {Prasetyo}, \citenamefont {Maulana}, \citenamefont {Ramadhan},\ and\ \citenamefont {Sulaksono}}]{prasetyo_26text_2021}%
  \BibitemOpen
  \bibfield  {author} {\bibinfo {author} {\bibfnamefont {I.}~\bibnamefont {Prasetyo}}, \bibinfo {author} {\bibfnamefont {H.}~\bibnamefont {Maulana}}, \bibinfo {author} {\bibfnamefont {H.}~\bibnamefont {Ramadhan}},\ and\ \bibinfo {author} {\bibfnamefont {A.}~\bibnamefont {Sulaksono}},\ }\bibfield  {title} {{\selectlanguage {en}\bibinfo {title} {\$2.6{\textbackslash}text\{ \}{\textbackslash}text\{ \}\{{M}\}\_\{{\textbackslash}ensuremath\{{\textbackslash}bigodot\vphantom{\{}\}\}\$ compact object and neutron stars in {Eddington}-inspired {Born}-{Infeld} theory of gravity\}, author = \{\vphantom{\}}{Prasetyo}, {I}. and {Maulana}, {H}. and {Ramadhan}, {H}. {S}. and {Sulaksono}, {A}.}},\ }\href {https://doi.org/10.1103/PhysRevD.104.084029} {\bibfield  {journal} {\bibinfo  {journal} {Physical Review D}\ }\textbf {\bibinfo {volume} {104}},\ \bibinfo {pages} {084029} (\bibinfo {year} {2021})}\BibitemShut {NoStop}%
\bibitem [{\citenamefont {Sham}\ \emph {et~al.}(2012)\citenamefont {Sham}, \citenamefont {Lin},\ and\ \citenamefont {Leung}}]{sham_radial_2012}%
  \BibitemOpen
  \bibfield  {author} {\bibinfo {author} {\bibfnamefont {Y.-H.}\ \bibnamefont {Sham}}, \bibinfo {author} {\bibfnamefont {L.-M.}\ \bibnamefont {Lin}},\ and\ \bibinfo {author} {\bibfnamefont {P.~T.}\ \bibnamefont {Leung}},\ }\bibfield  {title} {{\selectlanguage {en}\bibinfo {title} {Radial oscillations and stability of compact stars in {Eddington}-inspired {Born}-{Infeld} gravity}},\ }\href {https://doi.org/10.1103/PhysRevD.86.064015} {\bibfield  {journal} {\bibinfo  {journal} {Physical Review D}\ }\textbf {\bibinfo {volume} {86}},\ \bibinfo {pages} {064015} (\bibinfo {year} {2012})}\BibitemShut {NoStop}%
\bibitem [{\citenamefont {Banerjee}\ \emph {et~al.}(2017)\citenamefont {Banerjee}, \citenamefont {Shankar},\ and\ \citenamefont {Singh}}]{banerjee_constraints_2017}%
  \BibitemOpen
  \bibfield  {author} {\bibinfo {author} {\bibfnamefont {S.}~\bibnamefont {Banerjee}}, \bibinfo {author} {\bibfnamefont {S.}~\bibnamefont {Shankar}},\ and\ \bibinfo {author} {\bibfnamefont {T.~P.}\ \bibnamefont {Singh}},\ }\bibfield  {title} {{\selectlanguage {en}\bibinfo {title} {Constraints on modified gravity models from white dwarfs}},\ }\href {https://doi.org/10.1088/1475-7516/2017/10/004} {\bibfield  {journal} {\bibinfo  {journal} {Journal of Cosmology and Astroparticle Physics}\ }\textbf {\bibinfo {volume} {2017}}\bibinfo  {number} { (10)},\ \bibinfo {pages} {004}}\BibitemShut {NoStop}%
\bibitem [{\citenamefont {Casanellas}\ \emph {et~al.}(2012)\citenamefont {Casanellas}, \citenamefont {Pani}, \citenamefont {Lopes},\ and\ \citenamefont {Cardoso}}]{casanellas_testing_2012}%
  \BibitemOpen
\bibfield  {number} {  }\bibfield  {author} {\bibinfo {author} {\bibfnamefont {J.}~\bibnamefont {Casanellas}}, \bibinfo {author} {\bibfnamefont {P.}~\bibnamefont {Pani}}, \bibinfo {author} {\bibfnamefont {I.}~\bibnamefont {Lopes}},\ and\ \bibinfo {author} {\bibfnamefont {V.}~\bibnamefont {Cardoso}},\ }\bibfield  {title} {{\selectlanguage {en}\bibinfo {title} {Testing {Alternative} {Theories} of {Gravity} {Using} the {Sun}}},\ }\href {https://doi.org/10.1088/0004-637X/745/1/15} {\bibfield  {journal} {\bibinfo  {journal} {The Astrophysical Journal}\ }\textbf {\bibinfo {volume} {745}},\ \bibinfo {pages} {15} (\bibinfo {year} {2012})}\BibitemShut {NoStop}%
\bibitem [{\citenamefont {Das}\ and\ \citenamefont {Karmakar}(2024{\natexlab{b}})}]{das_solar_2024}%
  \BibitemOpen
  \bibfield  {author} {\bibinfo {author} {\bibfnamefont {S.}~\bibnamefont {Das}}\ and\ \bibinfo {author} {\bibfnamefont {P.~K.}\ \bibnamefont {Karmakar}},\ }\bibfield  {title} {{\selectlanguage {en}\bibinfo {title} {Solar {GES}-structure modified with {EiBI} gravity}},\ }\href {https://doi.org/10.1016/j.cjph.2024.05.008} {\bibfield  {journal} {\bibinfo  {journal} {Chinese Journal of Physics}\ }\textbf {\bibinfo {volume} {91}},\ \bibinfo {pages} {157} (\bibinfo {year} {2024}{\natexlab{b}})}\BibitemShut {NoStop}%
\bibitem [{\citenamefont {Will}(2016)}]{peron_gravity_2016}%
  \BibitemOpen
  \bibfield  {author} {\bibinfo {author} {\bibfnamefont {C.~M.}\ \bibnamefont {Will}},\ }\bibfield  {title} {{\selectlanguage {en}\bibinfo {title} {Gravity: {Newtonian}, {Post}-{Newtonian}, and {General} {Relativistic}}},\ }in\ \href {https://doi.org/10.1007/978-3-319-20224-2_2} {{\selectlanguage {en}\emph {\bibinfo {booktitle} {Gravity: {Where} {Do} {We} {Stand}?}}}},\ \bibinfo {editor} {edited by\ \bibinfo {editor} {\bibfnamefont {R.}~\bibnamefont {Peron}}, \bibinfo {editor} {\bibfnamefont {M.}~\bibnamefont {Colpi}}, \bibinfo {editor} {\bibfnamefont {V.}~\bibnamefont {Gorini}},\ and\ \bibinfo {editor} {\bibfnamefont {U.}~\bibnamefont {Moschella}}}\ (\bibinfo  {publisher} {Springer International Publishing},\ \bibinfo {address} {Cham},\ \bibinfo {year} {2016})\ pp.\ \bibinfo {pages} {9--72}\BibitemShut {NoStop}%
\bibitem [{\citenamefont {Vasyliunas}(1968)}]{carovillano_low-energy_1968}%
  \BibitemOpen
  \bibfield  {author} {\bibinfo {author} {\bibfnamefont {V.~M.}\ \bibnamefont {Vasyliunas}},\ }\bibfield  {title} {\bibinfo {title} {Low-{Energy} {Electrons} in the {Magnetosphere} as {Observed} by {OGO}-1 and {OGO}-3},\ }in\ \href {https://doi.org/10.1007/978-94-010-3467-8_22} {\emph {\bibinfo {booktitle} {Physics of the {Magnetosphere}}}},\ Vol.~\bibinfo {volume} {10},\ \bibinfo {editor} {edited by\ \bibinfo {editor} {\bibfnamefont {R.~L.}\ \bibnamefont {Carovillano}}, \bibinfo {editor} {\bibfnamefont {J.~F.}\ \bibnamefont {McClay}},\ and\ \bibinfo {editor} {\bibfnamefont {H.~R.}\ \bibnamefont {Radoski}}}\ (\bibinfo  {publisher} {Springer Netherlands},\ \bibinfo {address} {Dordrecht},\ \bibinfo {year} {1968})\ pp.\ \bibinfo {pages} {622--640},\ \bibinfo {note} {series Title: Astrophysics and Space Science Library}\BibitemShut {NoStop}%
\bibitem [{\citenamefont {Livadiotis}\ \emph {et~al.}(2011)\citenamefont {Livadiotis}, \citenamefont {McComas}, \citenamefont {Dayeh}, \citenamefont {Funsten},\ and\ \citenamefont {Schwadron}}]{livadiotis_first_2011}%
  \BibitemOpen
  \bibfield  {author} {\bibinfo {author} {\bibfnamefont {G.}~\bibnamefont {Livadiotis}}, \bibinfo {author} {\bibfnamefont {D.~J.}\ \bibnamefont {McComas}}, \bibinfo {author} {\bibfnamefont {M.~A.}\ \bibnamefont {Dayeh}}, \bibinfo {author} {\bibfnamefont {H.~O.}\ \bibnamefont {Funsten}},\ and\ \bibinfo {author} {\bibfnamefont {N.~A.}\ \bibnamefont {Schwadron}},\ }\bibfield  {title} {\bibinfo {title} {First {Sky} {Map} of the {Inner} {Heliosheath} {Temperature} {Using} \textit{{Ibex}} {Spectra}},\ }\href {https://doi.org/10.1088/0004-637X/734/1/1} {\bibfield  {journal} {\bibinfo  {journal} {The Astrophysical Journal}\ }\textbf {\bibinfo {volume} {734}},\ \bibinfo {pages} {1} (\bibinfo {year} {2011})}\BibitemShut {NoStop}%
\bibitem [{\citenamefont {Pierrard}\ \emph {et~al.}(1999)\citenamefont {Pierrard}, \citenamefont {Maksimovic},\ and\ \citenamefont {Lemaire}}]{pierrard_electron_1999}%
  \BibitemOpen
  \bibfield  {author} {\bibinfo {author} {\bibfnamefont {V.}~\bibnamefont {Pierrard}}, \bibinfo {author} {\bibfnamefont {M.}~\bibnamefont {Maksimovic}},\ and\ \bibinfo {author} {\bibfnamefont {J.}~\bibnamefont {Lemaire}},\ }\bibfield  {title} {{\selectlanguage {en}\bibinfo {title} {Electron velocity distribution functions from the solar wind to the corona}},\ }\href {https://doi.org/10.1029/1999JA900169} {\bibfield  {journal} {\bibinfo  {journal} {Journal of Geophysical Research: Space Physics}\ }\textbf {\bibinfo {volume} {104}},\ \bibinfo {pages} {17021} (\bibinfo {year} {1999})}\BibitemShut {NoStop}%
\bibitem [{\citenamefont {Pierrard}\ and\ \citenamefont {Lazar}(2010)}]{pierrard_kappa_2010}%
  \BibitemOpen
  \bibfield  {author} {\bibinfo {author} {\bibfnamefont {V.}~\bibnamefont {Pierrard}}\ and\ \bibinfo {author} {\bibfnamefont {M.}~\bibnamefont {Lazar}},\ }\bibfield  {title} {{\selectlanguage {en}\bibinfo {title} {Kappa {Distributions}: {Theory} and {Applications} in {Space} {Plasmas}}},\ }\href {https://doi.org/10.1007/s11207-010-9640-2} {\bibfield  {journal} {\bibinfo  {journal} {Solar Physics}\ }\textbf {\bibinfo {volume} {267}},\ \bibinfo {pages} {153} (\bibinfo {year} {2010})}\BibitemShut {NoStop}%
\bibitem [{\citenamefont {Sun}\ \emph {et~al.}(2025)\citenamefont {Sun}, \citenamefont {Oka}, \citenamefont {Øieroset}, \citenamefont {Turner}, \citenamefont {Phan}, \citenamefont {Cohen}, \citenamefont {Li}, \citenamefont {Huang}, \citenamefont {Smith}, \citenamefont {Slavin}, \citenamefont {Poh}, \citenamefont {Genestreti}, \citenamefont {Gershman}, \citenamefont {Dokgo}, \citenamefont {Le}, \citenamefont {Nakamura},\ and\ \citenamefont {Burch}}]{sun_relativistic_2025}%
  \BibitemOpen
  \bibfield  {author} {\bibinfo {author} {\bibfnamefont {W.}~\bibnamefont {Sun}}, \bibinfo {author} {\bibfnamefont {M.}~\bibnamefont {Oka}}, \bibinfo {author} {\bibfnamefont {M.}~\bibnamefont {Øieroset}}, \bibinfo {author} {\bibfnamefont {D.~L.}\ \bibnamefont {Turner}}, \bibinfo {author} {\bibfnamefont {T.}~\bibnamefont {Phan}}, \bibinfo {author} {\bibfnamefont {I.~J.}\ \bibnamefont {Cohen}}, \bibinfo {author} {\bibfnamefont {X.}~\bibnamefont {Li}}, \bibinfo {author} {\bibfnamefont {J.}~\bibnamefont {Huang}}, \bibinfo {author} {\bibfnamefont {A.~W.}\ \bibnamefont {Smith}}, \bibinfo {author} {\bibfnamefont {J.~A.}\ \bibnamefont {Slavin}}, \bibinfo {author} {\bibfnamefont {G.}~\bibnamefont {Poh}}, \bibinfo {author} {\bibfnamefont {K.~J.}\ \bibnamefont {Genestreti}}, \bibinfo {author} {\bibfnamefont {D.}~\bibnamefont {Gershman}}, \bibinfo {author} {\bibfnamefont {K.}~\bibnamefont {Dokgo}}, \bibinfo {author} {\bibfnamefont {G.}~\bibnamefont {Le}}, \bibinfo {author} {\bibfnamefont {R.}~\bibnamefont {Nakamura}},\
  and\ \bibinfo {author} {\bibfnamefont {J.~L.}\ \bibnamefont {Burch}},\ }\bibfield  {title} {\bibinfo {title} {Relativistic {Electron} {Acceleration} and the “{Ankle}” {Spectral} {Feature} in {Earth}’s {Magnetotail} {Reconnection}},\ }\href {https://doi.org/10.3847/2041-8213/ad9bb2} {\bibfield  {journal} {\bibinfo  {journal} {The Astrophysical Journal Letters}\ }\textbf {\bibinfo {volume} {978}},\ \bibinfo {pages} {L28} (\bibinfo {year} {2025})}\BibitemShut {NoStop}%
\bibitem [{\citenamefont {Li}\ \emph {et~al.}(2025)\citenamefont {Li}, \citenamefont {Jie}, \citenamefont {Liang}, \citenamefont {Jiang},\ and\ \citenamefont {Gao}}]{li_effect_2025}%
  \BibitemOpen
  \bibfield  {author} {\bibinfo {author} {\bibfnamefont {Z.-Z.}\ \bibnamefont {Li}}, \bibinfo {author} {\bibfnamefont {L.-Q.}\ \bibnamefont {Jie}}, \bibinfo {author} {\bibfnamefont {S.-D.}\ \bibnamefont {Liang}}, \bibinfo {author} {\bibfnamefont {K.}~\bibnamefont {Jiang}},\ and\ \bibinfo {author} {\bibfnamefont {D.-N.}\ \bibnamefont {Gao}},\ }\bibfield  {title} {{\selectlanguage {en}\bibinfo {title} {Effect of {Regularized} \$\${\textbackslash}kappa \$\$ {Distribution} and {Polarization} {Force} on the {Dust} {Acoustic} {Waves} in the {Mesosphere} {Region}}},\ }\href {https://doi.org/10.1007/s13538-024-01648-y} {\bibfield  {journal} {\bibinfo  {journal} {Brazilian Journal of Physics}\ }\textbf {\bibinfo {volume} {55}},\ \bibinfo {pages} {17} (\bibinfo {year} {2025})}\BibitemShut {NoStop}%
\bibitem [{\citenamefont {Heerikhuisen}\ \emph {et~al.}(2015)\citenamefont {Heerikhuisen}, \citenamefont {Zirnstein},\ and\ \citenamefont {Pogorelov}}]{heerikhuisen_kappadistributed_2015}%
  \BibitemOpen
  \bibfield  {author} {\bibinfo {author} {\bibfnamefont {J.}~\bibnamefont {Heerikhuisen}}, \bibinfo {author} {\bibfnamefont {E.}~\bibnamefont {Zirnstein}},\ and\ \bibinfo {author} {\bibfnamefont {N.}~\bibnamefont {Pogorelov}},\ }\bibfield  {title} {{\selectlanguage {en}\bibinfo {title} {kappa‐distributed protons in the solar wind and their charge‐exchange coupling to energetic hydrogen}},\ }\href {https://doi.org/10.1002/2014JA020636} {\bibfield  {journal} {\bibinfo  {journal} {Journal of Geophysical Research: Space Physics}\ }\textbf {\bibinfo {volume} {120}},\ \bibinfo {pages} {1516} (\bibinfo {year} {2015})}\BibitemShut {NoStop}%
\bibitem [{\citenamefont {Sarma}\ and\ \citenamefont {Karmakar}(2022)}]{sarma_solar_2022}%
  \BibitemOpen
  \bibfield  {author} {\bibinfo {author} {\bibfnamefont {P.}~\bibnamefont {Sarma}}\ and\ \bibinfo {author} {\bibfnamefont {P.~K.}\ \bibnamefont {Karmakar}},\ }\bibfield  {title} {{\selectlanguage {en}\bibinfo {title} {Solar plasma characterization in {Kappa}-modified polytropic turbomagnetic {GES}-model perspective}},\ }\href {https://doi.org/10.1093/mnras/stac3178} {\bibfield  {journal} {\bibinfo  {journal} {Monthly Notices of the Royal Astronomical Society}\ }\textbf {\bibinfo {volume} {519}},\ \bibinfo {pages} {2879} (\bibinfo {year} {2022})}\BibitemShut {NoStop}%
\bibitem [{\citenamefont {Dialynas}\ \emph {et~al.}(2009)\citenamefont {Dialynas}, \citenamefont {Krimigis}, \citenamefont {Mitchell}, \citenamefont {Hamilton}, \citenamefont {Krupp},\ and\ \citenamefont {Brandt}}]{dialynas_energetic_2009}%
  \BibitemOpen
  \bibfield  {author} {\bibinfo {author} {\bibfnamefont {K.}~\bibnamefont {Dialynas}}, \bibinfo {author} {\bibfnamefont {S.~M.}\ \bibnamefont {Krimigis}}, \bibinfo {author} {\bibfnamefont {D.~G.}\ \bibnamefont {Mitchell}}, \bibinfo {author} {\bibfnamefont {D.~C.}\ \bibnamefont {Hamilton}}, \bibinfo {author} {\bibfnamefont {N.}~\bibnamefont {Krupp}},\ and\ \bibinfo {author} {\bibfnamefont {P.~C.}\ \bibnamefont {Brandt}},\ }\bibfield  {title} {{\selectlanguage {en}\bibinfo {title} {Energetic ion spectral characteristics in the {Saturnian} magnetosphere using {Cassini}/{MIMI} measurements}},\ }\href {https://doi.org/10.1029/2008JA013761} {\bibfield  {journal} {\bibinfo  {journal} {Journal of Geophysical Research: Space Physics}\ }\textbf {\bibinfo {volume} {114}},\ \bibinfo {pages} {2008JA013761} (\bibinfo {year} {2009})}\BibitemShut {NoStop}%
\bibitem [{\citenamefont {Nicolaou}\ \emph {et~al.}(2014)\citenamefont {Nicolaou}, \citenamefont {McComas}, \citenamefont {Bagenal},\ and\ \citenamefont {Elliott}}]{nicolaou_properties_2014}%
  \BibitemOpen
  \bibfield  {author} {\bibinfo {author} {\bibfnamefont {G.}~\bibnamefont {Nicolaou}}, \bibinfo {author} {\bibfnamefont {D.~J.}\ \bibnamefont {McComas}}, \bibinfo {author} {\bibfnamefont {F.}~\bibnamefont {Bagenal}},\ and\ \bibinfo {author} {\bibfnamefont {H.~A.}\ \bibnamefont {Elliott}},\ }\bibfield  {title} {{\selectlanguage {en}\bibinfo {title} {Properties of plasma ions in the distant {Jovian} magnetosheath using {Solar} {Wind} {Around} {Pluto} data on {New} {Horizons}}},\ }\href {https://doi.org/10.1002/2013JA019665} {\bibfield  {journal} {\bibinfo  {journal} {Journal of Geophysical Research: Space Physics}\ }\textbf {\bibinfo {volume} {119}},\ \bibinfo {pages} {3463} (\bibinfo {year} {2014})}\BibitemShut {NoStop}%
\bibitem [{\citenamefont {Eyelade}\ \emph {et~al.}(2021)\citenamefont {Eyelade}, \citenamefont {Stepanova}, \citenamefont {Espinoza},\ and\ \citenamefont {Moya}}]{eyelade_relation_2021}%
  \BibitemOpen
  \bibfield  {author} {\bibinfo {author} {\bibfnamefont {A.~V.}\ \bibnamefont {Eyelade}}, \bibinfo {author} {\bibfnamefont {M.}~\bibnamefont {Stepanova}}, \bibinfo {author} {\bibfnamefont {C.~M.}\ \bibnamefont {Espinoza}},\ and\ \bibinfo {author} {\bibfnamefont {P.~S.}\ \bibnamefont {Moya}},\ }\bibfield  {title} {\bibinfo {title} {On the {Relation} between {Kappa} {Distribution} {Functions} and the {Plasma} {Beta} {Parameter} in the {Earth}’s {Magnetosphere}: {THEMIS} {Observations}},\ }\href {https://doi.org/10.3847/1538-4365/abdec9} {\bibfield  {journal} {\bibinfo  {journal} {The Astrophysical Journal Supplement Series}\ }\textbf {\bibinfo {volume} {253}},\ \bibinfo {pages} {34} (\bibinfo {year} {2021})}\BibitemShut {NoStop}%
\bibitem [{\citenamefont {Richardson}\ \emph {et~al.}(2022)\citenamefont {Richardson}, \citenamefont {Burlaga}, \citenamefont {Elliott}, \citenamefont {Kurth}, \citenamefont {Liu},\ and\ \citenamefont {Von~Steiger}}]{richardson_observations_2022}%
  \BibitemOpen
  \bibfield  {author} {\bibinfo {author} {\bibfnamefont {J.~D.}\ \bibnamefont {Richardson}}, \bibinfo {author} {\bibfnamefont {L.~F.}\ \bibnamefont {Burlaga}}, \bibinfo {author} {\bibfnamefont {H.}~\bibnamefont {Elliott}}, \bibinfo {author} {\bibfnamefont {W.~S.}\ \bibnamefont {Kurth}}, \bibinfo {author} {\bibfnamefont {Y.~D.}\ \bibnamefont {Liu}},\ and\ \bibinfo {author} {\bibfnamefont {R.}~\bibnamefont {Von~Steiger}},\ }\bibfield  {title} {{\selectlanguage {en}\bibinfo {title} {Observations of the {Outer} {Heliosphere}, {Heliosheath}, and {Interstellar} {Medium}}},\ }\href {https://doi.org/10.1007/s11214-022-00899-y} {\bibfield  {journal} {\bibinfo  {journal} {Space Science Reviews}\ }\textbf {\bibinfo {volume} {218}},\ \bibinfo {pages} {35} (\bibinfo {year} {2022})}\BibitemShut {NoStop}%
\bibitem [{\citenamefont {Zhang}\ \emph {et~al.}(2004)\citenamefont {Zhang}, \citenamefont {Liu}, \citenamefont {Wesson}, \citenamefont {Storey}, \citenamefont {Liu},\ and\ \citenamefont {Danziger}}]{zhang_electron_2004}%
  \BibitemOpen
  \bibfield  {author} {\bibinfo {author} {\bibfnamefont {Y.}~\bibnamefont {Zhang}}, \bibinfo {author} {\bibfnamefont {X.-W.}\ \bibnamefont {Liu}}, \bibinfo {author} {\bibfnamefont {R.}~\bibnamefont {Wesson}}, \bibinfo {author} {\bibfnamefont {P.~J.}\ \bibnamefont {Storey}}, \bibinfo {author} {\bibfnamefont {Y.}~\bibnamefont {Liu}},\ and\ \bibinfo {author} {\bibfnamefont {I.~J.}\ \bibnamefont {Danziger}},\ }\bibfield  {title} {{\selectlanguage {en}\bibinfo {title} {Electron temperatures and densities of planetary nebulae determined from the nebular hydrogen recombination spectrum and temperature and density variations}},\ }\href {https://doi.org/10.1111/j.1365-2966.2004.07838.x} {\bibfield  {journal} {\bibinfo  {journal} {Monthly Notices of the Royal Astronomical Society}\ }\textbf {\bibinfo {volume} {351}},\ \bibinfo {pages} {935} (\bibinfo {year} {2004})}\BibitemShut {NoStop}%
\bibitem [{\citenamefont {Nicholls}\ \emph {et~al.}(2012)\citenamefont {Nicholls}, \citenamefont {Dopita},\ and\ \citenamefont {Sutherland}}]{nicholls_resolving_2012}%
  \BibitemOpen
  \bibfield  {author} {\bibinfo {author} {\bibfnamefont {D.~C.}\ \bibnamefont {Nicholls}}, \bibinfo {author} {\bibfnamefont {M.~A.}\ \bibnamefont {Dopita}},\ and\ \bibinfo {author} {\bibfnamefont {R.~S.}\ \bibnamefont {Sutherland}},\ }\bibfield  {title} {{\selectlanguage {en}\bibinfo {title} {Resolving the {Electron} {Temperature} {Discrepancies} in {H} {Ii} {Regions} and {Planetary} {Nebulae}: {Kappa}-{Distributed} {Electrons}}},\ }\href {https://doi.org/10.1088/0004-637X/752/2/148} {\bibfield  {journal} {\bibinfo  {journal} {The Astrophysical Journal}\ }\textbf {\bibinfo {volume} {752}},\ \bibinfo {pages} {148} (\bibinfo {year} {2012})}\BibitemShut {NoStop}%
\bibitem [{\citenamefont {Zhang}\ \emph {et~al.}(2016)\citenamefont {Zhang}, \citenamefont {Zhang},\ and\ \citenamefont {Liu}}]{zhang_nonthermal_2016}%
  \BibitemOpen
  \bibfield  {author} {\bibinfo {author} {\bibfnamefont {Y.}~\bibnamefont {Zhang}}, \bibinfo {author} {\bibfnamefont {B.}~\bibnamefont {Zhang}},\ and\ \bibinfo {author} {\bibfnamefont {X.-W.}\ \bibnamefont {Liu}},\ }\bibfield  {title} {{\selectlanguage {en}\bibinfo {title} {On the {Nonthermal} {Kappa}-{Distributed} {Electrons} in {Planetary} {Nebulae} and {H} {Ii} {Regions}: {The} {Kappa} {Index} and {Its} {Correlations} with {Other} {Nebular} {Properties}}},\ }\href {https://doi.org/10.3847/0004-637X/817/1/68} {\bibfield  {journal} {\bibinfo  {journal} {The Astrophysical Journal}\ }\textbf {\bibinfo {volume} {817}},\ \bibinfo {pages} {68} (\bibinfo {year} {2016})}\BibitemShut {NoStop}%
\bibitem [{\citenamefont {Yao}\ and\ \citenamefont {Zhang}(2022)}]{yao_ultraviolet_2022}%
  \BibitemOpen
  \bibfield  {author} {\bibinfo {author} {\bibfnamefont {Z.-W.}\ \bibnamefont {Yao}}\ and\ \bibinfo {author} {\bibfnamefont {Y.}~\bibnamefont {Zhang}},\ }\bibfield  {title} {{\selectlanguage {en}\bibinfo {title} {The {Ultraviolet} {C} ii {Lines} as a {Diagnostic} of kappa-distributed {Electrons} in {Planetary} {Nebulae}}},\ }\href {https://doi.org/10.3847/1538-4357/ac8979} {\bibfield  {journal} {\bibinfo  {journal} {The Astrophysical Journal}\ }\textbf {\bibinfo {volume} {936}},\ \bibinfo {pages} {143} (\bibinfo {year} {2022})}\BibitemShut {NoStop}%
\bibitem [{\citenamefont {Sethi}\ \emph {et~al.}(2018)\citenamefont {Sethi}, \citenamefont {Singh},\ and\ \citenamefont {Saini}}]{sethi_effect_2018}%
  \BibitemOpen
  \bibfield  {author} {\bibinfo {author} {\bibfnamefont {P.}~\bibnamefont {Sethi}}, \bibinfo {author} {\bibfnamefont {K.}~\bibnamefont {Singh}},\ and\ \bibinfo {author} {\bibfnamefont {N.}~\bibnamefont {Saini}},\ }\bibfield  {title} {{\selectlanguage {en}\bibinfo {title} {Effect of {Superthermal} {Polarization} {Force} on {Dust} {Acoustic} {Nonlinear} {Structures}}},\ }\href {https://doi.org/10.1515/zna-2018-0079} {\bibfield  {journal} {\bibinfo  {journal} {Zeitschrift für Naturforschung A}\ }\textbf {\bibinfo {volume} {73}},\ \bibinfo {pages} {795} (\bibinfo {year} {2018})}\BibitemShut {NoStop}%
\bibitem [{\citenamefont {Hamaguchi}\ and\ \citenamefont {Farouki}(1994)}]{hamaguchi_polarization_1994}%
  \BibitemOpen
  \bibfield  {author} {\bibinfo {author} {\bibfnamefont {S.}~\bibnamefont {Hamaguchi}}\ and\ \bibinfo {author} {\bibfnamefont {R.~T.}\ \bibnamefont {Farouki}},\ }\bibfield  {title} {{\selectlanguage {en}\bibinfo {title} {Polarization force on a charged particulate in a nonuniform plasma}},\ }\href {https://doi.org/10.1103/PhysRevE.49.4430} {\bibfield  {journal} {\bibinfo  {journal} {Physical Review E}\ }\textbf {\bibinfo {volume} {49}},\ \bibinfo {pages} {4430} (\bibinfo {year} {1994})}\BibitemShut {NoStop}%
\bibitem [{\citenamefont {Prajapati}\ and\ \citenamefont {Bhakta}(2015)}]{prajapati_influence_2015}%
  \BibitemOpen
  \bibfield  {author} {\bibinfo {author} {\bibfnamefont {R.}~\bibnamefont {Prajapati}}\ and\ \bibinfo {author} {\bibfnamefont {S.}~\bibnamefont {Bhakta}},\ }\bibfield  {title} {{\selectlanguage {en}\bibinfo {title} {Influence of dust charge fluctuation and polarization force on radiative condensation instability of magnetized gravitating dusty plasma}},\ }\href {https://doi.org/10.1016/j.physleta.2015.08.007} {\bibfield  {journal} {\bibinfo  {journal} {Physics Letters A}\ }\textbf {\bibinfo {volume} {379}},\ \bibinfo {pages} {2723} (\bibinfo {year} {2015})}\BibitemShut {NoStop}%
\bibitem [{\citenamefont {Khrapak}\ \emph {et~al.}(2009)\citenamefont {Khrapak}, \citenamefont {Ivlev}, \citenamefont {Yaroshenko},\ and\ \citenamefont {Morfill}}]{khrapak_influence_2009}%
  \BibitemOpen
  \bibfield  {author} {\bibinfo {author} {\bibfnamefont {S.~A.}\ \bibnamefont {Khrapak}}, \bibinfo {author} {\bibfnamefont {A.~V.}\ \bibnamefont {Ivlev}}, \bibinfo {author} {\bibfnamefont {V.~V.}\ \bibnamefont {Yaroshenko}},\ and\ \bibinfo {author} {\bibfnamefont {G.~E.}\ \bibnamefont {Morfill}},\ }\bibfield  {title} {{\selectlanguage {en}\bibinfo {title} {Influence of a {Polarization} {Force} on {Dust} {Acoustic} {Waves}}},\ }\href {https://doi.org/10.1103/PhysRevLett.102.245004} {\bibfield  {journal} {\bibinfo  {journal} {Physical Review Letters}\ }\textbf {\bibinfo {volume} {102}},\ \bibinfo {pages} {245004} (\bibinfo {year} {2009})}\BibitemShut {NoStop}%
\bibitem [{\citenamefont {Asaduzzaman}\ and\ \citenamefont {Mamun}(2012)}]{asaduzzaman_effects_2012}%
  \BibitemOpen
  \bibfield  {author} {\bibinfo {author} {\bibfnamefont {M.}~\bibnamefont {Asaduzzaman}}\ and\ \bibinfo {author} {\bibfnamefont {A.~A.}\ \bibnamefont {Mamun}},\ }\bibfield  {title} {{\selectlanguage {en}\bibinfo {title} {Effects of nonthermal ions and polarization force on dust-acoustic waves in a density-varying dusty plasma}},\ }\href {https://doi.org/10.1103/PhysRevE.86.016409} {\bibfield  {journal} {\bibinfo  {journal} {Physical Review E}\ }\textbf {\bibinfo {volume} {86}},\ \bibinfo {pages} {016409} (\bibinfo {year} {2012})}\BibitemShut {NoStop}%
\bibitem [{\citenamefont {Bentabet}\ \emph {et~al.}(2017)\citenamefont {Bentabet}, \citenamefont {Mayout},\ and\ \citenamefont {Tribeche}}]{bentabet_generalized_2017}%
  \BibitemOpen
  \bibfield  {author} {\bibinfo {author} {\bibfnamefont {K.}~\bibnamefont {Bentabet}}, \bibinfo {author} {\bibfnamefont {S.}~\bibnamefont {Mayout}},\ and\ \bibinfo {author} {\bibfnamefont {M.}~\bibnamefont {Tribeche}},\ }\bibfield  {title} {{\selectlanguage {en}\bibinfo {title} {Generalized polarization force acting on dust grains in a dusty plasma}},\ }\href {https://doi.org/10.1016/j.physa.2016.09.055} {\bibfield  {journal} {\bibinfo  {journal} {Physica A: Statistical Mechanics and its Applications}\ }\textbf {\bibinfo {volume} {466}},\ \bibinfo {pages} {492} (\bibinfo {year} {2017})}\BibitemShut {NoStop}%
\bibitem [{\citenamefont {Dolai}\ and\ \citenamefont {Prajapati}(2020)}]{dolai_effects_2020}%
  \BibitemOpen
  \bibfield  {author} {\bibinfo {author} {\bibfnamefont {B.}~\bibnamefont {Dolai}}\ and\ \bibinfo {author} {\bibfnamefont {R.}~\bibnamefont {Prajapati}},\ }\bibfield  {title} {{\selectlanguage {en}\bibinfo {title} {Effects of dust-charge gradient and polarization forces on the waves and {Jeans} instability in strongly coupled dusty plasma}},\ }\href {https://doi.org/10.1016/j.physleta.2020.126462} {\bibfield  {journal} {\bibinfo  {journal} {Physics Letters A}\ }\textbf {\bibinfo {volume} {384}},\ \bibinfo {pages} {126462} (\bibinfo {year} {2020})}\BibitemShut {NoStop}%
\bibitem [{\citenamefont {Asaduzzaman}\ \emph {et~al.}(2011)\citenamefont {Asaduzzaman}, \citenamefont {Mamun},\ and\ \citenamefont {Ashrafi}}]{asaduzzaman_dust-acoustic_2011}%
  \BibitemOpen
  \bibfield  {author} {\bibinfo {author} {\bibfnamefont {M.}~\bibnamefont {Asaduzzaman}}, \bibinfo {author} {\bibfnamefont {A.~A.}\ \bibnamefont {Mamun}},\ and\ \bibinfo {author} {\bibfnamefont {K.~S.}\ \bibnamefont {Ashrafi}},\ }\bibfield  {title} {{\selectlanguage {en}\bibinfo {title} {Dust-acoustic waves in nonuniform dusty plasma in presence of polarization force}},\ }\href {https://doi.org/10.1063/1.3657432} {\bibfield  {journal} {\bibinfo  {journal} {Physics of Plasmas}\ }\textbf {\bibinfo {volume} {18}},\ \bibinfo {pages} {113704} (\bibinfo {year} {2011})}\BibitemShut {NoStop}%
\bibitem [{\citenamefont {El‐Labany}\ \emph {et~al.}(2020)\citenamefont {El‐Labany}, \citenamefont {El‐Taibany}, \citenamefont {El‐Tantawy},\ and\ \citenamefont {Zedan}}]{ellabany_effects_2020}%
  \BibitemOpen
  \bibfield  {author} {\bibinfo {author} {\bibfnamefont {S.}~\bibnamefont {El‐Labany}}, \bibinfo {author} {\bibfnamefont {W.}~\bibnamefont {El‐Taibany}}, \bibinfo {author} {\bibfnamefont {A.}~\bibnamefont {El‐Tantawy}},\ and\ \bibinfo {author} {\bibfnamefont {N.}~\bibnamefont {Zedan}},\ }\bibfield  {title} {{\selectlanguage {en}\bibinfo {title} {Effects of double spectral electron distribution and polarization force on dust acoustic waves in a negative dusty plasma}},\ }\href {https://doi.org/10.1002/ctpp.202000049} {\bibfield  {journal} {\bibinfo  {journal} {Contributions to Plasma Physics}\ }\textbf {\bibinfo {volume} {60}},\ \bibinfo {pages} {e202000049} (\bibinfo {year} {2020})}\BibitemShut {NoStop}%
\bibitem [{\citenamefont {Li}(2024)}]{li_transport-driven_2024}%
  \BibitemOpen
  \bibfield  {author} {\bibinfo {author} {\bibfnamefont {G.-X.}\ \bibnamefont {Li}},\ }\bibfield  {title} {{\selectlanguage {en}\bibinfo {title} {Transport-driven super-{Jeans} fragmentation in dynamical star-forming regions}},\ }\href {https://doi.org/10.1093/mnras/stae384} {\bibfield  {journal} {\bibinfo  {journal} {Monthly Notices of the Royal Astronomical Society}\ }\textbf {\bibinfo {volume} {528}},\ \bibinfo {pages} {7333} (\bibinfo {year} {2024})}\BibitemShut {NoStop}%
\bibitem [{\citenamefont {Khaled}\ \emph {et~al.}(2021)\citenamefont {Khaled}, \citenamefont {Shukri},\ and\ \citenamefont {Al-Shaibani}}]{khaled_modulational_2021}%
  \BibitemOpen
  \bibfield  {author} {\bibinfo {author} {\bibfnamefont {M.~A.~H.}\ \bibnamefont {Khaled}}, \bibinfo {author} {\bibfnamefont {M.~A.}\ \bibnamefont {Shukri}},\ and\ \bibinfo {author} {\bibfnamefont {A.~A.}\ \bibnamefont {Al-Shaibani}},\ }\bibfield  {title} {{\selectlanguage {en}\bibinfo {title} {Modulational {Instability} of {Dust} {Acoustic} {Waves} in an {Opposite} {Polarity} {Dusty} {Plasma} in the {Presence} of {Generalized} {Polarization} {Force} with {Superthermal} {Electrons} and {Ions}}},\ }\href {https://doi.org/10.1007/s13538-021-00920-9} {\bibfield  {journal} {\bibinfo  {journal} {Brazilian Journal of Physics}\ }\textbf {\bibinfo {volume} {51}},\ \bibinfo {pages} {1290} (\bibinfo {year} {2021})}\BibitemShut {NoStop}%
\bibitem [{\citenamefont {Mahmood}\ and\ \citenamefont {Ur-Rehman}(2023)}]{mahmood_existence_2023}%
  \BibitemOpen
  \bibfield  {author} {\bibinfo {author} {\bibfnamefont {S.}~\bibnamefont {Mahmood}}\ and\ \bibinfo {author} {\bibfnamefont {H.}~\bibnamefont {Ur-Rehman}},\ }\bibfield  {title} {{\selectlanguage {en}\bibinfo {title} {Existence and propagation characteristics of ion-acoustic {Kadomtsev}–{Petviashvili} ({KP}) solitons in nonthermal multi-ion plasmas with kappa distributed electrons}},\ }\href {https://doi.org/10.1016/j.chaos.2023.113225} {\bibfield  {journal} {\bibinfo  {journal} {Chaos, Solitons \& Fractals}\ }\textbf {\bibinfo {volume} {169}},\ \bibinfo {pages} {113225} (\bibinfo {year} {2023})}\BibitemShut {NoStop}%
\bibitem [{\citenamefont {Livadiotis}(2015)}]{livadiotis_kappa_2015}%
  \BibitemOpen
  \bibfield  {author} {\bibinfo {author} {\bibfnamefont {G.}~\bibnamefont {Livadiotis}},\ }\bibfield  {title} {{\selectlanguage {en}\bibinfo {title} {Kappa and q {Indices}: {Dependence} on the {Degrees} of {Freedom}}},\ }\href {https://doi.org/10.3390/e17042062} {\bibfield  {journal} {\bibinfo  {journal} {Entropy}\ }\textbf {\bibinfo {volume} {17}},\ \bibinfo {pages} {2062} (\bibinfo {year} {2015})}\BibitemShut {NoStop}%
\bibitem [{\citenamefont {Hakimi~Pajouh}\ and\ \citenamefont {Afshari}(2016)}]{hakimi_pajouh_influence_2016}%
  \BibitemOpen
  \bibfield  {author} {\bibinfo {author} {\bibfnamefont {H.}~\bibnamefont {Hakimi~Pajouh}}\ and\ \bibinfo {author} {\bibfnamefont {N.}~\bibnamefont {Afshari}},\ }\bibfield  {title} {{\selectlanguage {en}\bibinfo {title} {Influence of superthermal plasma particles on the {Jeans} instability in self-gravitating dusty plasmas with dust charge variations}},\ }\href {https://doi.org/10.1016/j.physleta.2016.09.039} {\bibfield  {journal} {\bibinfo  {journal} {Physics Letters A}\ }\textbf {\bibinfo {volume} {380}},\ \bibinfo {pages} {3810} (\bibinfo {year} {2016})}\BibitemShut {NoStop}%
\bibitem [{\citenamefont {Dhiman}\ and\ \citenamefont {Mahajan}(2024)}]{dhiman_radiation_2024}%
  \BibitemOpen
  \bibfield  {author} {\bibinfo {author} {\bibfnamefont {J.~S.}\ \bibnamefont {Dhiman}}\ and\ \bibinfo {author} {\bibfnamefont {M.}~\bibnamefont {Mahajan}},\ }\bibfield  {title} {{\selectlanguage {en}\bibinfo {title} {Radiation pressure and galactic cosmic rays-driven gravitational instability in rotating and magnetized viscoelastic fluids}},\ }\href {https://doi.org/10.1016/j.newast.2024.102251} {\bibfield  {journal} {\bibinfo  {journal} {New Astronomy}\ }\textbf {\bibinfo {volume} {111}},\ \bibinfo {pages} {102251} (\bibinfo {year} {2024})}\BibitemShut {NoStop}%
\bibitem [{\citenamefont {Sharma}\ \emph {et~al.}(2015)\citenamefont {Sharma}, \citenamefont {Argal}, \citenamefont {Tiwari},\ and\ \citenamefont {Prajapati}}]{sharma_jeans_2015}%
  \BibitemOpen
  \bibfield  {author} {\bibinfo {author} {\bibfnamefont {P.~K.}\ \bibnamefont {Sharma}}, \bibinfo {author} {\bibfnamefont {S.}~\bibnamefont {Argal}}, \bibinfo {author} {\bibfnamefont {A.}~\bibnamefont {Tiwari}},\ and\ \bibinfo {author} {\bibfnamefont {R.~P.}\ \bibnamefont {Prajapati}},\ }\bibfield  {title} {{\selectlanguage {en}\bibinfo {title} {Jeans {Instability} of {Rotating} {Viscoelastic} {Fluid} in the {Presence} of {Magnetic} {Field}}},\ }\href {https://doi.org/10.1515/zna-2014-0229} {\bibfield  {journal} {\bibinfo  {journal} {Zeitschrift für Naturforschung A}\ }\textbf {\bibinfo {volume} {70}},\ \bibinfo {pages} {39} (\bibinfo {year} {2015})}\BibitemShut {NoStop}%
\bibitem [{\citenamefont {Falco}\ \emph {et~al.}(2013)\citenamefont {Falco}, \citenamefont {Hansen}, \citenamefont {Wojtak},\ and\ \citenamefont {Mamon}}]{falco_why_2013}%
  \BibitemOpen
  \bibfield  {author} {\bibinfo {author} {\bibfnamefont {M.}~\bibnamefont {Falco}}, \bibinfo {author} {\bibfnamefont {S.~H.}\ \bibnamefont {Hansen}}, \bibinfo {author} {\bibfnamefont {R.}~\bibnamefont {Wojtak}},\ and\ \bibinfo {author} {\bibfnamefont {G.~A.}\ \bibnamefont {Mamon}},\ }\bibfield  {title} {{\selectlanguage {en}\bibinfo {title} {Why does the {Jeans} {Swindle} work?}},\ }\href {https://doi.org/10.1093/mnrasl/sls051} {\bibfield  {journal} {\bibinfo  {journal} {Monthly Notices of the Royal Astronomical Society: Letters}\ }\textbf {\bibinfo {volume} {431}},\ \bibinfo {pages} {L6} (\bibinfo {year} {2013})}\BibitemShut {NoStop}%
\bibitem [{\citenamefont {Kiessling}(2003)}]{kiessling_jeans_2003}%
  \BibitemOpen
  \bibfield  {author} {\bibinfo {author} {\bibfnamefont {M.~K.-H.}\ \bibnamefont {Kiessling}},\ }\bibfield  {title} {{\selectlanguage {en}\bibinfo {title} {The “{Jeans} swindle”}},\ }\href {https://doi.org/10.1016/S0196-8858(02)00556-0} {\bibfield  {journal} {\bibinfo  {journal} {Advances in Applied Mathematics}\ }\textbf {\bibinfo {volume} {31}},\ \bibinfo {pages} {132} (\bibinfo {year} {2003})}\BibitemShut {NoStop}%
\bibitem [{\citenamefont {Kalita}\ and\ \citenamefont {Karmakar}(2021{\natexlab{a}})}]{kalita_jeans_2021}%
  \BibitemOpen
  \bibfield  {author} {\bibinfo {author} {\bibfnamefont {D.}~\bibnamefont {Kalita}}\ and\ \bibinfo {author} {\bibfnamefont {P.~K.}\ \bibnamefont {Karmakar}},\ }\bibfield  {title} {{\selectlanguage {en}\bibinfo {title} {Jeans {Instability} in {Nonideal} {MHD} {Plasma} {Clouds} {With} {Geometric} {Curvature} {Effects}}},\ }\href {https://doi.org/10.1109/TPS.2021.3082811} {\bibfield  {journal} {\bibinfo  {journal} {IEEE Transactions on Plasma Science}\ }\textbf {\bibinfo {volume} {49}},\ \bibinfo {pages} {2042} (\bibinfo {year} {2021}{\natexlab{a}})}\BibitemShut {NoStop}%
\bibitem [{\citenamefont {Irwin}\ and\ \citenamefont {Nelms}(2022)}]{irwin_engineering_2022}%
  \BibitemOpen
  \bibfield  {author} {\bibinfo {author} {\bibfnamefont {J.~D.}\ \bibnamefont {Irwin}}\ and\ \bibinfo {author} {\bibfnamefont {R.~M.}\ \bibnamefont {Nelms}},\ }\href@noop {} {{\selectlanguage {eng}\emph {\bibinfo {title} {Engineering circuit analysis}}}},\ \bibinfo {edition} {12th}\ ed.\ (\bibinfo  {publisher} {John Wiley \& Sons Inc},\ \bibinfo {address} {New York},\ \bibinfo {year} {2022})\BibitemShut {NoStop}%
\bibitem [{\citenamefont {Patil}(2021)}]{hassanien_routh-hurwitz_2021}%
  \BibitemOpen
  \bibfield  {author} {\bibinfo {author} {\bibfnamefont {A.}~\bibnamefont {Patil}},\ }\bibfield  {title} {{\selectlanguage {en}\bibinfo {title} {Routh-{Hurwitz} {Criterion} for {Stability}: {An} {Overview} and {Its} {Implementation} on {Characteristic} {Equation} {Vectors} {Using} {MATLAB}}},\ }in\ \href {https://doi.org/10.1007/978-981-15-9927-9_32} {{\selectlanguage {en}\emph {\bibinfo {booktitle} {Emerging {Technologies} in {Data} {Mining} and {Information} {Security}}}}},\ Vol.\ \bibinfo {volume} {1286},\ \bibinfo {editor} {edited by\ \bibinfo {editor} {\bibfnamefont {A.~E.}\ \bibnamefont {Hassanien}}, \bibinfo {editor} {\bibfnamefont {S.}~\bibnamefont {Bhattacharyya}}, \bibinfo {editor} {\bibfnamefont {S.}~\bibnamefont {Chakrabati}}, \bibinfo {editor} {\bibfnamefont {A.}~\bibnamefont {Bhattacharya}},\ and\ \bibinfo {editor} {\bibfnamefont {S.}~\bibnamefont {Dutta}}}\ (\bibinfo  {publisher} {Springer Singapore},\ \bibinfo {address} {Singapore},\ \bibinfo {year} {2021})\ pp.\ \bibinfo {pages} {319--329},\
  \bibinfo {note} {series Title: Advances in Intelligent Systems and Computing}\BibitemShut {NoStop}%
\bibitem [{\citenamefont {Wood}\ and\ \citenamefont {Churchwell}(1989)}]{wood_morphologies_1989}%
  \BibitemOpen
  \bibfield  {author} {\bibinfo {author} {\bibfnamefont {D.~O.~S.}\ \bibnamefont {Wood}}\ and\ \bibinfo {author} {\bibfnamefont {E.}~\bibnamefont {Churchwell}},\ }\bibfield  {title} {\bibinfo {title} {The {Morphologies} and {Physical} {Properties} of {Ultracompact} {H} {II} {Regions}},\ }\href {https://doi.org/10.1086/191329} {\bibfield  {journal} {\bibinfo  {journal} {The Astrophysical Journal Supplement Series}\ }\textbf {\bibinfo {volume} {69}},\ \bibinfo {pages} {831} (\bibinfo {year} {1989})},\ \bibinfo {note} {publisher: IOP ADS Bibcode: 1989ApJS...69..831W}\BibitemShut {NoStop}%
\bibitem [{\citenamefont {Anderson}\ \emph {et~al.}(2010)\citenamefont {Anderson}, \citenamefont {Zavagno}, \citenamefont {Rodón}, \citenamefont {Russeil}, \citenamefont {Abergel}, \citenamefont {Ade}, \citenamefont {André}, \citenamefont {Arab}, \citenamefont {Baluteau}, \citenamefont {Bernard}, \citenamefont {Blagrave}, \citenamefont {Bontemps}, \citenamefont {Boulanger}, \citenamefont {Cohen}, \citenamefont {Compiègne}, \citenamefont {Cox}, \citenamefont {Dartois}, \citenamefont {Davis}, \citenamefont {Emery}, \citenamefont {Fulton}, \citenamefont {Gry}, \citenamefont {Habart}, \citenamefont {Huang}, \citenamefont {Joblin}, \citenamefont {Jones}, \citenamefont {Kirk}, \citenamefont {Lagache}, \citenamefont {Lim}, \citenamefont {Madden}, \citenamefont {Makiwa}, \citenamefont {Martin}, \citenamefont {Miville-Deschênes}, \citenamefont {Molinari}, \citenamefont {Moseley}, \citenamefont {Motte}, \citenamefont {Naylor}, \citenamefont {Okumura}, \citenamefont {Pinheiro~Gonçalves}, \citenamefont
  {Polehampton}, \citenamefont {Saraceno}, \citenamefont {Sauvage}, \citenamefont {Sidher}, \citenamefont {Spencer}, \citenamefont {Swinyard}, \citenamefont {Ward-Thompson},\ and\ \citenamefont {White}}]{anderson_physical_2010}%
  \BibitemOpen
  \bibfield  {author} {\bibinfo {author} {\bibfnamefont {L.~D.}\ \bibnamefont {Anderson}}, \bibinfo {author} {\bibfnamefont {A.}~\bibnamefont {Zavagno}}, \bibinfo {author} {\bibfnamefont {J.~A.}\ \bibnamefont {Rodón}}, \bibinfo {author} {\bibfnamefont {D.}~\bibnamefont {Russeil}}, \bibinfo {author} {\bibfnamefont {A.}~\bibnamefont {Abergel}}, \bibinfo {author} {\bibfnamefont {P.}~\bibnamefont {Ade}}, \bibinfo {author} {\bibfnamefont {P.}~\bibnamefont {André}}, \bibinfo {author} {\bibfnamefont {H.}~\bibnamefont {Arab}}, \bibinfo {author} {\bibfnamefont {J.-P.}\ \bibnamefont {Baluteau}}, \bibinfo {author} {\bibfnamefont {J.-P.}\ \bibnamefont {Bernard}}, \bibinfo {author} {\bibfnamefont {K.}~\bibnamefont {Blagrave}}, \bibinfo {author} {\bibfnamefont {S.}~\bibnamefont {Bontemps}}, \bibinfo {author} {\bibfnamefont {F.}~\bibnamefont {Boulanger}}, \bibinfo {author} {\bibfnamefont {M.}~\bibnamefont {Cohen}}, \bibinfo {author} {\bibfnamefont {M.}~\bibnamefont {Compiègne}}, \bibinfo {author} {\bibfnamefont
  {P.}~\bibnamefont {Cox}}, \bibinfo {author} {\bibfnamefont {E.}~\bibnamefont {Dartois}}, \bibinfo {author} {\bibfnamefont {G.}~\bibnamefont {Davis}}, \bibinfo {author} {\bibfnamefont {R.}~\bibnamefont {Emery}}, \bibinfo {author} {\bibfnamefont {T.}~\bibnamefont {Fulton}}, \bibinfo {author} {\bibfnamefont {C.}~\bibnamefont {Gry}}, \bibinfo {author} {\bibfnamefont {E.}~\bibnamefont {Habart}}, \bibinfo {author} {\bibfnamefont {M.}~\bibnamefont {Huang}}, \bibinfo {author} {\bibfnamefont {C.}~\bibnamefont {Joblin}}, \bibinfo {author} {\bibfnamefont {S.~C.}\ \bibnamefont {Jones}}, \bibinfo {author} {\bibfnamefont {J.~M.}\ \bibnamefont {Kirk}}, \bibinfo {author} {\bibfnamefont {G.}~\bibnamefont {Lagache}}, \bibinfo {author} {\bibfnamefont {T.}~\bibnamefont {Lim}}, \bibinfo {author} {\bibfnamefont {S.}~\bibnamefont {Madden}}, \bibinfo {author} {\bibfnamefont {G.}~\bibnamefont {Makiwa}}, \bibinfo {author} {\bibfnamefont {P.}~\bibnamefont {Martin}}, \bibinfo {author} {\bibfnamefont {M.-A.}\ \bibnamefont
  {Miville-Deschênes}}, \bibinfo {author} {\bibfnamefont {S.}~\bibnamefont {Molinari}}, \bibinfo {author} {\bibfnamefont {H.}~\bibnamefont {Moseley}}, \bibinfo {author} {\bibfnamefont {F.}~\bibnamefont {Motte}}, \bibinfo {author} {\bibfnamefont {D.~A.}\ \bibnamefont {Naylor}}, \bibinfo {author} {\bibfnamefont {K.}~\bibnamefont {Okumura}}, \bibinfo {author} {\bibfnamefont {D.}~\bibnamefont {Pinheiro~Gonçalves}}, \bibinfo {author} {\bibfnamefont {E.}~\bibnamefont {Polehampton}}, \bibinfo {author} {\bibfnamefont {P.}~\bibnamefont {Saraceno}}, \bibinfo {author} {\bibfnamefont {M.}~\bibnamefont {Sauvage}}, \bibinfo {author} {\bibfnamefont {S.}~\bibnamefont {Sidher}}, \bibinfo {author} {\bibfnamefont {L.}~\bibnamefont {Spencer}}, \bibinfo {author} {\bibfnamefont {B.}~\bibnamefont {Swinyard}}, \bibinfo {author} {\bibfnamefont {D.}~\bibnamefont {Ward-Thompson}},\ and\ \bibinfo {author} {\bibfnamefont {G.~J.}\ \bibnamefont {White}},\ }\bibfield  {title} {\bibinfo {title} {The physical properties of the dust in the
  {RCW} 120 {H} ii region as seen by \textit{{Herschel}}},\ }\href {https://doi.org/10.1051/0004-6361/201014657} {\bibfield  {journal} {\bibinfo  {journal} {Astronomy and Astrophysics}\ }\textbf {\bibinfo {volume} {518}},\ \bibinfo {pages} {L99} (\bibinfo {year} {2010})}\BibitemShut {NoStop}%
\bibitem [{\citenamefont {Shukla}\ and\ \citenamefont {Stenflo}(2006)}]{shukla_jeans_2006}%
  \BibitemOpen
  \bibfield  {author} {\bibinfo {author} {\bibfnamefont {P.~K.}\ \bibnamefont {Shukla}}\ and\ \bibinfo {author} {\bibfnamefont {L.}~\bibnamefont {Stenflo}},\ }\bibfield  {title} {{\selectlanguage {en}\bibinfo {title} {Jeans {Instability} in a {Self}-{Gravitating} {Dusty} {Plasma}}},\ }\href {http://www.jstor.org/stable/20208890} {\bibfield  {journal} {\bibinfo  {journal} {Proceedings: Mathematical, Physical and Engineering Sciences}\ }\textbf {\bibinfo {volume} {462}},\ \bibinfo {pages} {403} (\bibinfo {year} {2006})}\BibitemShut {NoStop}%
\bibitem [{\citenamefont {Kalita}\ and\ \citenamefont {Karmakar}(2021{\natexlab{b}})}]{kalita_adapted_2021}%
  \BibitemOpen
  \bibfield  {author} {\bibinfo {author} {\bibfnamefont {D.}~\bibnamefont {Kalita}}\ and\ \bibinfo {author} {\bibfnamefont {P.~K.}\ \bibnamefont {Karmakar}},\ }\bibfield  {title} {{\selectlanguage {en}\bibinfo {title} {Adapted instabilities excited in spherical magnetized viscoelastic astroclouds with extreme dust-fugacity moderations}},\ }\href {https://doi.org/10.1140/epjp/s13360-021-01479-9} {\bibfield  {journal} {\bibinfo  {journal} {The European Physical Journal Plus}\ }\textbf {\bibinfo {volume} {136}},\ \bibinfo {pages} {479} (\bibinfo {year} {2021}{\natexlab{b}})}\BibitemShut {NoStop}%
\bibitem [{\citenamefont {Karmakar}\ and\ \citenamefont {Dutta}(2017)}]{karmakar_evolutionary_2017}%
  \BibitemOpen
  \bibfield  {author} {\bibinfo {author} {\bibfnamefont {P.~K.}\ \bibnamefont {Karmakar}}\ and\ \bibinfo {author} {\bibfnamefont {P.}~\bibnamefont {Dutta}},\ }\bibfield  {title} {{\selectlanguage {en}\bibinfo {title} {Evolutionary pulsational mode dynamics in nonthermal turbulent viscous astrofluids}},\ }\href {https://doi.org/10.1007/s10509-017-3184-y} {\bibfield  {journal} {\bibinfo  {journal} {Astrophysics and Space Science}\ }\textbf {\bibinfo {volume} {362}},\ \bibinfo {pages} {203} (\bibinfo {year} {2017})}\BibitemShut {NoStop}%
\bibitem [{\citenamefont {Ashrafi}\ \emph {et~al.}(2014)\citenamefont {Ashrafi}, \citenamefont {Mamun},\ and\ \citenamefont {Shukla}}]{ashrafi_polarization_2014}%
  \BibitemOpen
  \bibfield  {author} {\bibinfo {author} {\bibfnamefont {K.~S.}\ \bibnamefont {Ashrafi}}, \bibinfo {author} {\bibfnamefont {A.~A.}\ \bibnamefont {Mamun}},\ and\ \bibinfo {author} {\bibfnamefont {P.~K.}\ \bibnamefont {Shukla}},\ }\bibfield  {title} {{\selectlanguage {en}\bibinfo {title} {Polarization force for different dusty plasma situations}},\ }\href {https://doi.org/10.1017/S0022377813000408} {\bibfield  {journal} {\bibinfo  {journal} {Journal of Plasma Physics}\ }\textbf {\bibinfo {volume} {80}},\ \bibinfo {pages} {1} (\bibinfo {year} {2014})}\BibitemShut {NoStop}%
\end{thebibliography}%

\end{document}